\documentclass[aps, prb, reprint, superscriptaddress, longbibliography, footinbib]{revtex4-2}

\usepackage{amsmath, amssymb, graphicx, xcolor}
\usepackage[colorlinks=true, breaklinks=true, allcolors=blue]{hyperref}
\hypersetup{pdftitle={}, pdfauthor={}, pdfsubject={}, pdfkeywords={}, pdfcreator={}, pdfproducer={}}
\ifdefined\pdfinfoomitdate\pdfinfoomitdate=1\fi
\ifdefined\pdfsuppressptexinfo\pdfsuppressptexinfo=-1\fi
\ifdefined\pdftrailerid\pdftrailerid{}\fi

\newcommand{\Ai}{\operatorname{Ai}}
\newcommand{\Li}{\operatorname{Li}}
\newcommand{\re}{\operatorname{Re}}

\begin{document}
	
	\title{One Measurement, Many Fronts in Kitaev and Cluster Chains}
	\author{Nicholas Musat}
	\email{musat@usc.edu}
	\affiliation{Department of Physics and Astronomy, University of Southern California, Los Angeles, California 90089-0484, USA}
	\author{Stephan Haas}
	\email{shaas@usc.edu}
	\affiliation{Department of Physics and Astronomy, University of Southern California, Los Angeles, California 90089-0484, USA}
	\author{Rishabh Jha}
	\email{rishabh.jha@usc.edu}
	\affiliation{Department of Physics and Astronomy, University of Southern California, Los Angeles, California 90089-0484, USA}
	
	\begin{abstract}
		A projective measurement of one site in the ground state of a quantum chain disturbs the chain in two ways: learning the outcome shifts distant expectation values at once, through ground-state correlations, and the measurement erases the correlations between the measured site and the rest. Both changes travel outward. We solve this problem exactly for the Kitaev chain, the cluster chain in a transverse field, and the Kitaev chain with power-law couplings, and we find that far from the measured site every quadratic observable responds through a front. Where the front is and how fast it broadens depend only on the quasiparticle dispersion. How strong it is depends on two operators: how strongly the measured one creates the fastest quasiparticles and how strongly the observed one detects them; if either factor vanishes, the front is weaker and decays faster. One measurement therefore launches many fronts, one per observable, with a common position and width but different decay laws. In the Kitaev chain the density and the bond energy, the two terms of its nearest-neighbor Hamiltonian, respond to one density measurement with decay laws that trade places between the topological and the trivial phase; in the cluster chain, measuring a symmetry-odd spin instead of a symmetry-even one changes the critical decay law at a fixed Hamiltonian; and with power-law couplings the width law itself changes. All decay laws and profiles are obtained in closed form.
	\end{abstract}
	\maketitle
	
	\textit{Introduction.---}How fast, and in what form, does a local disturbance travel through a quantum many-body system? Short-range interactions give an effective Lieb--Robinson light cone~\cite{LiebRobinson1972, Bravyi2006}, and behind it stands a simple picture: the disturbance creates quasiparticles, each moves at its group velocity, and the fastest ones draw a light cone within which correlations build up~\cite{CalabreseCardy2005, CalabreseCardy2006, AlbaCalabrese2017, Cheneau2012, Jurcevic2014}. But the edge of the cone is not sharp: where the group velocity is largest the rays of many momenta arrive together and form a caustic, which in free-fermion chains after inhomogeneous quenches can be described by Airy functions, broaden as the cube root of time, and have a universal shape~\cite{Hunyadi2004, EislerRacz2013, Viti2016, Eisler2016, Kormos2017, PerfettoGambassi2017}; local quenches and local excitations launch fronts as well~\cite{EislerPeschel2007, StephanDubail2011, EislerMaislinger2020, Yoshinaga2021}.
	
	A projective measurement of one site is a local disturbance of a different kind: the elementary step of measurement-based quantum computation on cluster states~\cite{BriegelRaussendorf2001, RaussendorfBriegel2001, DohertyBartlett2009} and the subject of measurement quenches~\cite{Bayat2018}, of post-measurement entanglement~\cite{Rajabpour2015, NajafiRajabpour2016, Rajabpour2016}, and of monitored dynamics~\cite{LiChenFisher2018, Skinner2019, CaoTilloyDeLuca2019, Alberton2021}. Unlike a unitary kick, a measurement has an outcome, and learning it shifts distant expectation values at once; the measurement also erases the correlations of the measured site with the rest of the chain. What arrives far away, and how strongly, is not known.
	
	Here we answer this question exactly for the Kitaev chain, the cluster chain in a transverse field, and the Kitaev chain with power-law couplings~\cite{Vodola2014, Vodola2016, Lepori2016, Lepori2017jstat, Viyuela2016, VanRegemortel2016, Jager2020, JhaMenzler2026}, a setting in which light cones soften or disappear~\cite{Hauke2013, Eisert2013, Richerme2014, FossFeig2015, Tran2020, Kuwahara2020, Defenu2023}. The quasiparticle dispersion alone fixes where a front is and how fast it broadens; its height is fixed by how strongly the measurement creates the fastest quasiparticles times how strongly the observable detects them. One measurement therefore launches many fronts, one per observable, with a common position and width but different decay laws: in the Kitaev chain the density and bond-energy laws trade places between the topological and the trivial phase, in the cluster chain a bulk symmetry-odd measurement~\footnote{With $P_{\mathrm e}=\prod_{j\,\mathrm{even}}X_j$ and $P_{\mathrm o}=\prod_{j\,\mathrm{odd}}X_j$, the symmetries of Eq.~\eqref{eq:cluster}, an operator $O$ is symmetry-even if $P_{\mathrm e}OP_{\mathrm e}=P_{\mathrm o}OP_{\mathrm o}=O$ and symmetry-odd if $P_{\mathrm e}OP_{\mathrm e}=-O$ or $P_{\mathrm o}OP_{\mathrm o}=-O$. Since $XZX=-Z$ on a site, $X_j$ and $K_j=Z_{j-1}X_jZ_{j+1}$ are even and $Z_j$ is odd. With the Jordan--Wigner Majorana operators $a_j=\Pi_jZ_j$ and $b_j=\Pi_jY_j$, where $\Pi_j=\prod_{l<j}X_l$, each symmetry is the fermion parity of one sublattice, $P_{\mathrm e}=\prod_{j\,\mathrm{even}}ia_jb_j$ and $P_{\mathrm o}=\prod_{j\,\mathrm{odd}}ia_jb_j$, so an even operator contains an even number of Majorana operators of each sublattice, as $X_j=ia_jb_j$ and $K_j=ib_{j-1}a_{j+1}$ do, while $Z_j=\Pi_ja_j$ is a single Majorana operator times the string $\Pi_j$.} changes the critical decay law, and in the long-range chain the width law itself changes at the known free-particle threshold, inverse-square couplings~\cite{Buyskikh2016, Cevolani2016, Tran2020} (Table~\ref{tab:summary}).
	
	\begin{table}[t]
		\caption{Width and decay laws of the fronts launched by one measurement. Upper block: decay of the front height for the measured (left of the arrow) and observed (right) operators, the density $n_j$, bond energy $B_j$, and stabilizer $K_j$, in the nearest-neighbor Kitaev chain [Eq.~\eqref{eq:kitaev}] with $J=\Delta$, $\mu\neq0$, and the cluster chain [Eq.~\eqref{eq:cluster}] with $\lambda>0$; all these fronts broaden as $\ell\propto t^{1/3}$. A subscript $0$ marks a measurement on the first site of a half-line; it shares a row with the bulk measurement because their powers coincide. Lower block: width $\ell(t)$ and decay of the density front, as powers of $\ell$, for the long-range chain [Eq.~\eqref{eq:kitaev}] with $J=\Delta$ and $f_r=r^{-\alpha}$, after a bulk ($n_{j_0}$) and an edge ($n_0$) measurement. Closing rows have the gap closed at the momentum given; in the interior row $0<|k_*|<\pi$, and on the line $\theta_{\mathrm b}(\alpha)$ of the text both decay entries are $\ell^{-4}$. Here $\beta=\alpha-1$, $z_{\mathrm c}=\min(\alpha-1, 3)$, $\nu_{\mathrm c}=\min(\alpha-2, 1)$, and $\alpha_\pi\simeq2.283$; the row $\alpha=2$, $k\to0$ applies when these modes are the fastest.}
		\label{tab:summary}
		\begin{ruledtabular}
			\begin{tabular}{llll}
				Measured $\to$ observed & Topological & Critical & Trivial\\
				$|\mu|/(2J)$ or $\lambda$ & $<1$ & $=1$ & $>1$\\
				\hline
				Kitaev: $n_{j_0}$, $n_0\to n_j$ & $t^{-4/3}$ & $t^{-1}$ & $t^{-2/3}$\\
				Kitaev: $n_{j_0}$, $n_0\to B_j$ & $t^{-2/3}$ & $t^{-1}$ & $t^{-4/3}$\\
				Cluster: $X_{j_0}$, $X_0$, $Z_0\to K_j$ & $t^{-2/3}$ & $t^{-1}$ & $t^{-4/3}$\\
				Cluster: $Z_{j_0}\to K_j$ & $t^{-2/3}$ & $t^{-2/3}$ & $t^{-4/3}$\\
				\hline
				Long-range Kitaev & Width $\ell$ & $n_{j_0}\to n_j$ & $n_0\to n_j$\\
				\hline
				Gapped, $1<\alpha<2$ & $t^{1/\beta}$ & $\ell^{-2}$ & $\ell^{-2-\beta}$\\
				Gapped, $\alpha=2$, $k\to0$ & $\sqrt{t}\log t$ & $\ell^{-2}$ & $\ell^{-3}$\\
				Gapped, $\alpha\geq2$, interior & $t^{1/3}$ & $\ell^{-2}$ & $\ell^{-2}$\\
				Closing at $k=0$, $1<\alpha<2$ & $t^{1/\beta}$ & $\ell^{-2}$ & $\ell^{-2}$\\
				Closing at $k=0$, $\alpha>2$ & $t^{1/z_{\mathrm c}}$ & $\ell^{-2-\nu_{\mathrm c}}$ & $\ell^{-2-\nu_{\mathrm c}}$\\
				Closing at $k=\pi$, $\alpha>\alpha_\pi$ & $t^{1/3}$ & $\ell^{-3}$ & $\ell^{-3}$\\
			\end{tabular}
		\end{ruledtabular}
	\end{table}
	
\begin{figure}[t]
		\centering
		\includegraphics[width=\columnwidth]{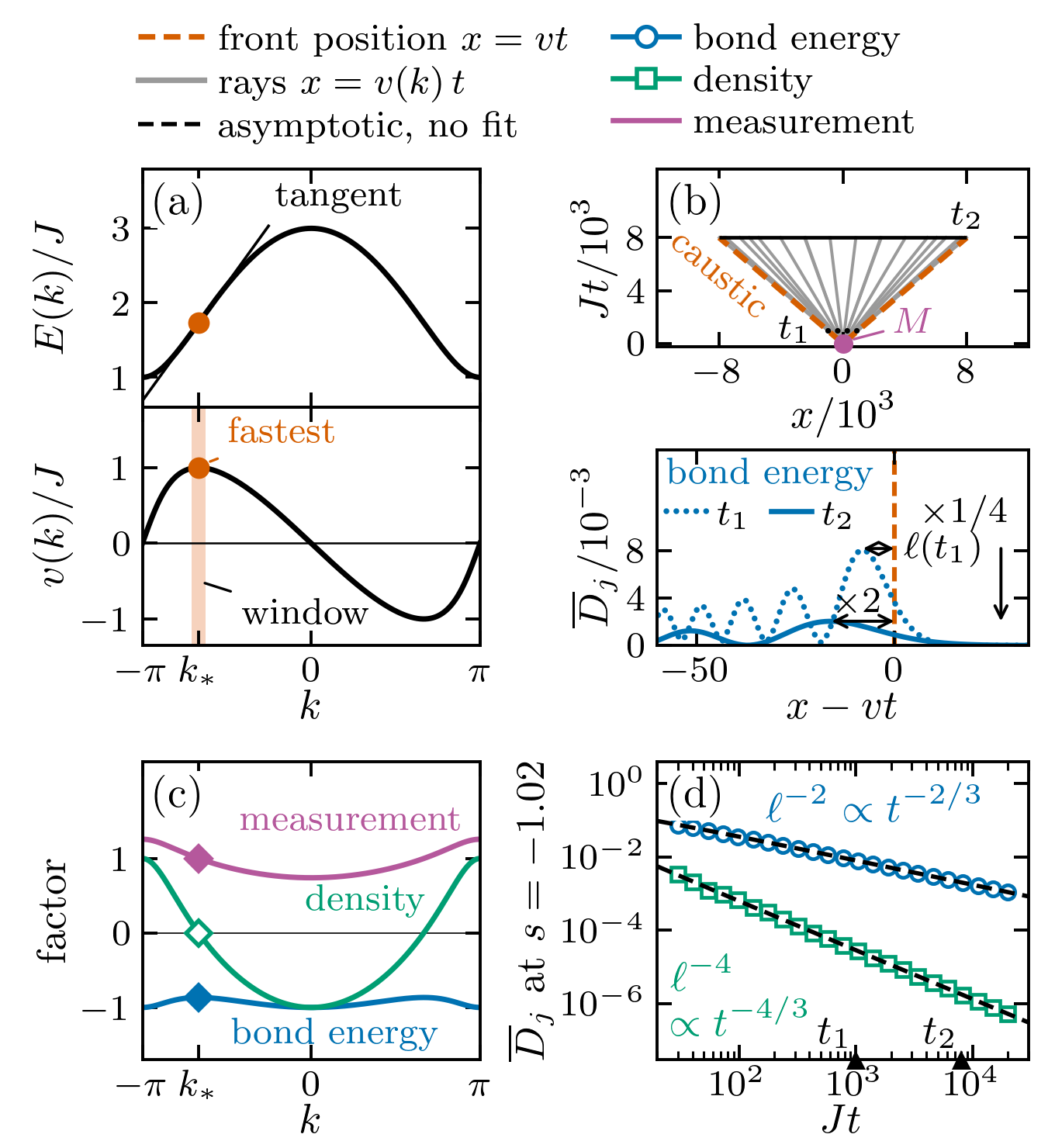}
		\caption{Where a front is and how strong it is, for one density measurement in the topological nearest-neighbor Kitaev chain with $\mu=\Delta=J$. (a) Quasiparticle energy $E(k)$ with its tangent at the fastest momentum $k_*=-2\pi/3$, and group velocity $v(k)=E'(k)$; shading: the window $|k-k_*|\lesssim\ell^{-1}$ at $Jt_1=1000$. (b) Top: rays $x=v(k)t$ for equally spaced $k$ crowd toward the caustics $x=\pm vt$, $v=J$; horizontal lines: times $t_1$, $t_2$; $M$: the measurement. Bottom: the bond-energy front near $x=vt$ at $t_1$ and $t_2=8t_1$; its width $\ell=(Jt/2)^{1/3}$ doubles and its height, $\propto\ell^{-2}$, drops fourfold. (c) Measurement factor, $\propto1-\langle M\rangle_0\re g(k)$, and observable factors $\re g(k)$ of the density and $\re[e^{-ik}g(k)]$ of the bond energy, with $g(k)=-(\mu+2Je^{ik})/E(k)$; diamonds mark the values at $k_*$; the open diamond is the zero of the density factor there. (d) Heights $\overline D_j$ at fixed $s=(x-vt)/\ell=-1.02$, the maximum of $\Ai(s)^2$: $\ell^{-2}\propto t^{-2/3}$ for the bond energy and $\ell^{-4}\propto t^{-4/3}$ for the density, and the derived asymptotes (dashed). All data are exact [Eq.~\eqref{eq:exact}] on a ring of $2^{16}$ sites; $\overline D_j$ is the root mean square of $D_j$ over its three-site period, and $x=j-j_0$ for sites, $j+\frac12-j_0$ for bonds.}
		\label{fig:fronts}
	\end{figure}

	\textit{The response to one measurement.---}Consider a chain of $N$ sites $j=0, \ldots, N-1$, with lattice spacing and $\hbar$ set to one, in its ground state (or the equal mixture of degenerate ground states). At $t=0$ we measure a local Hermitian operator $M$ with $M^2=1$ at site $j_0$, obtaining $\eta=\pm1$, and then let the chain evolve. Outcome $\eta$ occurs with probability $p_\eta=(1+\eta m)/2$, where $m=\langle M\rangle_0$ and the subscript $0$ denotes the initial state. For an observable $O_j$ we record the change conditioned on the outcome, $\delta O_{j, \eta}(t)=\langle O_j(t)\rangle_\eta-\langle O_j\rangle_0$, and summarize the outcomes with $p_\eta>0$ by the root-mean-square response
	\begin{equation}
		D_j(t)=\Bigl[\sum_{\eta}p_\eta\,\delta O_{j, \eta}(t)^2\Bigr]^{1/2}.
		\label{eq:response}
	\end{equation}
	We write $x=j-j_0$. The measured site is either in the bulk, far from any end, or the first site, $j_0=0$, of a half-line; a subscript $0$ on an operator, as in $n_0$, marks the latter. Learning the outcome shifts distant expectation values at once, through the ground-state correlations of $O_j$ with $M$; for observables that commute with $M$ the two shifts average to zero, so no signal is sent, but $D_j$ records them. The measurement also erases the correlations of the measured site with the rest of the chain. The Hamiltonian carries both changes outward, and far from $j_0$ they arrive as a front [Fig.~\ref{fig:fronts}(b)]: a region of appreciable $D_j$ that moves at a speed $v$, broadens to a width $\ell(t)$, has a height that decays as a power of $t$, and keeps a fixed shape in the scaled position $s=(x-vt)/\ell$.
	
	A front raises two questions: where is it, and how strong is it? The measurement creates quasiparticles of every momentum near $j_0$, and the fastest lead. Position and width, we show, depend only on their dispersion; the height depends on how strongly $M$ creates the fastest quasiparticles and how strongly $O_j$ detects them, and if either factor vanishes the front is built from slightly slower quasiparticles, is weaker, and decays faster.
	
	\textit{Exact solution.---}For quadratic Hamiltonians both questions have an exact answer. Write each fermion as two Hermitian Majorana operators, $a_j\equiv\gamma_{2j}=c_j+c_j^\dagger$ and $b_j\equiv\gamma_{2j+1}=i(c_j-c_j^\dagger)$. The ground state is Gaussian~\cite{Bravyi2005}: by Wick's theorem all its correlations follow from the real antisymmetric matrix $\Gamma_{pq}=(i/2)\langle[\gamma_p, \gamma_q]\rangle$. Every observable below is quadratic in the Majorana operators, so its expectation value is a fixed linear function $\mathcal L_j$ of $\Gamma$; for instance $ia_jb_j=1-2n_j$ gives $\langle n_j\rangle=(1-\Gamma_{a_jb_j})/2$. The measurement changes the ground-state matrix $\Gamma_0$ by a matrix $S_\eta$, which the Supplemental Material (SM~\cite{SM}) obtains exactly from the projector $(1+\eta M)/2$, even when the state after the measurement is not Gaussian. The Hamiltonian rotates the Majorana operators, $\gamma(t)=R(t)\gamma(0)$, with a real orthogonal matrix $R(t)$, the propagator, whose element $R_{jl}(t)$ is the amplitude for a Majorana operator to travel from site $l$ to site $j$ in a time $t$. Since the ground state is stationary, $R\Gamma_0R^{\mathrm T}=\Gamma_0$, so at all times
	\begin{equation}
		\delta O_{j, \eta}(t)=\mathcal L_j\bigl[R(t)S_\eta R(t)^{\mathrm T}\bigr].
		\label{eq:exact}
	\end{equation}
	Equation~\eqref{eq:exact} agrees with many-body exact diagonalization of short chains to within machine precision.
	
	\textit{Where the front is.---}On a ring, $R_{jl}(t)$ is a superposition of plane waves $e^{i[k(j-l)-\omega t]}$ with $\omega=\pm E(k)$, where $E(k)\geq0$ is the energy of the Bogoliubov quasiparticle of momentum $k$. At large $t$ only momenta with a stationary phase contribute: a wave packet of momentum $k$ travels along the ray $x=v(k)t$ with the group velocity $v(k)=E'(k)$, and the fastest packets, at $k_*$ with speed $v=v(k_*)$, lead the front [Fig.~\ref{fig:fronts}(a,b)]. Because $v(k)$ is stationary at its maximum, the rays of a whole range of momenta around $k_*$ arrive together near $x=vt$. Such a crowding of rays is a caustic, like the bright edge of a rainbow, for which Sir G.~B.~Airy introduced the function $\Ai$~\cite{Airy1838}. If $E(k)$ departs from its tangent $E(k_*)+v(k-k_*)$ at order $|k-k_*|^z$, waves near $x=vt$ stay in phase as long as $t|k-k_*|^z\lesssim1$, which selects a window of momenta of width $\ell^{-1}$ around $k_*$ with $\ell\propto t^{1/z}$. The window builds a front $\ell$ sites wide, and each propagator, an integral over the window of an integrand of order one, is of order $\ell^{-1}$. At a smooth maximum of $v(k)$, $E''(k_*)=0$ and generically $E'''(k_*)\neq0$, so $z=3$, the width $\ell=[|E'''(k_*)|t/2]^{1/3}$ grows as $t^{1/3}$, and near the front $R_{jl}(t)\simeq\ell^{-1}\Ai(s)$ times a phase (SM~\cite{SM}).
	
	\textit{How strong the front is.---}Equation~\eqref{eq:exact} multiplies two propagators, so the response is a double sum over pairs of waves, one for each Majorana index of $S_\eta$, and near the front both waves lie in the window, which gives the generic height $\ell^{-2}$. Each pair enters with a weight $W(k, k')$ fixed by $S_\eta$ and $\mathcal L_j$, which at leading order, with both waves among the fastest, is the product (SM~\cite{SM}) of a measurement factor, how strongly $M$ creates quasiparticles at $k_*$, and an observable factor, how strongly $O_j$ detects them [diamonds in Fig.~\ref{fig:fronts}(c)]. If this product vanishes, the front is built from pairs whose momenta are offset from $k_*$ by about $\ell^{-1}$, and each power of an offset in the weight costs the height a factor $\ell^{-1}$. Let $\nu$ be the lowest order in $\ell^{-1}$ at which the weight survives the integration over both momenta, counting both the offsets and the small phases from higher terms of $E(k)$. On a half-line the waves are standing waves and each propagator carries their first-site amplitude; if it vanishes as $|k-k_*|^\sigma$, each propagator loses a factor $\ell^{-\sigma}$ ($\sigma=0$ in the bulk). For every $S_\eta$ whose Fourier transform is bounded near $k_*$, and for a single fastest momentum, provided the contribution from two front waves dominates the response, we show (SM~\cite{SM}) that
	\begin{equation}
		\overline D_j(t)\simeq\ell^{-2-\nu-2\sigma}\,\Phi(s), \qquad s=\frac{x-vt}{\ell},
		\label{eq:rule}
	\end{equation}
	where $\Phi(s)$ is a fixed profile with a prefactor set by $M$, $j_0$, and $O_j$, profile and prefactor obtained in closed form in the SM~\cite{SM}, and the overline is a root-mean-square average over lattice-scale oscillations, when present. The height decays as $t^{-(2+\nu+2\sigma)/z}$: $z$ comes from the dispersion, $\nu$ from the two operators, and $\sigma$ from the edge. For every measurement below the measurement factor is nonzero (SM~\cite{SM}), so the observable factor alone decides whether the leading term survives. When it vanishes at $k_*$ inside a gapped phase, the observable is blind to the fastest quasiparticles, and we call the loss of the leading term a cancellation. At a smooth maximum of $v(k)$ a cancellation generically gives $\nu=2$, because the term linear in the offsets is odd under exchange of the two waves of a pair and integrates to zero. At a linear gap closing, $E\simeq v|k-k_*|$, right-moving waves exist on one side of $k_*$ only, the linear term survives, and the observable factor, which vanishes as $|k-k_*|^\nu$, sets $\nu$ directly.
	
	Most protocols below have $z=3$ and $\sigma=0$, and their profiles are $\Ai(s)^2$ for $\nu=0$, $\mathcal Q(s)$ for $\nu=1$, and $\mathcal B(s)$ for $\nu=2$, where
	\begin{equation}
		\mathcal Q(s)=\int_s^\infty\Ai(u)\,du, \qquad \mathcal B(s)=\int_s^\infty\Ai(u)^2\,du.
		\label{eq:shapes}
	\end{equation}
	Both are positive and vanish ahead of the front. Behind it $\mathcal Q$ oscillates about $1$ and $\mathcal B=\Ai'^2-s\Ai^2$ rises like a staircase; $\mathcal B$ also gives the density at the front of free fermions released from a filled half-chain~\cite{EislerRacz2013}.
	
	\begin{figure*}[t]
		\centering
		\includegraphics[width=\textwidth]{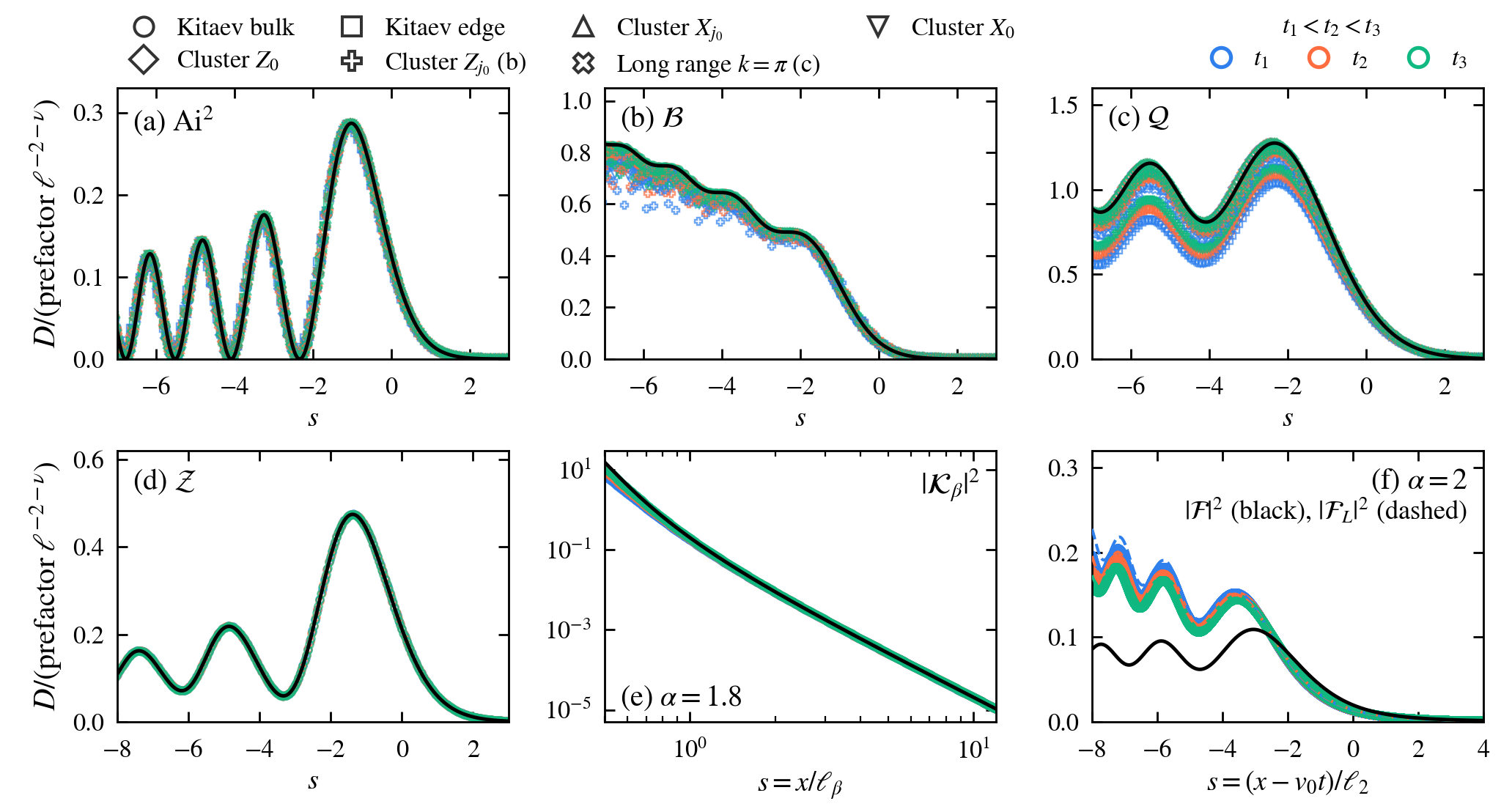}
		\caption{Rescaled fronts against the derived profiles: each curve is divided by its derived height and prefactor and plotted against the derived scaled position, with no fitted parameter; deviations shrink as inverse powers of $\ell$, or of $\log\ell$ at $\alpha=2$ (SM~\cite{SM}). (a)--(c) $\Ai^2$, $\mathcal B$, and $\mathcal Q$ at $\mu/J=3$, $1$, and $2$ for the nearest-neighbor Kitaev density after bulk ($n_{j_0}$) and edge ($n_0$) measurements, and at $\lambda=0.5$, $2$, and $1$ for the cluster protocols $X_{j_0}$, $X_0$, and $Z_0$. Panel (b) adds bulk $Z_{j_0}$ at $\lambda=2$, and (c) the long-range closing at $k=\pi$ for $\alpha=3$. (d) Critical bulk $Z_{j_0}$ against $\mathcal Z(s)$ of Eq.~\eqref{eq:criticalZ}. (e) Bulk profile at $\alpha=1.8$ and $\theta=1$ against $|\mathcal K_\beta|^2$, $\mathcal K_\beta(s)=\int_{-\infty}^\infty\frac{du}{2\pi}e^{ius+i|u|^\beta}$ (SM~\cite{SM}). (f) Modes at $k\to0$ for $\alpha=2$ and $\theta=1$ against $|\mathcal F_L|^2$, the Fresnel kernel $\mathcal F(s)=\int_0^\infty\frac{du}{2\pi}e^{i(us+u^2)}$ with the logarithmic corrections at each $L=\log\ell_2$ (dashed, same colors; SM~\cite{SM}), and its limit $|\mathcal F|^2$ (black), reached only as $\log t\to\infty$. Colors mark three times $(t_1, t_2, t_3)$: (a) $(1600, 6400, 12800)$; (b) the same for the Kitaev chain, $(800, 3200, 6400)$ for $X_{j_0}$, $X_0$, and $Z_0$, and $(150, 300, 500)$ for $Z_{j_0}$; (c) $(1600, 6400, 12800)$ for the Kitaev bulk, $(800, 3200, 6400)$ for $X_{j_0}$, $(6400, 12800, 25600)$ for the edges, of which $n_0$ and $X_0$ lie a correction $O(\ell^{-1})$ below $\mathcal Q$, and $(2000, 8000, 32000)$ at $k=\pi$; (d) $(1600, 3200, 6400)$; (e) $(128, 256, 512)$; (f) $(8000, 16000, 32000)$. In (a)--(c), $J=\Delta=1$. Rings have $N=2^{18}$ sites for the Kitaev chains in (a)--(c), $2^{17}$ for the cluster chains, $2^{22}$ in (e), and $2^{19}$ in (f), and are antiperiodic at gap closings; half-lines use $2^{20}$ standing waves, bulk $Z_{j_0}$ at $\lambda=2$ an open chain with $N=3000$ and $j_0=750$, and (d) the infinite chain. At $\mu=J$, and for bulk $Z_{j_0}$ and edge $X_0$ at $\lambda=2$, $\overline D$ averages over three sites or cells.}
		\label{fig:collapse}
	\end{figure*}
	
	\textit{Kitaev chain: one width, three decay laws.---}The Kitaev chain is the simplest model of a topological superconductor~\cite{Kitaev2001}. With occupations $n_j=c_j^\dagger c_j$, its Hamiltonian is
	\begin{equation}
		\begin{aligned}
			H_{\mathrm K}&=-\sum_{j<l}f_{l-j}\bigl[Jc_j^\dagger c_l+\Delta c_jc_l+\mathrm{H.c.}\bigr]\\
			&\quad-\mu\sum_j\bigl(n_j-\tfrac12\bigr),
		\end{aligned}
		\label{eq:kitaev}
	\end{equation}
	with hopping $J>0$, real pairing $\Delta$, chemical potential $\mu$, and relative coupling $f_r$ at distance $r$. We measure the occupation of site $j_0$, $M=1-2n_{j_0}$ ($\eta=+1$ for an empty, $-1$ for a filled site), and observe the density, $O_j=n_j$, or the bond energy, $O_j=B_j=c_j^\dagger c_{j+1}+c_jc_{j+1}+\mathrm{H.c.}$, the hopping and pairing energy of the bond $(j, j+1)$ in units of $-J$: with nearest neighbors and $J=\Delta$, $H_{\mathrm K}=-J\sum_jB_j-\mu\sum_j(n_j-\tfrac12)$.
	
	For nearest-neighbor couplings ($f_1=1$, $f_{r>1}=0$) and $J=\Delta$, the quasiparticle energy is $E(k)=|\mu+2Je^{ik}|$, the distance from the origin of a point on a circle of radius $2J$ about $\mu$; the chain is topological for $|\mu|<2J$, where the circle encloses the origin, trivial for $|\mu|>2J$, and critical at $|\mu|=2J$. A quasiparticle is a superposition of a particle and a hole, and the density detects it through the particle weight minus the hole weight, $\re g(k)$ with $g(k)=-(\mu+2Je^{ik})/E(k)$ of modulus one; the bond energy detects it through $\re[e^{-ik}g(k)]$. Differentiating $E^2=\mu^2+4J^2+4\mu J\cos k$ gives $v(k)=-2\mu J\sin k/E(k)=\mu\operatorname{Im}g(k)$, so speed and density weight are complementary at every $k$,
	\begin{equation}
		v(k)^2+\mu^2\bigl[\re g(k)\bigr]^2=\mu^2.
		\label{eq:complement}
	\end{equation}
	No quasiparticle is faster than $|\mu|$, and one that reaches this bound, possible only for $|\mu|\leq2J$, carries no density weight. The bond-energy weight obeys Eq.~\eqref{eq:complement} with $2J$ in place of $\mu$ (SM~\cite{SM}): no quasiparticle is faster than $2J$, and one that moves at $2J$ carries no bond-energy weight. Hence for every $\mu\neq0$ the front moves at $v=\min(|\mu|, 2J)$, and since $E(k)$ departs from a straight line at cubic order even at the kink of the gap closing, the width grows as $t^{1/3}$ in all three cases, $\ell=(|\mu|t/2)^{1/3}$, $(Jt/4)^{1/3}$, and $(Jt)^{1/3}$ for $|\mu|<2J$, $|\mu|=2J$, and $|\mu|>2J$. At $\mu=0$ the band is flat and nothing propagates (SM~\cite{SM}).
	
	In the topological phase the fastest quasiparticles reach $v=|\mu|$ and by Eq.~\eqref{eq:complement} carry no density weight: the density is blind to them [Fig.~\ref{fig:fronts}(c)]. The measurement creates them (SM~\cite{SM}), but the $\ell^{-2}$ term of the density front is absent and the term linear in the momentum offsets is odd. At second order the quadratic terms of the weight survive, as does the odd linear term times the quartic phase correction $tE''''(k_*)(k-k_*)^4/24$, itself odd under exchange of the two waves and of order $\ell^{-1}$: $\nu=2$ and $\overline D_j\propto\ell^{-4}\mathcal B(s)$ [Fig.~\ref{fig:collapse}(b)], a $t^{-4/3}$ decay. The bond energy detects them, with an observable factor of modulus $(4J^2-\mu^2)^{1/2}/(2J)$, so its front keeps $\nu=0$ and its height $\ell^{-2}\Ai(s)^2$ exceeds the density front by a factor of order $\ell^2$ [Fig.~\ref{fig:fronts}(d)]. In the trivial phase the fastest quasiparticles move at $v=2J<|\mu|$ and carry density but no bond-energy weight, so the two laws exchange: the density decays as $t^{-2/3}$ with the profile $\Ai^2$ and the bond energy as $t^{-4/3}$ with $\mathcal B$; the observed operator decides which phase decays faster. At $|\mu|=2J$ both weights vanish as $|k-k_*|$ at the closing, so $\nu=1$ for both observables: both fronts are $\ell^{-3}\mathcal Q(s)$, decaying as $t^{-1}$ [Table~\ref{tab:summary}; Figs.~\ref{fig:collapse}(a)--\ref{fig:collapse}(c)]. Ahead of this critical front the density keeps an algebraic tail, $\propto(x^2-v^2t^2)^{-1}$, present from $t=0$ as a ground-state correlation with the outcome; it grows as the front approaches and carries no signal (SM~\cite{SM}).
	
	The cancellation survives at an open end: measuring the first site of a half-line, $n_0$, gives the same three powers for either observable, since the standing waves carry the bulk observable factor and a nonzero first-site amplitude ($\sigma=0$; SM~\cite{SM}). This cancellation belongs to the band structure, not to the phase. For $\Delta\neq J$ the point $\mu+2J\cos k+2i\Delta\sin k$ runs around an ellipse, Eq.~\eqref{eq:complement} fails, the fastest quasiparticle acquires the small density weight $-2(J-\Delta)E(k_*)/(J\mu)$, and the $\ell^{-2}\Ai^2$ term returns and overtakes the $\ell^{-4}\mathcal B$ term once this weight times $\ell^2$ is of order one, at a crossover time $t_\times\propto J^{1/2}|J-\Delta|^{-3/2}$: inside the topological phase the density front decays as $t^{-4/3}$ before $t_\times$ and as $t^{-2/3}$ after it (SM~\cite{SM}).
	
	\textit{Cluster chain: the observed operator and the string set the decay.---} The cluster chain maps onto two Kitaev chains and offers measurements with a Jordan--Wigner string. With Pauli operators $X_j$, $Y_j$, $Z_j$ and the stabilizer $K_j=Z_{j-1}X_jZ_{j+1}$ of the cluster state~\cite{BriegelRaussendorf2001}, the cluster chain in a transverse field is
	\begin{equation}
		H_{\mathrm C}=-\sum_{j=1}^{N-2}K_j-\lambda\sum_{j=0}^{N-1}X_j,
		\label{eq:cluster}
	\end{equation}
	with the field $\lambda\geq0$ in units of the stabilizer coupling. For $\lambda<1$ the chain is in the cluster phase, a symmetry-protected topological phase~\cite{SonAmicoVedral2012, DohertyBartlett2009} whose protecting symmetries are the products of $X_j$ over the even and over the odd sites; for $\lambda>1$ it is a paramagnet; the gap closes at $\lambda=1$~\cite{PachosPlenio2004, ZonzoGiampaolo2018}. We measure the symmetry-even $M=X_{j_0}$ or the symmetry-odd $M=Z_{j_0}$ and observe the stabilizers, $O_j=K_j$.
	
	The Jordan--Wigner Majorana operators $a_j=\Pi_jZ_j$ and $b_j=\Pi_jY_j$, with the string $\Pi_j=\prod_{l<j}X_l$, give $X_j=ia_jb_j$ and $K_j=ib_{j-1}a_{j+1}$~\cite{LSM1961}. Since $K_j$ couples sites two apart, $H_{\mathrm C}$ splits into independent quadratic chains on the even and on the odd sites. Each is the nearest-neighbor Kitaev chain of Eq.~\eqref{eq:kitaev} with $J=\Delta=1$ and $\mu=-2\lambda$, read from right to left, in which $X_j$ plays the role of the local parity $1-2n_j$ and $K_j$ that of the bond energy. The front moves at $v=4\min(\lambda, 1)$ sites per unit time and broadens as $\ell=(8\lambda t)^{1/3}$, $(2t)^{1/3}$, and $(8t)^{1/3}$ for $\lambda<1$, $\lambda=1$, and $\lambda>1$. A measurement of $X_{j_0}$ changes only the chain of $j_0$, so only its stabilizers, those with odd $j-j_0$, respond.
	
	With $K_j$ in the role of the bond energy, the cancellation moves to the paramagnet: by Eq.~\eqref{eq:complement} read in each chain, the fastest quasiparticles of the paramagnet, which move at the bound set by the stabilizer coupling, carry no stabilizer weight. Those of the cluster phase, which move at the bound set by the field, carry no $X_j$ weight but a nonzero stabilizer weight. The stabilizer response to an $X_{j_0}$ measurement therefore follows the laws of the Kitaev bond energy: $t^{-2/3}$, $t^{-1}$, and $t^{-4/3}$ in the cluster phase, at $\lambda=1$, and in the paramagnet [Table~\ref{tab:summary} and Figs.~\ref{fig:collapse}(a)--\ref{fig:collapse}(c)].
	
	A measurement of $Z_{j_0}=\Pi_{j_0}a_{j_0}$ includes the string and acts on both chains. Conjugation by $Z_{j_0}$ reverses the sign of $b_{j_0}$ and of every Majorana operator to its right, so the measurement erases every two-point correlation across this cut and nothing else, and both outcomes give the same stabilizer response (SM~\cite{SM}). In either gapped phase the erased correlations decay exponentially, $S_\eta$ stays localized, and the decay laws are those of $X_{j_0}$: in the cluster phase each chain shows $D_j\simeq4\epsilon\,\ell^{-2}\Ai(s)^2$ with $s=(x-4\lambda t)/\ell$, where $\epsilon$ is the energy the measurement deposits in that chain, and in the paramagnet $\delta K_j$ oscillates on the lattice scale under an $\ell^{-4}\mathcal B$ envelope [Fig.~\ref{fig:collapse}(b)]. A five-site term $-\lambda'\sum_jZ_jX_{j+1}X_{j+2}X_{j+3}Z_{j+4}$, which respects both symmetries and does not close the gap, gives the fastest quasiparticles a stabilizer weight $\propto\lambda'$, so that, as for $J\neq\Delta$, the $t^{-4/3}$ decay turns into $t^{-2/3}$ after $t_\times\propto|\lambda'|^{-3/2}$ (SM~\cite{SM}).
	
	At $\lambda=1$ the erased correlations decay only as the inverse distance, the Fourier transform of $S_\eta$ diverges, and Eq.~\eqref{eq:rule} does not apply. A direct evaluation of Eq.~\eqref{eq:exact} gives instead (SM~\cite{SM})
	\begin{equation}
		D_j(t)\simeq\frac{\mathcal Z(s)}{\pi\ell^2}, \qquad \mathcal Z(s)=\int_s^\infty\mathcal Q(u)[1-\mathcal Q(u)]\,du,
		\label{eq:criticalZ}
	\end{equation}
	with $\ell=(2t)^{1/3}$ and $s=(x-4t)/\ell$; $\mathcal Z$ is positive, with maximum $0.47$ at $s=-1.38$ [Fig.~\ref{fig:collapse}(d)]. The antisymmetry of $S_\eta$ leaves $\ell^{-3}$ instead of $\ell^{-2}$ from the two propagators, each erased correlation contributes its inverse distance $\ell^{-1}$, and the two sums over the erased correlations, each over a range $\ell$, contribute $\ell^2$, in all $\ell^{-2}$. Measuring $Z_{j_0}$ instead of $X_{j_0}$ thus changes the critical decay from $t^{-1}$ to $t^{-2/3}$ at the same Hamiltonian and front width.
	
	At the first site of a half-line the string is empty, so $Z_0=a_0$ is a single Majorana operator, and the measurement, with nothing left of the cut, erases only the correlations of $a_0$; measured there, $X_0$ and $Z_0$ both launch stabilizer fronts with the bulk $X_{j_0}$ laws of Table~\ref{tab:summary} (SM~\cite{SM}). In the cluster phase the edge also keeps a record of the outcome: its zero-energy Majorana mode has weight $1-\lambda^2$ on $a_0$ and does not evolve~\cite{MenzlerJha2026}, so $\langle Z_0(t)\rangle_\eta\to\eta(1-\lambda^2)$ while the stabilizer front leaves (SM~\cite{SM}).
	
	\textit{Long-range couplings change the width.---}Power-law couplings keep the density as the measured and the observed operator but reshape the dispersion. For $f_r=r^{-\alpha}$ with $\alpha>1$ and $J=\Delta$, the quasiparticle energy becomes
	\begin{equation}
		E_\alpha(k)=\left|\mu+2J\Li_\alpha(e^{ik})\right|,
		\label{eq:longdisp}
	\end{equation}
	with the polylogarithm $\Li_\alpha(e^{ik})=\sum_{r\geq1}e^{ikr}/r^\alpha$. The chain is topological for $-2J\zeta(\alpha)<\mu<2J(1-2^{1-\alpha})\zeta(\alpha)$, with $\zeta$ the Riemann zeta function, and the gap closes only at the ends of this interval, at $k=0$ and $k=\pi$ (SM~\cite{SM}); elsewhere we call the chain gapped. The long-range tail makes $\Li_\alpha$ nonanalytic at $k=0$, with a term of order $|k|^{\alpha-1}$ beside the integer powers~\cite{DLMFPolylog}, and this term controls the front whenever it is the leading departure of $E_\alpha$ from a straight line. Examples use $J=\sin\theta$, $\mu=-2\cos\theta$, $0<\theta<\pi$.
	
	For a gapped chain with $1<\alpha<2$, $E_\alpha(k)\simeq E_\alpha(0)+c_\beta|k|^\beta$ with $\beta=\alpha-1<1$, so $z=\beta$: the group velocity, $\propto|k|^{\beta-1}$, diverges as $k\to0$, and there is no finite speed. The waves near $k=0$ still stay in phase for $t|k|^\beta\lesssim1$, so the response spreads over $\ell_\beta=(|c_\beta|t)^{1/\beta}$, which grows faster than $t$, and Eq.~\eqref{eq:rule} holds with $v=0$ and $s=x/\ell_\beta$ [Fig.~\ref{fig:collapse}(e)]. The bulk response falls as $\ell_\beta^{-2}$ with a peakless profile. At an edge the first-site amplitude vanishes as $|k|^{\beta/2}$, so $\sigma=\beta/2$ and the response falls as $\ell_\beta^{-2-\beta}$; its standing waves are those of a fractional Laplacian~\cite{Kwasnicki2011}. Closing the gap at $k=0$ keeps $\ell_\beta\propto t^{1/\beta}$ but gives $\sigma=0$, so bulk and edge both decay as $\ell_\beta^{-2}$.
	
	At $\alpha=2$ the modes at $k\to0$ have the finite speed $v_0=\pi J$, and $E_2$ departs from its tangent by $Bk^2\log^2(1/|k|)$ with $B=2J^2/E_2(0)$. The width $\ell_2$ is the larger solution of
	\begin{equation}
		\frac{\ell_2}{\log\ell_2}=\sqrt{Bt}, \qquad \ell_2\simeq\tfrac12\sqrt{Bt}\log(Bt),
		\label{eq:marginal}
	\end{equation}
	the rescaled phase is quadratic, and the profile is a Fresnel integral, approached only as $\log t\to\infty$ [Fig.~\ref{fig:collapse}(f)]; at an edge $\sigma=1/2$. These modes are the fastest for $\theta>\theta_{\mathrm d}\simeq0.6467$, so the width law switches from $t^{1/3}$ to $\sqrt t\log t$ inside the topological phase; at $\theta_{\mathrm d}$ itself the front needs two scaling variables (SM~\cite{SM}).
	
	For a gapped chain with $\alpha>2$ the speed at $k\to0$ vanishes, an interior maximum leads, and $\ell\propto t^{1/3}$ again. The tail deforms the circle of the nearest-neighbor chain, Eq.~\eqref{eq:complement} fails, the density generically keeps its leading term, and its front decays as $t^{-2/3}$, after a crossover where the density weight is small; the cancellation survives only on a line $\theta_{\mathrm b}(\alpha)$ inside the topological phase, present for $\alpha\gtrsim2.11$, with $\theta_{\mathrm b}\simeq1.082$ at $\alpha=3$ (SM~\cite{SM}).
	
	Closing the gap at $k=0$, with $\mu=-2J\zeta(\alpha)$, gives for $\alpha>2$ a front at the finite speed $v_{\mathrm c}=2J\zeta(\alpha-1)$. For $2<\alpha<4$ the departure of $E_\alpha$ from the line $v_{\mathrm c}|k|$ is of order $|k|^{\alpha-1}$, so $\ell\propto t^{1/(\alpha-1)}$, an exponent that varies continuously from $1$ to $\frac13$; for $\alpha\geq4$, $\ell\propto t^{1/3}$ up to a logarithm at $\alpha=4$. The density weight vanishes as $|k|^{\nu_{\mathrm c}}$ with $\nu_{\mathrm c}=\min(\alpha-2, 1)$, up to a logarithm at $\alpha=3$, so the contribution from two front waves has height $\ell^{-2-\nu_{\mathrm c}}$. For $2<\alpha<3$ a second term of the same order, the product of the right-moving front with the algebraic tail of the left-moving wave, changes the profile but not the power (SM~\cite{SM}).
	
	At the upper end of the interval the gap closes at $k=\pi$, where $\Li_\alpha$ is analytic, $E_\alpha\simeq v_\pi|k-\pi|$ with $v_\pi=2J(1-2^{2-\alpha})\zeta(\alpha-1)$, and the density weight vanishes linearly. These modes are the fastest for $\alpha>\alpha_\pi\simeq2.283$ and reproduce the nearest-neighbor critical front $\ell^{-3}\mathcal Q(s)$ [Fig.~\ref{fig:collapse}(c)]; below $\alpha_\pi$ the gapped rows of Table~\ref{tab:summary} apply, except at an edge for $1<\alpha<2$: the closing leaves the correlation $|\langle ia_jb_0\rangle_0|\simeq2/(\pi j)$, which multiplies the edge front and raises its height from $\ell^{-2-\beta}$ to $\ell^{-2-\beta/2}$ (SM~\cite{SM}).
	
	\textit{Conclusion.---} A single parity or string measurement in the ground state of a free-fermion chain launches, for every quadratic observable, a front whose position and width are fixed by the quasiparticle dispersion and whose height is set by both operators through their weights on the fastest quasiparticles. In the Kitaev chain, one density measurement gives density and bond-energy decay laws that trade places between the phases; a string measurement at the critical cluster point lowers the decay exponent from one to two-thirds at a fixed Hamiltonian; an edge mode keeps a one-site record of the outcome; and power-law couplings change the width law. The quantity to measure is the decay of the height: in the topological Kitaev chain the density and bond-energy fronts at one site differ in height by a factor of order $\ell^2$.
	
	The experimental ingredients exist: a projective readout of one site that leaves the rest of the chain coherent, available as a mid-circuit measurement in neutral-atom arrays~\cite{Deist2022, Graham2023, Lis2023}, trapped ions~\cite{Pino2021}, and superconducting processors~\cite{Koh2023}; evolution of the transverse-field Ising chain, the spin image of the Kitaev chain~\cite{LSM1961, Pfeuty1970}, accessible on ion and Rydberg platforms~\cite{Monroe2021, Browaeys2020} and realized in kicked form on a qubit chain~\cite{Mi2022}; and site-resolved readout of a density or of the single-site factors of a bond or a stabilizer~\cite{Gross2021, Fukuhara2013}, with the runs sorted by the measurement outcome. Two- and three-site Kitaev chains exist in coupled quantum dots~\cite{Dvir2023, tenHaaf2025}, and longer ones would reach the fermionic model itself. Interactions can prolong edge memory after quenches~\cite{HangHaasJha2026}. Two questions remain: whether the height factorization survives interactions and temperature, and how the fronts change under repeated measurements or in periodically driven chains exhibiting temporal entanglement transitions~\cite{GadgePremJha2026} or magic revivals~\cite{jha2026magicentanglementexactrevivals}.
	
	\textit{Acknowledgments.---} R.J. acknowledges partial support, and S.H. acknowledges support, from the U.S. Department of Energy, Office of Science, Office of Advanced Scientific Computing Research, through the Exploratory Research for Extreme Scale Science (EXPRESS) program under Award No. DE-SC0026337. Claude Fable 5.1 was used to create the schematics and aided in writing the code for this paper.

	\bibliography{refs_measurement}
	
\end{document}


\title{Supplemental Material for ``One Measurement, Many Fronts in Kitaev and Cluster Chains''}
	\author{Nicholas Musat}
	\email{musat@usc.edu}
	\affiliation{Department of Physics and Astronomy, University of Southern California, Los Angeles, California 90089-0484, USA}
	\author{Stephan Haas}
	\email{shaas@usc.edu}
	\affiliation{Department of Physics and Astronomy, University of Southern California, Los Angeles, California 90089-0484, USA}
	\author{Rishabh Jha}
	\email{rishabh.jha@usc.edu}
	\affiliation{Department of Physics and Astronomy, University of Southern California, Los Angeles, California 90089-0484, USA}
	
	\maketitle
	
	\tableofcontents
	
	\section{The question}
	\label{sec:question}
	
\begin{figure}[!htb]
		\centering
		\includegraphics[width=\textwidth]{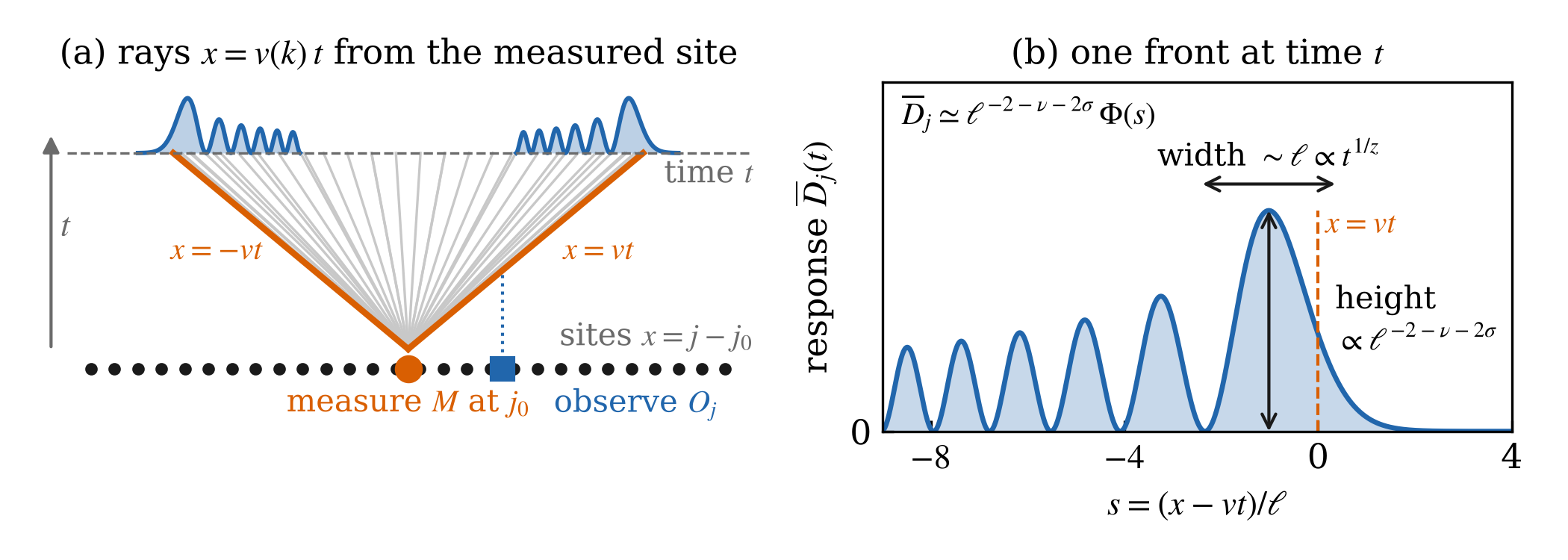}
		\caption{(a) Measuring $M$ at $j_0$ (orange disk) at $t=0$ launches quasiparticles along rays $x=v(k)t$ (gray), one per momentum $k$; the outermost, $x=\pm vt$ (orange), carry the two fronts (blue, at a later time). An observer reads $O_j$ (blue square). (b) One front, $\overline D_j(t)$ against $s$ [Eq.~(3) of the main text], for the bond energy of the running example, Sec.~\ref{sec:example}.}
		\label{fig:question}
	\end{figure}

	A chain of fermions or spins is in its ground state. At $t=0$ we measure a local operator $M$, with $M^2=1$ and outcome $\eta=\pm1$, once at site $j_0$, and let the chain evolve with its own Hamiltonian. The main example is the occupation parity $M=1-2n_{j_0}$, $+1$ on an empty site and $-1$ on a filled one. At distance $x=j-j_0$ we watch another local operator $O_j$ (Fig.~\ref{fig:question}).
	
	The measurement does two things. Learning the outcome changes expectation values everywhere at once, because the ground state already correlates $O_j$ with $M$; nothing travels, we only learn more. It also cuts the measured site's correlations with the rest of the chain. The Hamiltonian carries both changes outward, and far from $j_0$ they arrive as a front [Fig.~\ref{fig:question}(b)], a bump that moves at speed $v$, widens, weakens as a power of $t$, and keeps its shape. Its position and strength follow from Eq.~(3) of the main text,
	\begin{equation}
		\overline D_j(t)\simeq\ell^{-2-\nu-2\sigma}\,\Phi(s), \qquad s=\frac{x-vt}{\ell}, \qquad \ell\propto t^{1/z}.
		\label{eq:rule}
	\end{equation}
	Here $D_j$ is the root-mean-square response over the outcomes, Eq.~\eqref{eq:D}; $\ell$ is the front width and $\Phi(s)$ is the scaled front profile, including its time-independent prefactor. The quasiparticle energy fixes the front speed $v$ and the broadening exponent $z$. The exponent $\nu$ counts the additional inverse powers of $\ell$ contributed by the measurement and observable factors of Sec.~\ref{sec:factor} after momentum integration. At an open end, $\sigma$ is the scaling exponent of the first-site wave amplitude in the momentum offset from the fastest quasiparticles, producing the factor $\ell^{-2\sigma}$; $\sigma=0$ in the bulk. These notes derive Eq.~\eqref{eq:rule}, Table~I, and every other statement of the main text from first principles.
	
	\textit{What was known.} With short-range couplings, commutators of distant operators, and hence signals, are exponentially small outside a light cone set by the Lieb--Robinson velocity~\cite{LiebRobinson1972, Bravyi2006}. In the simplest picture, a local disturbance creates quasiparticles of all momenta, each at its group velocity, and the fastest mark the cone's edge~\cite{CalabreseCardy2005, CalabreseCardy2006, AlbaCalabrese2017, Cheneau2012, Jurcevic2014}. The edge is not sharp: near the largest group velocity many momenta move at almost the same speed and pile up into a caustic, like light at a rainbow's bright edge, for which Airy introduced his function~\cite{Airy1838}. For free fermions released from a domain wall (left half filled, right half empty), the edge density is given by Airy functions, with universal shape and width $\propto t^{1/3}$~\cite{Hunyadi2004, EislerRacz2013, Viti2016, Eisler2016, Kormos2017, PerfettoGambassi2017}. Local quenches (sudden local changes of the Hamiltonian) and single added excitations also launch fronts~\cite{EislerPeschel2007, StephanDubail2011, EislerMaislinger2020, Yoshinaga2021}.
	
	A unitary kick, such as a spin flip, changes the state in one definite way; a measurement has a random outcome on which the later state depends. Single-site measurements are the elementary step of measurement-based quantum computation, which measures the qubits of a cluster state in sequence~\cite{BriegelRaussendorf2001, RaussendorfBriegel2001, DohertyBartlett2009}. Earlier work studied the dynamics after a sudden measurement (measurement quench)~\cite{Bayat2018}, the entanglement left by measuring part of a chain~\cite{Rajabpour2015, NajafiRajabpour2016, Rajabpour2016}, and chains measured repeatedly while they evolve (monitored dynamics)~\cite{LiChenFisher2018, Skinner2019, CaoTilloyDeLuca2019, Alberton2021}. What one measurement sends far away, and how strongly, had not been computed. Couplings decaying as $r^{-\alpha}$ soften or remove the light cone~\cite{Hauke2013, Eisert2013, Richerme2014, FossFeig2015, Tran2020, Kuwahara2020, Defenu2023}; the main text and Sec.~\ref{sec:lr} treat them too.
	
	\textit{Organization.} Sections~\ref{sec:eq} to \ref{sec:lr} follow the experiment in time: the chain at equilibrium, its quasiparticles, their speeds, and what they carry (Sec.~\ref{sec:eq}); the ground state and its correlations (Sec.~\ref{sec:before}); what the measurement does to them, and the exact solution, Eq.~(2) of the main text (Sec.~\ref{sec:measure}); the fronts (Sec.~\ref{sec:fronts}); and the same four steps for long-range couplings, where only $E(k)$ and the states near an open end change (Sec.~\ref{sec:lr}). Sec.~\ref{sec:numerics} explains the numerics and lists each entry of Table~I with the equation deriving it. One example, Fig.~1 of the main text, runs through Secs.~\ref{sec:eq} to \ref{sec:lr}: the Kitaev chain at $J=\Delta=\mu$, measured in its density, observed in its density and bond energy. Every other case modifies it, and we say what changes.
	
	\textit{Conventions.} We set $\hbar=1$ and the lattice spacing to one. As in the main text, a subscript $0$ marks the initial state ($\langle\cdot\rangle_0$, $\Gamma_0$), the first site of a half-line ($n_0$, $X_0$, $Z_0$), or the band edge $k\to0$ ($v_0$). A bar is complex conjugation, but over $D$ the local average of Sec.~\ref{sec:pairs}. The letter $l$ is a site, $\ell$ a front width. 
	
	\section{Equilibrium, the chains and their quasiparticles}
	\label{sec:eq}
	
	Before the measurement the chain is in its ground state, whose quasiparticles have definite momentum and energy; their speeds and what they carry fix where every front is and which observables see the fastest.
	
	\subsection{Majorana operators}
	\label{sec:majorana}
	
	Let $c_j$ annihilate a spinless fermion at site $j$, with $\{c_j, c_l^\dagger\}=\delta_{jl}$, $\{c_j, c_l\}=0$, and $n_j=c_j^\dagger c_j$; split each into two Hermitian Majorana operators,
	\begin{equation}
		a_j=c_j+c_j^\dagger, \qquad b_j=i(c_j-c_j^\dagger), \qquad c_j=\frac{a_j-ib_j}2,
		\label{eq:majorana}
	\end{equation}
	which square to one, $a_j^2=b_j^2=1$, and anticommute when different: with $\gamma_{2j}=a_j$ and $\gamma_{2j+1}=b_j$ as in the main text, $\{\gamma_p, \gamma_q\}=2\delta_{pq}$. Multiplying out,
	\begin{equation}
		a_jb_j=i(c_j+c_j^\dagger)(c_j-c_j^\dagger)=i(c_j^\dagger c_j-c_jc_j^\dagger)=i(2n_j-1), \qquad ia_jb_j=1-2n_j,
		\label{eq:parity}
	\end{equation}
	so $ia_jb_j$ is the occupation parity, $\pm1$ for an empty or filled site. In the one-site basis $(|0\rangle, |1\rangle)$,
	\begin{equation}
		c=\begin{pmatrix}0&1\\0&0\end{pmatrix}, \qquad a=\begin{pmatrix}0&1\\1&0\end{pmatrix}, \qquad b=\begin{pmatrix}0&i\\-i&0\end{pmatrix}, \qquad iab=\begin{pmatrix}1&0\\0&-1\end{pmatrix}=1-2n.
		\label{eq:onesite}
	\end{equation}
	So $a$, $-b$, and $iab$ are the Pauli matrices of this basis (Sec.~\ref{sec:cluster} labels spins differently, with $iab=X$). Majoranas pay off once the Hamiltonian enters.
	
	\subsection{Quadratic Hamiltonians rotate Majoranas}
	\label{sec:rotate}
	
	Every Hamiltonian here is quadratic in the fermions, hence in the Majoranas,
	\begin{equation}
		H=\frac i4\sum_{p, q}A_{pq}\gamma_p\gamma_q, \qquad A^{\mathrm T}=-A, \qquad A_{pq}\ \text{real},
		\label{eq:Hmaj}
	\end{equation}
	up to a constant, with $A$ of size $2N\times2N$; $H$ is Hermitian as $(i\gamma_p\gamma_q)^\dagger=-i\gamma_q\gamma_p=i\gamma_p\gamma_q$ for $p\neq q$. The Heisenberg equation $d\gamma_p/dt=i[H, \gamma_p]$ needs only $[\gamma_q\gamma_r, \gamma_p]=\gamma_q\{\gamma_r, \gamma_p\}-\{\gamma_q, \gamma_p\}\gamma_r=2\delta_{rp}\gamma_q-2\delta_{qp}\gamma_r$, and with $A_{qp}=-A_{pq}$ gives
	\begin{equation}
		\frac{d\gamma_p}{dt}=\sum_qA_{pq}\gamma_q, \qquad \gamma(t)=R(t)\gamma(0), \qquad R(t)=e^{At},
		\label{eq:R}
	\end{equation}
	with $\gamma$ the column of Majoranas. As $A$ is real antisymmetric, $R(t)$ is real orthogonal, $RR^{\mathrm T}=1$: the Majoranas rotate into each other, and the main text's propagator $R_{pq}(t)$ is the amplitude for $\gamma_q$ to reach $\gamma_p$ in time $t$. Instead of $2^N$ amplitudes we rotate $2N$ operators.
	
\begin{figure}[!htb]
		\centering
		\includegraphics[width=\textwidth]{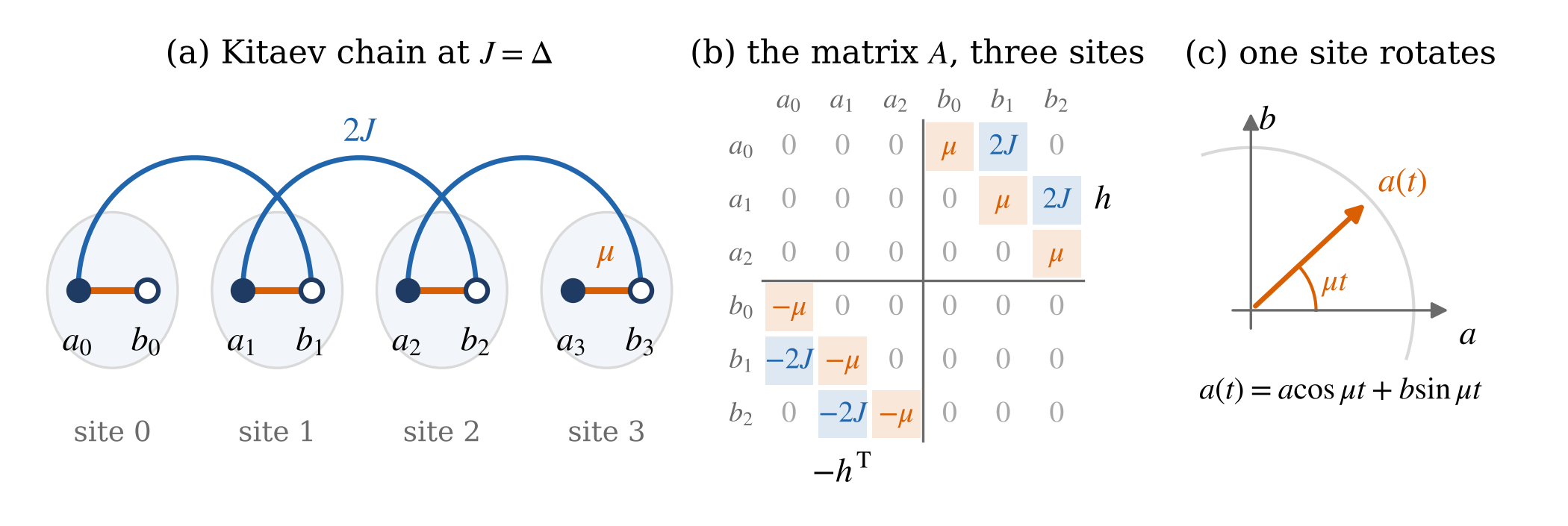}
		\caption{Rotating Majoranas, nearest-neighbor Kitaev chain at $J=\Delta$. (a) $a_j$ (filled) and $b_j$ (open), coupled by $\mu$ on site (orange) and by $2J$ from $a_j$ to $b_{j+1}$ (blue), Eq.~\eqref{eq:HKmaj}. (b) $A$ of Eq.~\eqref{eq:Hmaj} for three sites, Eq.~\eqref{eq:A3}. (c) One site: $a$ and $b$ rotate by $\mu t$, Eq.~\eqref{eq:onesiterot}.}
		\label{fig:rotation}
	\end{figure}

	\textit{Example: one site.} For $H=-\mu(n-\frac12)$, Eq.~\eqref{eq:parity} gives $H=\frac\mu2iab$, so $A_{ab}=\mu=-A_{ba}$, and Eq.~\eqref{eq:R} reads $\dot a=\mu b$, $\dot b=-\mu a$, solved by
	\begin{equation}
		a(t)=a\cos\mu t+b\sin\mu t, \qquad b(t)=b\cos\mu t-a\sin\mu t,
		\label{eq:onesiterot}
	\end{equation}
	a rotation by $\mu t$ in the $(a, b)$ plane [Fig.~\ref{fig:rotation}(c)], or $c(t)=e^{i\mu t}c$: each fermion mode rotates at a frequency equal to its energy, and every $R(t)$ here is built from such rotations.
	
	\subsection{The Kitaev chain}
	\label{sec:kitaev}
	
	The Kitaev chain, Eq.~(5) of the main text, is
	\begin{equation}
		H_{\mathrm K}=-\sum_{j<l}f_{l-j}\bigl[Jc_j^\dagger c_l+\Delta c_jc_l+\mathrm{H.c.}\bigr]-\mu\sum_j\bigl(n_j-\tfrac12\bigr),
		\label{eq:HK}
	\end{equation}
	with hopping $J>0$, real pairing $\Delta$, chemical potential $\mu$, and relative coupling $f_r$ at distance $r$. Pairing removes two fermions at once, conserving only particle-number parity: the simplest superconductor~\cite{Kitaev2001}. With Eq.~\eqref{eq:majorana}, the $aa$ and $bb$ products cancel and, for $j\neq l$,
	\begin{equation}
		c_j^\dagger c_l+c_l^\dagger c_j=\frac i2(b_ja_l-a_jb_l), \qquad c_jc_l+c_l^\dagger c_j^\dagger=-\frac i2(a_jb_l+b_ja_l).
		\label{eq:bondmaj}
	\end{equation}
	At $J=\Delta$ the $b_ja_l$ terms of hopping and pairing cancel, each $a$ couples only to $b$'s on its right, and for nearest neighbors ($f_1=1$, $f_{r>1}=0$),
	\begin{equation}
		H_{\mathrm K}=iJ\sum_ja_jb_{j+1}+\frac{i\mu}2\sum_ja_jb_j=-J\sum_jB_j-\mu\sum_j\bigl(n_j-\tfrac12\bigr), \qquad B_j=-ia_jb_{j+1}=c_j^\dagger c_{j+1}+c_jc_{j+1}+\mathrm{H.c.}
		\label{eq:HKmaj}
	\end{equation}
	Here $B_j$, the main text's bond energy, is the hopping and pairing energy of bond $(j, j+1)$ in units of $-J$ [Fig.~\ref{fig:rotation}(a)]. In the form of Eq.~\eqref{eq:Hmaj}, ordered $(a_0, a_1, a_2; b_0, b_1, b_2)$, three sites give
	\begin{equation}
		A=\begin{pmatrix}0&h\\-h^{\mathrm T}&0\end{pmatrix}, \qquad h=\begin{pmatrix}\mu&2J&0\\0&\mu&2J\\0&0&\mu\end{pmatrix},
		\label{eq:A3}
	\end{equation}
	and in general $H_{\mathrm K}=\frac i2\sum_{j, l}h_{jl}a_jb_l$ with $h_{jj}=\mu$ and $h_{j, j+r}=2Jf_r$ at $J=\Delta$. At $\mu=0$ each $a_j$ pairs with $b_{j+1}$, leaving $b_0$ (left end) and $a_{N-1}$ (right end) unpaired, with vanishing rows and columns in Eq.~\eqref{eq:A3}. These zero modes cost no energy, make the chain topological, and survive, smeared over a few sites, while $|\mu|<2J$ (Sec.~\ref{sec:edgemode}).
	
	\subsection{Plane waves: the energy as a distance}
	\label{sec:circle}
	
\begin{figure}[!htb]
		\centering
		\includegraphics[width=\textwidth]{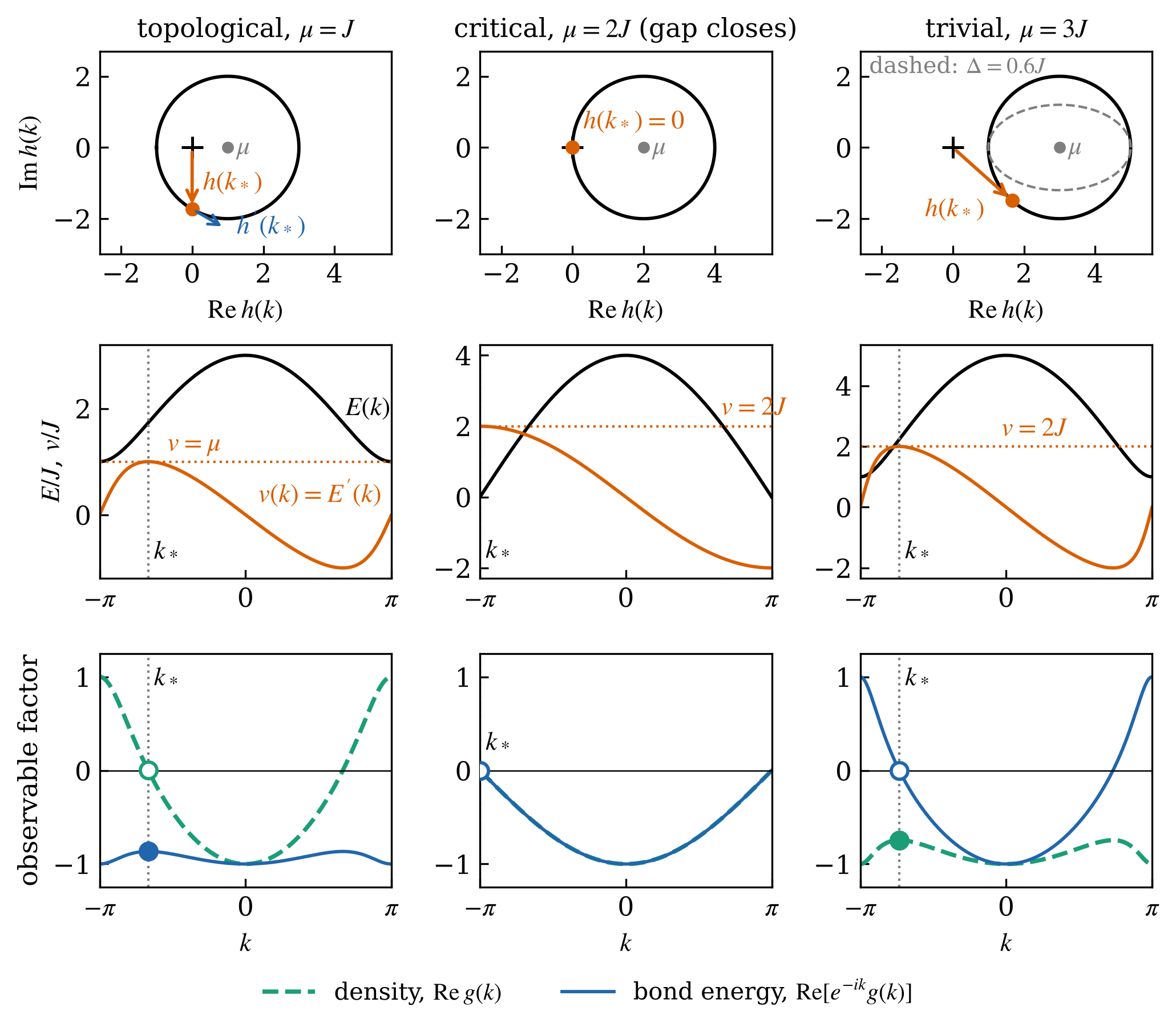}
		\caption{Kitaev chain at $J=\Delta$, three phases. Top: circle $h(k)=\mu+2Je^{ik}$, Eq.~\eqref{eq:hk}, at distance $E(k)$ from the origin ($+$), Eq.~\eqref{eq:EK}; orange arrow, $h(k_*)$ at the fastest momentum, Eq.~\eqref{eq:kstar}; dashed ellipse, $\Delta=0.6J$ (trivial). Middle: $E(k)$ and speed $v(k)=E'(k)$, maximal at $k_*$ (dotted). Bottom: density and bond-energy weights $\re g(k)$ and $\re[e^{-ik}g(k)]$, Eqs.~\eqref{eq:uv} and \eqref{eq:complement}; open circles mark factors vanishing at $k_*$.}
		\label{fig:circle}
	\end{figure}

	On a ring of $N$ sites $h_{jl}$ depends only on $l-j$, so plane waves $e^{ikj}$ diagonalize it,
	\begin{equation}
		\sum_lh_{jl}e^{ikl}=h(k)e^{ikj}, \qquad h(k)=\mu+2J\sum_{r\geq1}f_r\cos kr+2i\Delta\sum_{r\geq1}f_r\sin kr,
		\label{eq:hk}
	\end{equation}
	with $h(k)=\mu+2Je^{ik}$ for nearest neighbors at $J=\Delta$. Per momentum, $A$ of Eq.~\eqref{eq:Hmaj} reduces to a $2\times2$ block coupling the $a$ and $b$ waves,
	\begin{equation}
		A(k)=\begin{pmatrix}0&h(k)\\-\overline{h(k)}&0\end{pmatrix}, \qquad \text{eigenvalues}\ \pm iE(k), \qquad E(k)=|h(k)|.
		\label{eq:Ak}
	\end{equation}
	Majorana waves thus oscillate at $\pm E(k)$, with $E(k)\geq0$ the quasiparticle energy. The nearest-neighbor $h(k)$ runs on a circle of radius $2J$ about $\mu$, and $E$ is its distance from the origin,
	\begin{equation}
		E(k)=|\mu+2Je^{ik}|, \qquad E(k)^2=\mu^2+4J^2+4\mu J\cos k.
		\label{eq:EK}
	\end{equation}
	If the circle (Fig.~\ref{fig:circle}, top) encloses the origin, $|\mu|<2J$, the chain is topological; if not, $|\mu|>2J$, trivial; if it passes through the origin, $|\mu|=2J$, critical, with the gap closed at one momentum. For $\Delta\neq J$ it becomes an ellipse (dashed in Fig.~\ref{fig:circle}; Sec.~\ref{sec:whatif}). Rotating $h(k)$ by $\pi$ and normalizing gives the phase
	\begin{equation}
		g(k)=-\frac{h(k)}{E(k)}=-\frac{\mu+2Je^{ik}}{E(k)}, \qquad |g(k)|=1,
		\label{eq:g}
	\end{equation}
	used throughout the main text; it fixes every ground-state correlation (Sec.~\ref{sec:before}).
	
	\subsection{What a quasiparticle carries}
	\label{sec:carry}
	
\begin{figure}[!htb]
		\centering
		\includegraphics[width=\textwidth]{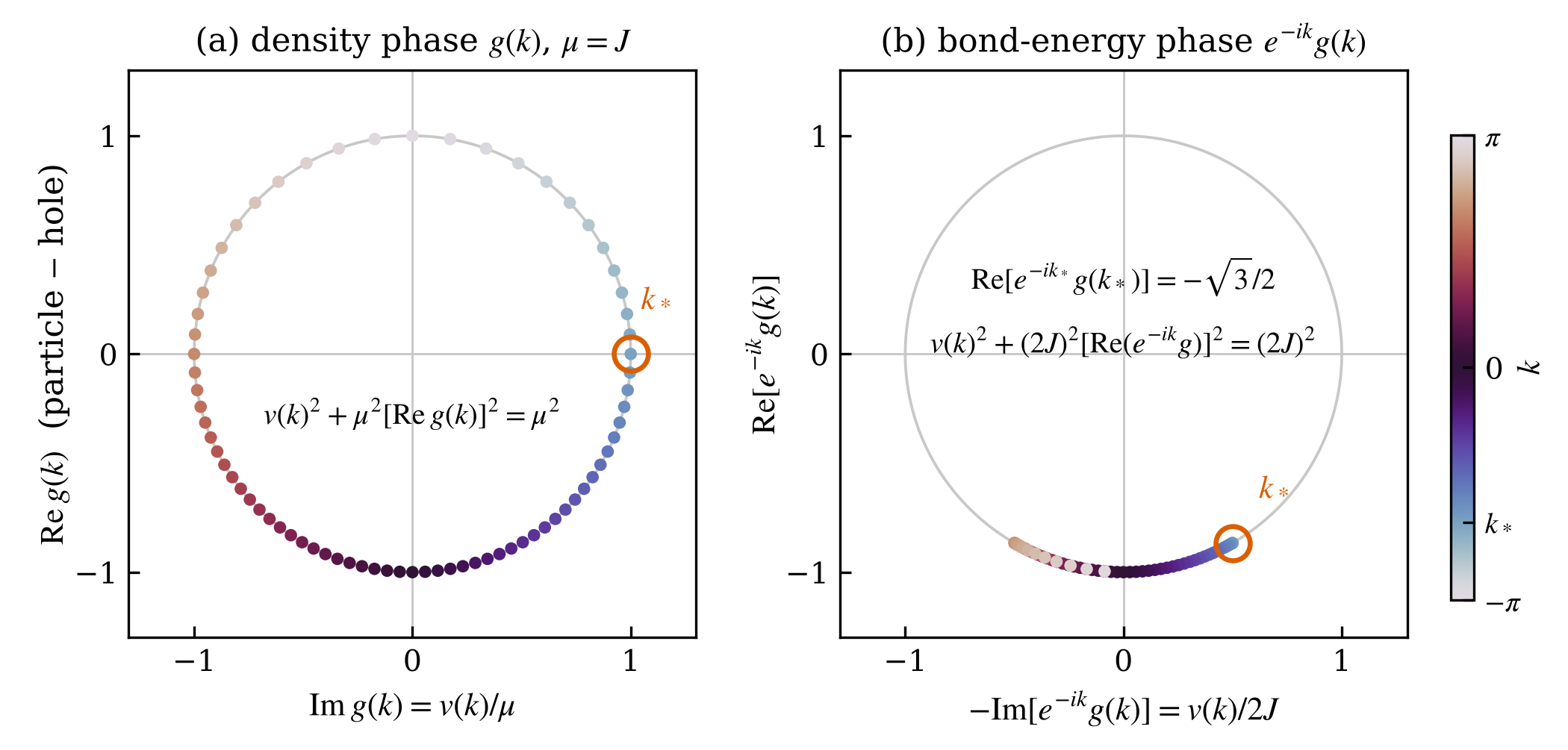}
		\caption{Speed and weight of every quasiparticle, Kitaev chain at $J=\Delta=\mu$. (a) Unit spin direction $(\im g, \re g)$ of Eq.~\eqref{eq:bdg}, one dot per momentum: vertical, density weight, Eq.~\eqref{eq:uv}; horizontal, speed in units of $\mu$, Eq.~\eqref{eq:vmu}; orange ring, the fastest, $k_*=-2\pi/3$, with no density weight, Eq.~\eqref{eq:complement}. (b) The same for the bond-energy phase $e^{-ik}g(k)$, speed in units of $2J$, at most $\frac12$: the fastest keeps bond weight $-\sqrt3/2$.}
		\label{fig:quasi}
	\end{figure}

	A quasiparticle of a superconductor is part particle, part hole. With $c_k=N^{-1/2}\sum_je^{-ikj}c_j$, pairing couples the particle at $k$ only to the hole at $-k$, so each $k$ gives a $2\times2$ matrix on $(c_k, c_{-k}^\dagger)$,
	\begin{equation}
		H_{\mathrm K}=\frac12\sum_k\begin{pmatrix}c_k^\dagger&c_{-k}\end{pmatrix}\begin{pmatrix}-\re h(k)&i\im h(k)\\-i\im h(k)&\re h(k)\end{pmatrix}\begin{pmatrix}c_k\\c_{-k}^\dagger\end{pmatrix}+\text{const}.
		\label{eq:bdg}
	\end{equation}
	This matrix, $-\re h\,\sigma_z-\im h\,\sigma_y$ with Pauli matrices on particle and hole, is a spin one-half in a field $E(k)$ along the unit vector $\bigl(0, -\im h/E, -\re h/E\bigr)=\bigl(0, \im g, \re g\bigr)$, with eigenvalues $\pm E(k)$. The quasiparticle of energy $+E(k)$, spin along the field, is $u_k|\text{particle}\rangle+v_k|\text{hole}\rangle$, and its $z$ component is particle minus hole weight:
	\begin{equation}
		|u_k|^2-|v_k|^2=\langle\sigma_z\rangle=\re g(k).
		\label{eq:uv}
	\end{equation}
	It is a pure particle or hole for $\re g=\pm1$, an equal mixture for $\re g=0$. The density, counting particles minus holes, sees it with weight $\re g(k)$, the vertical axis in Fig.~\ref{fig:quasi}(a), and is blind when $\re g=0$: observables detect quasiparticles with weights that can vanish, the main text's first idea.
	
	\subsection{How fast: speed and weight are complementary}
	\label{sec:speed}
	
\begin{figure}[!htb]
		\centering
		\includegraphics[width=\textwidth]{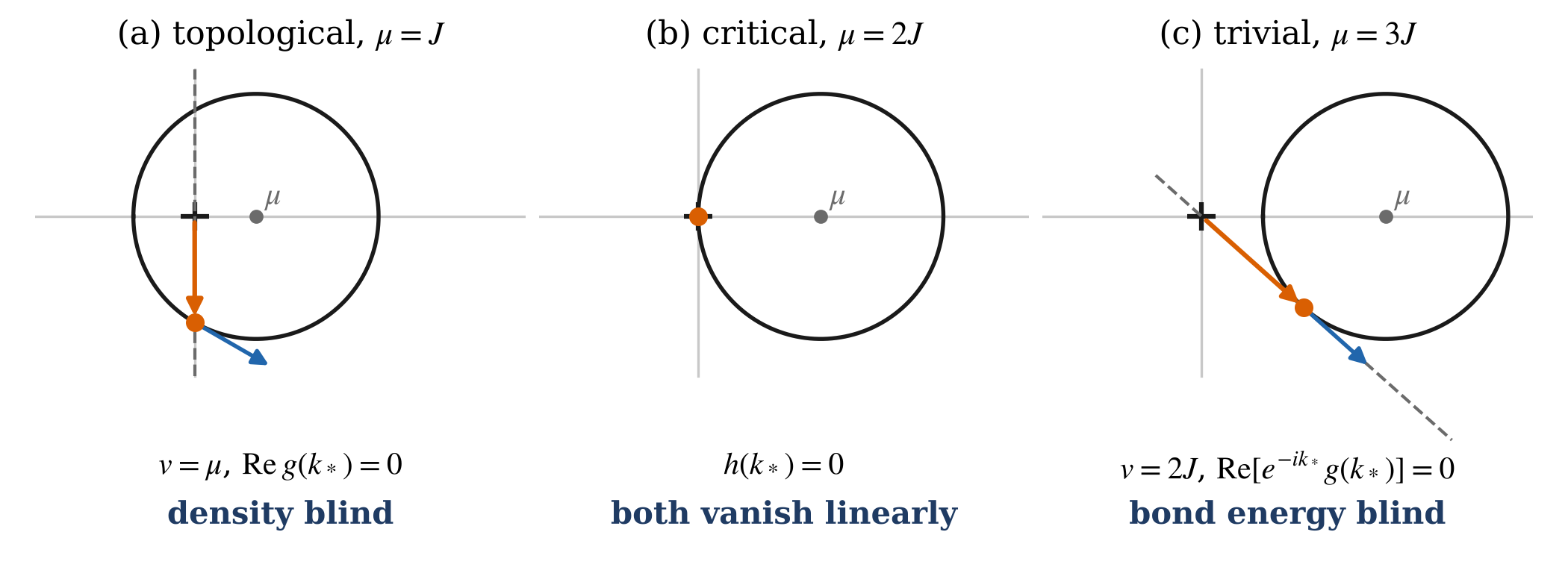}
		\caption{The geometry of Eq.~\eqref{eq:complement}, Kitaev chain at $J=\Delta=1$: circle $h(k)$, orange arrow of length $E(k_*)$, blue tangent $h'(k_*)$. (a) Topological, $\mu=J$: top speed $v=\mu$ on the imaginary axis (dashed). (b) Critical, $\mu=2J$: the circle meets the origin, closing the gap at $k_*=\pi$. (c) Trivial, $\mu=3J$: top speed $v=2J$ where a line from the origin touches the circle, so $\re[e^{-ik_*}g(k_*)]=0$.}
		\label{fig:circles}
	\end{figure}

	A wave packet moves at the group velocity $v(k)=E'(k)$. Differentiating $E^2$ of Eq.~\eqref{eq:EK} gives $EE'=-2\mu J\sin k$, and with $\im g=-2J\sin k/E$ from Eq.~\eqref{eq:g},
	\begin{equation}
		v(k)=-\frac{2\mu J\sin k}{E(k)}=\mu\im g(k).
		\label{eq:vmu}
	\end{equation}
	This is $\mu$ times the horizontal component in Fig.~\ref{fig:quasi}(a); since $|g|=1$, speed and density weight cannot both be large:
	\begin{equation}
		v(k)^2+\mu^2\bigl[\re g(k)\bigr]^2=\mu^2, \qquad v(k)^2+(2J)^2\bigl[\re\bigl(e^{-ik}g(k)\bigr)\bigr]^2=(2J)^2.
		\label{eq:complement}
	\end{equation}
	The first is Eq.~(6) of the main text; the second follows as the bond energy sees a quasiparticle through $e^{-ik}g(k)=-(\mu e^{-ik}+2J)/E$ (Sec.~\ref{sec:factor}), of imaginary part $\mu\sin k/E$, so $v=-2J\im(e^{-ik}g)$. On the circle (Fig.~\ref{fig:circles}), the tangent $h'(k)=2iJe^{ik}=i(h-\mu)$ splits into $ih$, perpendicular to $h$, which only turns it, and $-i\mu$, which alone changes $E=|h|$. So the speed peaks where the circle crosses the imaginary axis, if it does, with $h$ imaginary and $\re g=0$.
	
	Three consequences follow. First, no quasiparticle is faster than $|\mu|$ or $2J$, and the smaller bound is reached: the fastest, at momentum $k_*$, move with
	\begin{equation}
		v=\min(|\mu|, 2J), \qquad \cos k_*=\begin{cases}-\mu/(2J), &|\mu|<2J, \\-2J/\mu, &|\mu|>2J, \end{cases} \qquad E(k_*)=\sqrt{|4J^2-\mu^2|}, \qquad E'''(k_*)=-v,
		\label{eq:kstar}
	\end{equation}
	with $\sin k_*<0$ for right movers. Taking $\mu>0$ costs nothing: the map $c_j\to(-1)^jc_j^\dagger$ sends $\mu$ to $-\mu$ and $k$ to $k+\pi$, keeps $B_j$, and turns $n_j$ into $1-n_j$, which swaps density outcomes and flips the sign of $\delta n_j$, leaving the response $D_j$ unchanged; at the closing we use $\mu=-2J$ (gap at $k=0$), where the parities $m$ and $m_{\mathrm e}$ of Sec.~\ref{sec:before} change sign. For $E'''(k_*)$ (Sec.~\ref{sec:fronts}), differentiate $EE'=-2\mu J\sin k$ twice more, $E'^2+EE''=-2\mu J\cos k$ and $EE'''+3E'E''=2\mu J\sin k$; as $E''=0$ at $k_*$, $E\,E'''=2\mu J\sin k_*=-E\,v$. Second, in the topological phase the fastest reach $|\mu|$, so $\re g(k_*)=0$ by Eq.~\eqref{eq:complement}: the density is blind to them, and their bond weight is $|\re[e^{-ik_*}g(k_*)]|=E(k_*)/(2J)=(4J^2-\mu^2)^{1/2}/(2J)$. In the trivial phase they reach $2J$ and the roles exchange: the bond energy is blind, and the density weight is $|\re g(k_*)|=E(k_*)/\mu$. Third, at $|\mu|=2J$ they sit at the gap closing, where $E(k)$ has a kink, $E=2J|\kappa|-\frac J{12}|\kappa|^3+\ldots$ for $\mu=-2J$ and $k=\kappa\to0$, and both weights vanish linearly in $|\kappa|$ [Fig.~\ref{fig:circle}, bottom].
	
	\textit{The running example.} At $\mu=\Delta=J$ Eq.~\eqref{eq:kstar} gives $\cos k_*=-\frac12$, so $k_*=-2\pi/3$, $v=J$, $E(k_*)=\sqrt3J$, and $E'''(k_*)=-J$. There $h(k_*)=J+2Je^{-2\pi i/3}=-i\sqrt3J$ and $g(k_*)=i$: zero density weight, bond weight $\re(e^{2\pi i/3}i)=-\sqrt3/2$ [Fig.~\ref{fig:quasi}]. This predicts the main text's Fig.~1: after a density measurement the bond energy shows a strong front, the density a weak one.
	
	\textit{What if $\mu=0$?} Then $h(k)=2Je^{ik}$ circles the origin and $E(k)=2J$ for all $k$: the band is flat, every speed vanishes, and the response stays at the measured site (Sec.~\ref{sec:measure}).
	
	\subsection{An open end and its zero mode}
	\label{sec:edgemode}
	
\begin{figure}[!htb]
		\centering
		\includegraphics[width=\textwidth]{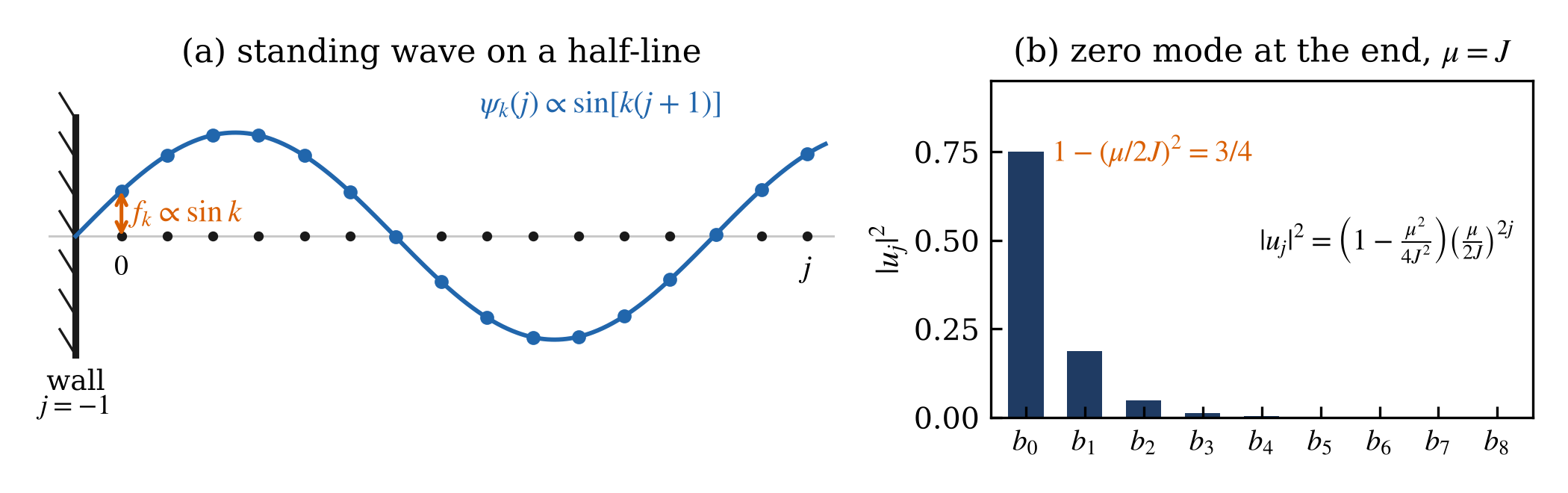}
		\caption{A half-line. (a) Standing wave $\psi_k(j)\propto\sin[k(j+1)]$, zero at the wall $j=-1$, with first-site amplitude $f^a_k\propto\sin k$, Eq.~\eqref{eq:hardwall}. (b) Majorana zero mode at $\mu=J=\Delta$: weights $|u_j|^2$ on $b_j$ decay as $(\mu/2J)^{2j}$ from $1-\mu^2/4J^2=\frac34$, Eq.~\eqref{eq:zeromode}.}
		\label{fig:halfline}
	\end{figure}

	On a half-line, $j\geq0$, plane waves become standing waves (Sec.~\ref{sec:before}), and the topological phase gains an end state. A Majorana $\Psi=\sum_lu_lb_l$ costs no energy if it commutes with $H$; by Eq.~\eqref{eq:R}, $d\Psi/dt=\sum_lu_l\sum_qA_{b_lq}\gamma_q=-\sum_j\bigl(\sum_lh_{jl}u_l\bigr)a_j$, so it needs $\sum_lh_{jl}u_l=0$, or $\mu u_j+2Ju_{j+1}=0$ for nearest neighbors at $J=\Delta$. The solution $u_j=(-\mu/2J)^j$ decays into the chain exactly when $|\mu|<2J$ and, normalized, has weight
	\begin{equation}
		|u_0|^2=1-\frac{\mu^2}{4J^2}
		\label{eq:zeromode}
	\end{equation}
	on $b_0$ [Fig.~\ref{fig:halfline}(b)]; at $\mu=0$ it is one: the zero mode is the unpaired $b_0$ of Sec.~\ref{sec:kitaev}.
	
	\subsection{The cluster chain is two Kitaev chains}
	\label{sec:cluster}
	
	With Pauli operators $X_j$, $Y_j$, $Z_j$, the cluster-chain stabilizers
	\begin{equation}
		K_j=Z_{j-1}X_jZ_{j+1}
		\label{eq:stab}
	\end{equation}
	square to one and commute: $K_j$ and $K_{j+1}$ overlap in two anticommuting pairs, $X_j$ with $Z_j$ and $Z_{j+1}$ with $X_{j+1}$, whose signs cancel. The Hamiltonian, Eq.~(7) of the main text, is
	\begin{equation}
		H_{\mathrm C}=-\sum_{j=1}^{N-2}K_j-\lambda\sum_{j=0}^{N-1}X_j, \qquad \lambda\geq0.
		\label{eq:HC}
	\end{equation}
	At $\lambda=0$ the ground state, the common $+1$ eigenstate of all $K_j$, is the cluster state, the resource of measurement-based quantum computation~\cite{BriegelRaussendorf2001,RaussendorfBriegel2001}. There a flipped stabilizer costs energy $2$ anywhere, so the band is flat and nothing moves; hence $\lambda>0$ in Table~I of the main text. The products $P_{\mathrm e}=\prod_{j\,\mathrm{even}}X_j$ and $P_{\mathrm o}=\prod_{j\,\mathrm{odd}}X_j$ commute with $H_{\mathrm C}$; $O$ is symmetry-even if $P_{\mathrm e}OP_{\mathrm e}=P_{\mathrm o}OP_{\mathrm o}=O$ and symmetry-odd if either conjugation gives $-O$. Since $XZX=-Z$, $Z_j$ is odd, while $X_j$ and $K_j$ are even, as both $Z$'s of $K_j$ sit on the other sublattice. For $\lambda<1$ the chain is in the symmetry-protected topological cluster phase: while both symmetries hold it cannot reach a product state without closing the gap, and its ends carry protected degrees of freedom~\cite{DohertyBartlett2009, SonAmicoVedral2012}. For $\lambda>1$ it is a paramagnet, connected without gap closing to the product state with every $X_j=+1$; the gap closes at $\lambda=1$.
	
\begin{figure}[!htb]
		\centering
		\includegraphics[width=\textwidth]{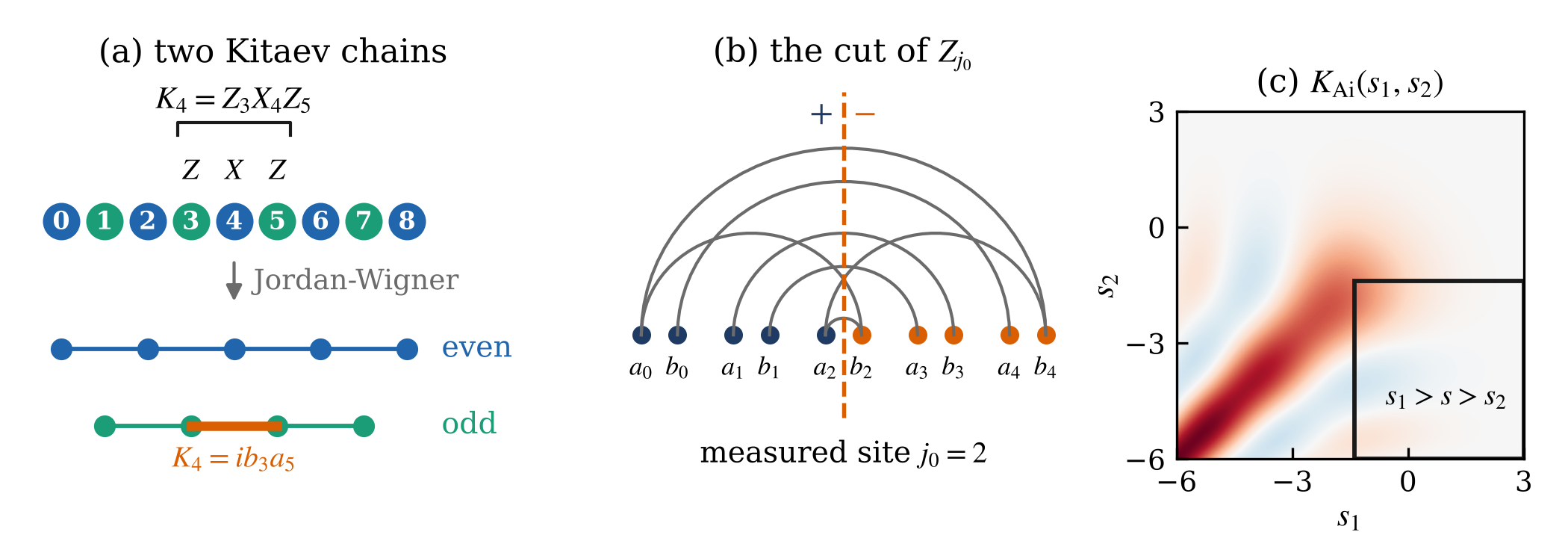}
		\caption{The cluster chain. (a) Spins $j=0, \ldots, 8$ and the stabilizer $K_4=Z_3X_4Z_5=ib_3a_5$, Eq.~\eqref{eq:XK}, an odd-chain bond; even (blue) and odd (green) sites form separate Kitaev chains. (b) The string measurement $Z_{j_0}$, $j_0=2$: conjugation by $Z_{j_0}$ keeps the signs of $a_j$ for $j\leq j_0$ and $b_j$ for $j<j_0$ (navy), flipping the others (orange); the measurement erases all correlations across the cut (gray arcs), Eq.~\eqref{eq:SZ}. (c) The Airy kernel $K_{\mathrm{Ai}}(s_1, s_2)=[\Ai(s_1)\Ai'(s_2)-\Ai'(s_1)\Ai(s_2)]/(s_1-s_2)$, red positive, blue negative; its integral over the framed quadrant $s_1>s>s_2$, here $s=-1.384$, gives the critical response of Sec.~\ref{sec:Zcrit}.}
		\label{fig:cluster}
	\end{figure}

	The Jordan--Wigner map~\cite{LSM1961} turns spins into fermions by attaching a string of $X$'s to each site,
	\begin{equation}
		a_j=\Pi_jZ_j, \qquad b_j=\Pi_jY_j, \qquad \Pi_j=\prod_{l<j}X_l.
		\label{eq:JW}
	\end{equation}
	On one site $a=Z$ and $b=Y$; the strings make Majoranas on different sites anticommute, as $a_0=Z_0$ and $a_1=X_0Z_1$ do through $Z_0$ and $X_0$. Since $\Pi_j^2=1$ and $\Pi_j$ commutes with site-$j$ operators, $a_jb_j=Z_jY_j=-iX_j$, and with $\Pi_{j+1}=\Pi_{j-1}X_{j-1}X_j$ and $Y_{j-1}X_{j-1}=-iZ_{j-1}$, also $b_{j-1}a_{j+1}=-iZ_{j-1}X_jZ_{j+1}$. Hence
	\begin{equation}
		X_j=ia_jb_j, \qquad K_j=ib_{j-1}a_{j+1}.
		\label{eq:XK}
	\end{equation}
	So $P_{\mathrm e}=\prod_{j\,\mathrm{even}}ia_jb_j$ and $P_{\mathrm o}=\prod_{j\,\mathrm{odd}}ia_jb_j$ are the fermion parities of the two sublattices: an even operator contains an even number of Majoranas of each sublattice, as $X_j$ and $K_j$ do, while $Z_j=\Pi_ja_j$ is one Majorana times the string. $H_{\mathrm C}$ is quadratic and couples only Majoranas two sites apart, so even and odd sites form independent chains [Fig.~\ref{fig:cluster}(a)]. With $j=2q+p$, cell $q$ and sublattice $p\in\{0, 1\}$, each carries
	\begin{equation}
		H_p=-i\sum_qb_{2q+p}a_{2q+2+p}-i\lambda\sum_qa_{2q+p}b_{2q+p}.
		\label{eq:Hsector}
	\end{equation}
	Compared with Eq.~\eqref{eq:HKmaj}, the on-site terms agree if $\mu=-2\lambda$, and the bond term $-ib_qa_{q+1}=ia_{q+1}b_q$, with $a_q$ and $b_q$ the Majoranas of cell $q$, is the Kitaev bond $ia_qb_{q+1}$ read right to left. Each sublattice is thus the Kitaev chain of Eq.~\eqref{eq:HKmaj} in a mirror, with the dictionary
	\begin{equation}
		X_j\leftrightarrow1-2n, \qquad K_j\leftrightarrow B, \qquad \mu\leftrightarrow-2\lambda, \qquad J=\Delta\leftrightarrow1, \qquad \text{one cell}\leftrightarrow\text{two sites},
		\label{eq:dictionary}
	\end{equation}
	and the cluster, critical, and paramagnetic phases match the topological, critical, and trivial Kitaev phases. All Kitaev results carry over: the energy per cell is $E(k)=2|\lambda-e^{ik}|$, with $k$ the momentum per cell; the fastest speed is $\min(2\lambda, 2)$ cells, or
	\begin{equation}
		v=4\min(\lambda, 1)\ \text{sites per unit time};
		\label{eq:vcluster}
	\end{equation}
	and the zero mode of Sec.~\ref{sec:edgemode} becomes, mirrored, one on the left-end $a$ Majoranas with weight $1-\lambda^2$ on $a_0=Z_0$ (empty string), returning in Sec.~\ref{sec:clusteredge} as the edge's record of a measurement.
	
	Applied to the Kitaev chain, Eq.~\eqref{eq:JW} gives $1-2n_j=ia_jb_j=X_j$ and $B_j=-ia_jb_{j+1}=Y_jY_{j+1}$, so Eq.~\eqref{eq:HKmaj} becomes $-J\sum_jY_jY_{j+1}+\frac\mu2\sum_jX_j$, the transverse-field Ising chain~\cite{LSM1961, Pfeuty1970} with coupling $J$ and field $|\mu|/2$, whose ordered phase and paramagnet are the topological and trivial phases; the Kitaev density is thus a one-spin and the bond energy a two-spin observable.

	\section{Before the measurement, the ground state}
	\label{sec:before}
	
	The measurement acts on the ground state, fixed for quadratic Hamiltonians by one matrix of two-point correlations.
	
	\subsection{The correlation matrix}
	\label{sec:Gamma}
	
	For any state, Majorana pair correlations form the real antisymmetric matrix of the main text,
	\begin{equation}
		\Gamma_{pq}=\frac i2\bigl\langle[\gamma_p, \gamma_q]\bigr\rangle=i\langle\gamma_p\gamma_q\rangle\quad(p\neq q), \qquad \Gamma_{pp}=0.
		\label{eq:Gamma}
	\end{equation}
	It is real because $(i\gamma_p\gamma_q)^\dagger=i\gamma_p\gamma_q$ and antisymmetric because different Majoranas anticommute. Indices are numbers $p, q$ or operators, as in $\Gamma_{a_jb_l}$. Every observable of the main text is a product of two Majoranas, so its expectation value is a linear function $\mathcal L_j$ of $\Gamma$; by Eqs.~\eqref{eq:parity}, \eqref{eq:HKmaj}, and \eqref{eq:XK},
	\begin{equation}
		\langle n_j\rangle=\frac{1-\Gamma_{a_jb_j}}2, \qquad \langle B_j\rangle=-\Gamma_{a_jb_{j+1}}, \qquad \langle X_j\rangle=\Gamma_{a_jb_j}, \qquad \langle K_j\rangle=\Gamma_{b_{j-1}a_{j+1}}.
		\label{eq:Lmap}
	\end{equation}
	When the operators rotate, $\gamma(t)=R(t)\gamma(0)$, any state's correlations rotate with them:
	\begin{equation}
		\Gamma(t)=R(t)\Gamma(0)R(t)^{\mathrm T}.
		\label{eq:Gammat}
	\end{equation}
	
	\subsection{Gaussian states and Wick's theorem}
	\label{sec:wick}
	
\begin{figure}[!htb]
		\centering
		\includegraphics[width=\textwidth]{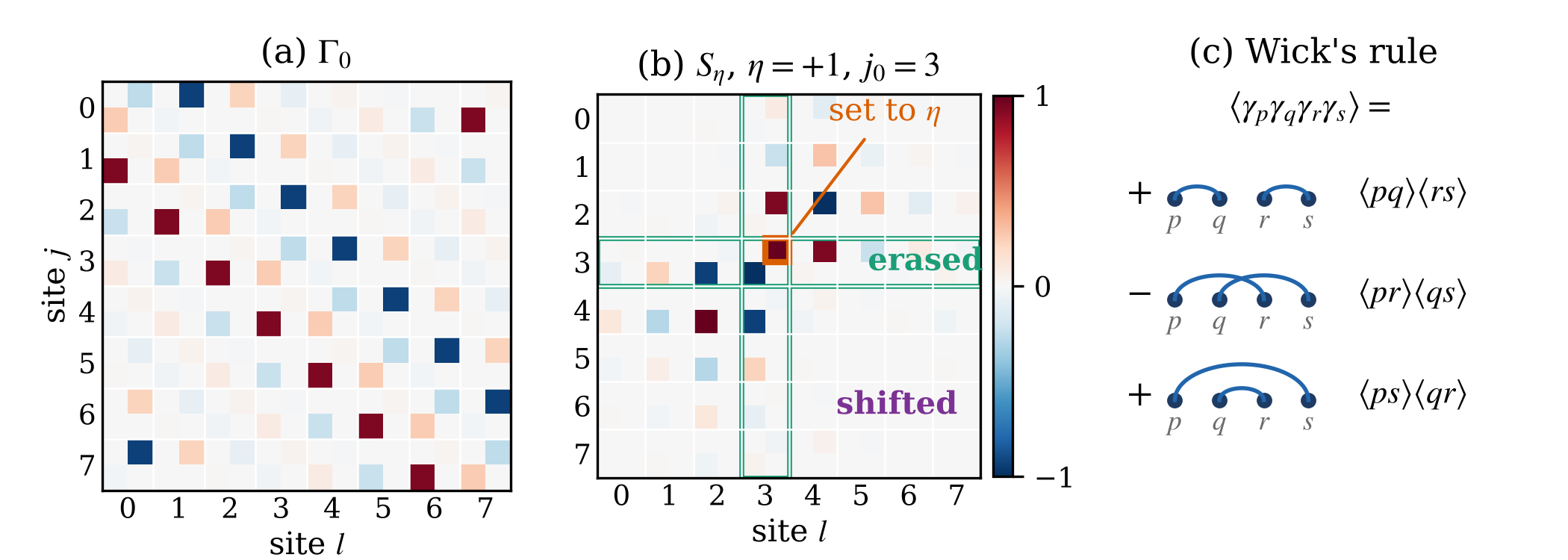}
		\caption{Correlation matrix, Kitaev ring of $8$ sites at $J=\Delta=\mu=1$, Majoranas ordered $(a_0, b_0, a_1, b_1, \ldots)$. (a) Ground state $\Gamma_0$, Eq.~\eqref{eq:Gamma0ring}. (b) Change $S_\eta=\Gamma_\eta-\Gamma_0$ after measuring $ia_3b_3$ with $\eta=+1$, Sec.~\ref{sec:parity}: measured entry of $\Gamma_\eta$ set to $\eta$, giving $S_{\eta,a_3b_3}=\eta-m$ (orange), rest of rows and columns $a_3$, $b_3$ erased (green frames), other entries shifted, strongest near the measured site. (c) The three pairings of Wick's theorem, Eq.~\eqref{eq:wick}.}
		\label{fig:kick}
	\end{figure}

	A state is Gaussian if all Majorana correlation functions follow from $\Gamma$ alone (equivalently, up to limits, if its density matrix is the exponential of a quadratic operator); for four different Majoranas, the only case we need, this is Wick's theorem,
	\begin{equation}
		\langle\gamma_p\gamma_q\gamma_r\gamma_s\rangle=\langle\gamma_p\gamma_q\rangle\langle\gamma_r\gamma_s\rangle-\langle\gamma_p\gamma_r\rangle\langle\gamma_q\gamma_s\rangle+\langle\gamma_p\gamma_s\rangle\langle\gamma_q\gamma_r\rangle, \qquad \langle\gamma_p\gamma_q\rangle=-i\Gamma_{pq},
		\label{eq:wick}
	\end{equation}
	one term per pairing, the crossed one with a minus sign [Fig.~\ref{fig:kick}(c)]. Ground and thermal states of quadratic Hamiltonians are Gaussian~\cite{Bravyi2005}: in the mode basis they are products of independent one-mode states, so averages factorize, up to anticommutation signs. Two-site check: with both sites empty, only $\Gamma_{a_0b_0}=\Gamma_{a_1b_1}=1$ are nonzero, and Eq.~\eqref{eq:wick} gives $\langle a_0b_0a_1b_1\rangle=(-i)^2=-1$; directly, $a_0b_0a_1b_1=(-i)(1-2n_0)(-i)(1-2n_1)=-(1-2n_0)(1-2n_1)$, which is $-1$ here.
	
	\subsection{The ground state: every mode empty}
	\label{sec:ground}
	
	A rotation $\mathcal O$ brings every real antisymmetric matrix to $2\times2$ blocks,
	\begin{equation}
		\mathcal OA\mathcal O^{\mathrm T}=\bigoplus_n\begin{pmatrix}0&E_n\\-E_n&0\end{pmatrix}, \qquad E_n\geq0,
		\label{eq:normal}
	\end{equation}
	because $iA$ is Hermitian with eigenvalues in pairs $\pm E_n$, and the real and imaginary parts of its eigenvectors span the blocks. In the rotated Majoranas $\gamma'=\mathcal O\gamma$, $H=\sum_n\frac{E_n}2i\gamma'_{2n}\gamma'_{2n+1}$ splits into independent one-site terms, each lowest when $i\gamma'_{2n}\gamma'_{2n+1}=-1$: every mode is in its lower state, and no quasiparticle is present. Each ground-state correlation block, $\bigl(\begin{smallmatrix}0&-1\\1&0\end{smallmatrix}\bigr)$, is the block of $A$ divided by $-E_n$. Rotating back, with $-A^2=E_n^2$ on each block,
	\begin{equation}
		\Gamma_0=-A\,(-A^2)^{-1/2}.
		\label{eq:Gamma0}
	\end{equation}
	As a function of $A$ alone, $\Gamma_0$ commutes with $A$ and hence with $R(t)=e^{At}$, so $R\Gamma_0R^{\mathrm T}=\Gamma_0$: the ground state is stationary. If some $E_n=0$, we take, as in the main text, the equal mixture of the degenerate ground states, with zero correlation block for those modes.
	
	On a ring Eq.~\eqref{eq:Gamma0} holds at each momentum, and by Eq.~\eqref{eq:Ak}, $-A(k)/E(k)$ has entries $-h/E=g$ and $\bar h/E=-\bar g$, so each $a$ is correlated only with $b$'s, through the Fourier coefficients of the phase $g(k)$ of Eq.~\eqref{eq:g}:
	\begin{equation}
		\Gamma_{0, a_jb_l}=g_{j-l}=\int_{-\pi}^\pi\frac{dk}{2\pi}g(k)e^{ik(j-l)}, \qquad \Gamma_{0, a_ja_l}=\Gamma_{0, b_jb_l}=0, \qquad m=\langle1-2n_j\rangle_0=g_0.
		\label{eq:Gamma0ring}
	\end{equation}
	The density sees $\re g$, the particle-minus-hole weight of Eq.~\eqref{eq:uv}; Fig.~\ref{fig:kick}(a) shows the running example.
	
	\textit{Three examples.} At $\mu=0$, $g(k)=-e^{ik}$ has one coefficient, $g_{-1}=-1$: each $a_j$ is locked to $b_{j+1}$, $\Gamma_{0, a_jb_{j+1}}=-1$, every bond is in its lowest state, $\langle B_j\rangle=1$, and nothing else is correlated. In the gapped chain the coefficients $g_r$ of the smooth periodic $g(k)$ decay exponentially with $|r|$; for the running example $\mu=J$, $g_{-1}=-0.934$, $g_0=m=-0.2587$, and $g_1=0.033$, so the bond pairing still dominates. At the gap closing $\mu=-2J$ the circle passes through the origin at $k=0$, $h(k)=2J(e^{ik}-1)=4iJe^{ik/2}\sin(k/2)$, and $g(k)=-ie^{ik/2}\sgn(k)$ on $-\pi<k<\pi$ jumps at $k=0$. Integrating,
	\begin{equation}
		g_r=\frac1{\pi(r+\frac12)}, \qquad m=g_0=\frac2\pi,
		\label{eq:critcorr}
	\end{equation}
	and the jump of $g$ makes the correlations decay only as $1/r$. This returns as a response at $t=0$ (Sec.~\ref{sec:t0}), a tail ahead of the critical front (Sec.~\ref{sec:critical}), and a long-range change from the string measurement (Sec.~\ref{sec:Zcrit}).
	
	\subsection{The ground state on a half-line}
	\label{sec:halfline}
	
	For nearest neighbors at $J=\Delta$ on a half-line, the squared $a$-mode energies are the eigenvalues of $hh^{\mathrm T}$, the $a$ block of $-A^2$, with $\mu^2+4J^2$ on the diagonal, $2\mu J$ beside it, and no correction at the end, since row $0$ of $h$ is like every other row. Its eigenvectors are standing waves vanishing at the wall $j=-1$ [Fig.~\ref{fig:halfline}(a)]; normalized per unit momentum, $\sum_j\psi_k(j)\psi_{k'}(j)=\delta(k-k')$,
	\begin{equation}
		\psi_k(j)=\sqrt{\frac2\pi}\sin[(j+1)k], \qquad 0<k<\pi, \qquad f^a_k=\sqrt{\frac2\pi}\sin k, \qquad f^b_k=\frac\mu{E(k)}f^a_k,
		\label{eq:hardwall}
	\end{equation}
	with the same $E(k)$ as on the ring; $f^a_k$ and $f^b_k$ are the first-site amplitudes on $a_0$ and $b_0$, the second from $\psi^b_k=h^{\mathrm T}\psi_k/E(k)$, as the first column of $h$ holds only $\mu$. They couple a first-site measurement to the waves and, by Sec.~\ref{sec:fronts}, are all an open end changes. The first-site occupation parity follows as on the ring,
	\begin{equation}
		m_{\mathrm e}=\langle1-2n_0\rangle_0=-\int_0^\pi dk\,f^a_kf^b_k=-\frac{2\mu}\pi\int_0^\pi\frac{\sin^2k}{E(k)}dk,
		\label{eq:me}
	\end{equation}
	which is $-0.48384$, $-8/(3\pi)$, and $-0.94082$ for $\mu/J=1$, $2$, and $3$ (e for edge).
	
	\subsection{The cluster chain}
	\label{sec:clusterground}
	
	By the dictionary~\eqref{eq:dictionary}, each cluster sublattice has the ground-state correlations of a Kitaev chain with $\mu=-2\lambda$. We need $\langle X_j\rangle_0$ and $\langle K_j\rangle_0$, both positive for $\lambda>0$. For couplings $-\lambda_K\sum_jK_j-\lambda\sum_jX_j$ each sublattice has $E(k)=2|\lambda_Ke^{ik}-\lambda|$ per cell, and the ground-state energy, minus half the sum of mode energies, is $-\sum_k|\lambda_Ke^{ik}-\lambda|$. By the Hellmann--Feynman theorem, $\partial E_0/\partial\lambda=\langle\partial H/\partial\lambda\rangle_0$, $\langle X_j\rangle_0$ and $\langle K_j\rangle_0$ are the $\lambda$ and $\lambda_K$ derivatives of the mean distance between the point $\lambda$ and the circle of radius $\lambda_K$, which, even and convex in each variable, has positive derivatives away from zero, in particular at $\lambda_K=1$, $\lambda>0$. At $\lambda=2$, $\langle X_j\rangle_0=0.93422$ and $\langle K_j\rangle_0=0.25866$.

	\section{The measurement}
	\label{sec:measure}
	
	For observables that are products of two Majoranas, the whole effect of the measurement at $t=0$ is a change $S_\eta$ of the correlation matrix, carried along the chain by the propagator.
	
	\subsection{What is recorded, and why the root mean square}
	\label{sec:rms}
	
	Measuring $M$, with $M^2=1$, projects the initial state $\rho_0$ onto an eigenspace:
	\begin{equation}
		P_\eta=\frac{1+\eta M}2, \qquad p_\eta=\mathrm{Tr}(P_\eta\rho_0)=\frac{1+\eta m}2, \qquad m=\langle M\rangle_0, \qquad \rho_\eta=\frac{P_\eta\rho_0P_\eta}{p_\eta}.
		\label{eq:proj}
	\end{equation}
	The state then evolves, $\rho_\eta(t)=e^{-iHt}\rho_\eta e^{iHt}$, and we record an observable's change given the outcome and its root mean square over outcomes, Eq.~(1) of the main text,
	\begin{equation}
		\delta O_{j, \eta}(t)=\mathrm{Tr}[\rho_\eta(t)O_j]-\langle O_j\rangle_0, \qquad D_j(t)=\Bigl[\sum_\eta p_\eta\,\delta O_{j, \eta}(t)^2\Bigr]^{1/2}.
		\label{eq:D}
	\end{equation}
	Since the ground state is stationary, $\langle O_j\rangle_0$ is time independent; sums over $\eta$ run over outcomes with $p_\eta>0$. If $O$ commutes with $M$, then $P_\eta OP_\eta=P_\eta O$, and since $P_++P_-=1$,
	\begin{equation}
		\sum_\eta p_\eta\mathrm{Tr}(\rho_\eta O)=\sum_\eta\mathrm{Tr}(\rho_0P_\eta OP_\eta)=\mathrm{Tr}(\rho_0O), \qquad \text{so}\qquad p_+\delta O_+=-p_-\delta O_-\quad\text{at }t=0.
		\label{eq:nosignal}
	\end{equation}
	So at $t=0$ a distant observer ignorant of the outcome sees no change: no signal is sent. Each outcome alone does change $\langle O\rangle$, by opposite amounts: the instant shift from learning $\eta$, real only for someone who knows it, which the average misses and $D_j$ keeps. Experimentally, one sorts repeated runs by outcome, averages $O_j$ within each group, and combines the groups as in Eq.~\eqref{eq:D}. For $t>0$ the outcome average is in general nonzero: the measurement also erases correlations, and the Hamiltonian spreads this change.
	
	The measured site is in the bulk, far from any end, or is the first site $j_0=0$ of a half-line.
	
	\subsection{Measuring a local parity}
	\label{sec:parity}
	
	The Kitaev density, $M=1-2n_{j_0}=ia_{j_0}b_{j_0}$, and the cluster field, $M=X_{j_0}=ia_{j_0}b_{j_0}$, are both local parities. For a Gaussian initial state, $m=\Gamma_{0, a_{j_0}b_{j_0}}$, and three rules fix $\Gamma_\eta$.
	
	(i) $M$ has the value $\eta$ in $\rho_\eta$, so $\Gamma_{\eta, a_{j_0}b_{j_0}}=\eta$.
	
	(ii) The measured Majoranas lose every other correlation. Since $a_{j_0}$ anticommutes with $M$, $P_\eta a_{j_0}=a_{j_0}P_{-\eta}$, while any other Majorana $\gamma$ commutes with $P_\eta$. Hence $P_\eta a_{j_0}\gamma P_\eta=a_{j_0}\gamma P_{-\eta}P_\eta=0$ and $\Gamma_{\eta, a_{j_0}\gamma}=0$; likewise for $b_{j_0}$.
	
\begin{figure}[!htb]
		\centering
		\includegraphics[width=\textwidth]{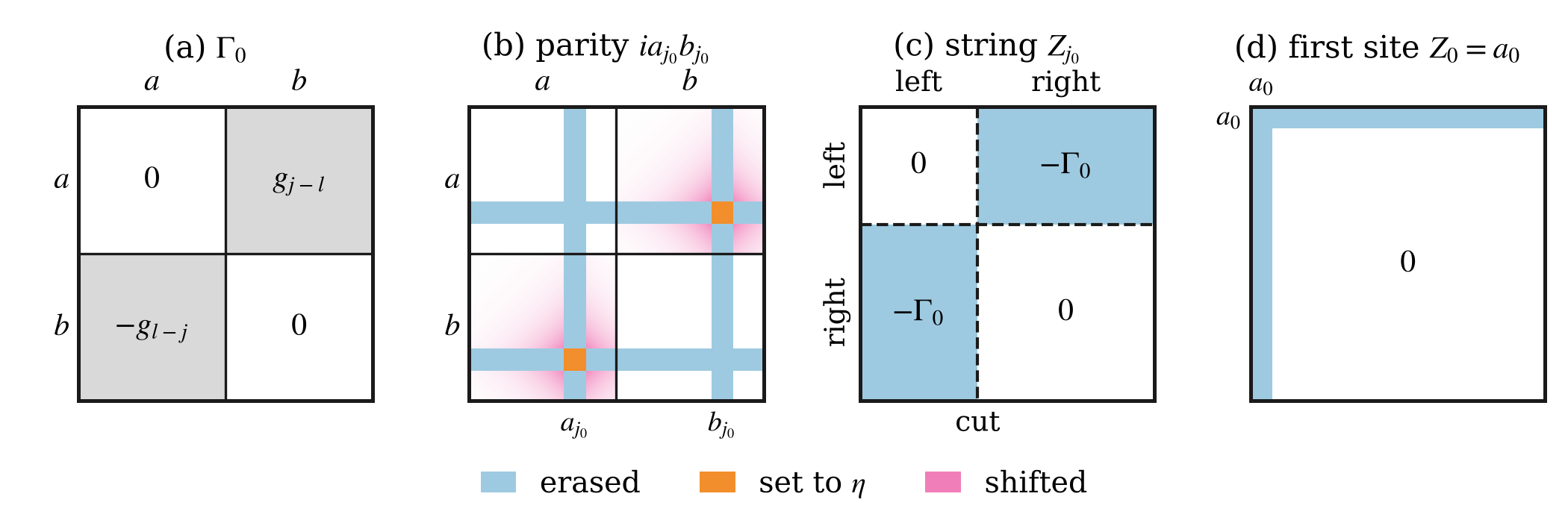}
		\caption{Effect of each measurement on the correlation matrix. (a) Ground state, all $a$ Majoranas first and then all $b$, Eq.~\eqref{eq:Gamma0ring}. (b) Parity measurement $ia_{j_0}b_{j_0}$, Eq.~\eqref{eq:update}: erased rows and columns (blue), entry set to $\eta$ (orange), shifted entries (pink). (c) String measurement $Z_{j_0}$, Majoranas ordered along the chain, Eq.~\eqref{eq:SZ}. (d) $Z_0=a_0$ on the first site, Eq.~\eqref{eq:SZ0}.}
		\label{fig:blocks}
	\end{figure}

	(iii) All other correlations shift. For two other Majoranas $\gamma$ and $\gamma'$, $O=i\gamma\gamma'$ commutes with $P_\eta$, so $\mathrm{Tr}(P_\eta\rho_0P_\eta O)=\mathrm{Tr}(\rho_0P_\eta O)=\frac12(\langle O\rangle_0+\eta\langle MO\rangle_0)$. By Wick's theorem~\eqref{eq:wick}, $\langle MO\rangle_0=m\Gamma_{0, \gamma\gamma'}-\Gamma_{0, a_{j_0}\gamma}\Gamma_{0, b_{j_0}\gamma'}+\Gamma_{0, a_{j_0}\gamma'}\Gamma_{0, b_{j_0}\gamma}$, and dividing by $p_\eta=(1+\eta m)/2$,
	\begin{equation}
		\Gamma_{\eta, \gamma\gamma'}=\Gamma_{0, \gamma\gamma'}-\frac\eta{2p_\eta}\bigl(\Gamma_{0, a_{j_0}\gamma}\Gamma_{0, b_{j_0}\gamma'}-\Gamma_{0, a_{j_0}\gamma'}\Gamma_{0, b_{j_0}\gamma}\bigr).
		\label{eq:update}
	\end{equation}
	Rules (i) and (ii) are the erasure, cutting the measured site out of the correlation network and fixing its parity; rule (iii) is the learning, shifting distant correlations by products of two correlations with the measured site. A projector built from a quadratic operator keeps the state Gaussian, so these rules fix the new state completely. The change $S_\eta=\Gamma_\eta-\Gamma_0$, built from the columns of $\Gamma_0$ for $a_{j_0}$ and $b_{j_0}$ and the unit vectors of these two Majoranas, has rank at most four, though the columns span the whole chain [Figs.~\ref{fig:kick}(b) and \ref{fig:blocks}(b)].
	
	\textit{Example: three sites at $\mu=0$.} For $H=iJ(a_0b_1+a_1b_2)$ the ground state locks $a_0$ to $b_1$ and $a_1$ to $b_2$, $\Gamma_{0, a_0b_1}=\Gamma_{0, a_1b_2}=-1$, leaving the zero modes $b_0$ and $a_2$. We measure the middle-site parity, with $m=\Gamma_{0, a_1b_1}=0$ and $p_\eta=\frac12$: rule (ii) breaks both locked pairs, and rule (iii) with $\gamma=b_2$ and $\gamma'=a_0$ gives $\Gamma_{\eta, b_2a_0}=-\eta\,\Gamma_{0, a_1b_2}\Gamma_{0, b_1a_0}=\eta$, locking the previously uncorrelated outer Majoranas with an outcome-dependent sign.
	
	\subsection{The response at $t=0$}
	\label{sec:t0}
	
	Equation~\eqref{eq:update} already changes distant observables at $t=0$. For the density at $x\neq0$ on a ring, take $\gamma=a_j$ and $\gamma'=b_j$ in Eq.~\eqref{eq:update}, with $\delta n_j=-\frac12\delta\Gamma_{a_jb_j}$ and, from Eq.~\eqref{eq:Gamma0ring}, $\Gamma_{0, a_{j_0}b_j}=g_{-x}$, $\Gamma_{0, b_{j_0}a_j}=-g_x$, and vanishing $aa$ and $bb$ correlations. With the outcome sums
	\begin{equation}
		\sum_\eta\frac{p_\eta}{(2p_\eta)^2}=\frac1{1-m^2}, \qquad \sum_\eta\frac{p_\eta\,\eta}{2p_\eta}=0,
		\label{eq:sums}
	\end{equation}
	this gives
	\begin{equation}
		\delta n_{j, \eta}(0)=\frac{\eta\,g_xg_{-x}}{4p_\eta}, \qquad D_j(0)=\frac{|g_xg_{-x}|}{2\sqrt{1-m^2}}.
		\label{eq:Dt0}
	\end{equation}
	This product of two correlations linking $j_0$ to $j$ decays exponentially with $|x|$ in a gapped chain; at the gap closing Eq.~\eqref{eq:critcorr} makes it algebraic,
	\begin{equation}
		D_j(0)=\frac1{2\pi^2\sqrt{1-m^2}\,|x^2-\frac14|}, \qquad m=\frac2\pi,
		\label{eq:tailt0}
	\end{equation}
	whose outcome average vanishes by Eq.~\eqref{eq:sums}, so it carries no signal, only the correlation of the recorded outcome with the distant density. Section~\ref{sec:critical} follows this tail in time.
	
	\subsection{Measuring a Majorana times a string}
	\label{sec:string}
	
	By Eq.~\eqref{eq:JW}, $Z_{j_0}=\Pi_{j_0}a_{j_0}$ is one Majorana times the string $\Pi_{j_0}$. Its projector is not the exponential of a quadratic operator, so the measured state is not Gaussian and Wick's theorem fails. Still, our observables are quadratic and need only two-point functions.
	
	\textit{Conjugation is a sign flip.} The operator $Z_{j_0}$ commutes with $a_j$ and $b_j$ for $j<j_0$ and with $a_{j_0}$, and anticommutes with $b_{j_0}=\Pi_{j_0}Y_{j_0}$ and every Majorana to its right, whose string contains $X_{j_0}$. Hence $Z_{j_0}\gamma Z_{j_0}=F\gamma$, with $F$ diagonal, $+1$ on the left and $-1$ on the right [Fig.~\ref{fig:cluster}(b)]; the boundary, between $a_{j_0}$ and $b_{j_0}$, is the cut of the main text.
	
	\textit{Both outcomes give the same response.} Expanding the projectors, $\mathrm{Tr}[P_\eta\rho_0P_\eta O(t)]=\frac14\bigl(\langle O(t)\rangle_0+\langle Z_{j_0}O(t)Z_{j_0}\rangle_0+\eta\langle\{Z_{j_0}, O(t)\}\rangle_0\bigr)$, with $O(t)$ the Heisenberg operator. The total fermion parity $\prod_jia_jb_j$ commutes with $H$, every quadratic $O(t)$, and the ground state, but anticommutes with $Z_{j_0}=\bigl(\prod_{l<j_0}ia_lb_l\bigr)a_{j_0}$, an odd product of Majoranas. So $\langle\{Z_{j_0}, O(t)\}\rangle_0=0$ and $m=\langle Z_{j_0}\rangle_0=0$, $p_\eta=\frac12$, and both outcomes respond like the state $\frac12(\rho_0+Z_{j_0}\rho_0Z_{j_0})$. For a quadratic observable, $Z_{j_0}OZ_{j_0}$ is the same observable with $\gamma\to F\gamma$, whose correlation matrix is $F\Gamma_0F$. Hence
	\begin{equation}
		S_Z=\frac12\bigl(F\Gamma_0F-\Gamma_0\bigr), \qquad S_{Z, \gamma\gamma'}=\begin{cases}-\Gamma_{0, \gamma\gamma'}, &\gamma, \gamma'\text{ on opposite sides of the cut}, \\0, &\text{otherwise}, \end{cases}
		\label{eq:SZ}
	\end{equation}
	the $S_\eta$ of the main text for this measurement, the same for both outcomes. Only correlations crossing the cut are erased, in both sublattices [Fig.~\ref{fig:blocks}(c)], and nothing is removed from the chain. The number of erased pairs at separation $r$ grows as $|r|$, as such a pair straddles the cut in $|r|$ positions [Fig.~\ref{fig:cluster}(b)]; the erased correlations decay exponentially in a gapped phase, concentrating $S_Z$ near $j_0$, but only as $1/r$ at $\lambda=1$, Eq.~\eqref{eq:critcorr}, so $S_Z$ reaches far. \textit{Example.} At $\lambda=0$ only $K_{j_0}$ anticommutes with $Z_{j_0}$, so $\langle K_{j_0}\rangle$ drops from $1$ to $0$ and, the band being flat, stays there: $D_{j_0}(t)=1$.
	
	\textit{A single Majorana at the first site.} At the left end the string is empty, $Z_0=a_0$, and conjugation by $a_0$ keeps $a_0$ and flips every other Majorana, so only the correlations of $a_0$ are erased [Fig.~\ref{fig:blocks}(d)],
	\begin{equation}
		S_{Z_0, a_0\gamma}=-\Gamma_{0, a_0\gamma}, \qquad S_{Z_0, \gamma\gamma'}=0\quad(\gamma, \gamma'\neq a_0),
		\label{eq:SZ0}
	\end{equation}
	one pair per separation instead of $|r|$, as for a parity measurement. Also $\langle a_0\rangle_\eta=\eta$, while every other Majorana, anticommuting with $a_0$, has zero mean. These means, the edge's record of the outcome (Sec.~\ref{sec:clusteredge}), rotate like the operators, $\langle\gamma(t)\rangle_\eta=\eta R(t)e_{a_0}$, with $e_{a_0}$ the unit vector of $a_0$.
	
	\subsection{The exact solution}
	\label{sec:exact}
	
	By Eq.~\eqref{eq:Gammat}, $\Gamma_0+S_\eta$ evolves into $R(\Gamma_0+S_\eta)R^{\mathrm T}$, and $R\Gamma_0R^{\mathrm T}=\Gamma_0$, so at every time and in a chain of any length
	\begin{equation}
		\delta O_{j, \eta}(t)=\mathcal L_j\bigl[R(t)S_\eta R(t)^{\mathrm T}\bigr],
		\label{eq:exact}
	\end{equation}
	which is Eq.~(2) of the main text, with $S_\eta$ from Eq.~\eqref{eq:update}, \eqref{eq:SZ}, or \eqref{eq:SZ0}. It requires only a quadratic observable, not a Gaussian measured state, and every figure evaluates it. On six to twelve sites Eq.~\eqref{eq:exact} matches a direct calculation in the $2^N$-dimensional Hilbert space to machine precision, for every measurement of the main text and both outcomes. Since the two sublattices of the cluster chain are independent, measuring $X_{j_0}=ia_{j_0}b_{j_0}$ changes only the sublattice of $j_0$, and $K_j=ib_{j-1}a_{j+1}$ belongs to that of $j\pm1$, so only stabilizers with odd $j-j_0$ respond.
	
\begin{figure}[!htb]
		\centering
		\includegraphics[width=\textwidth]{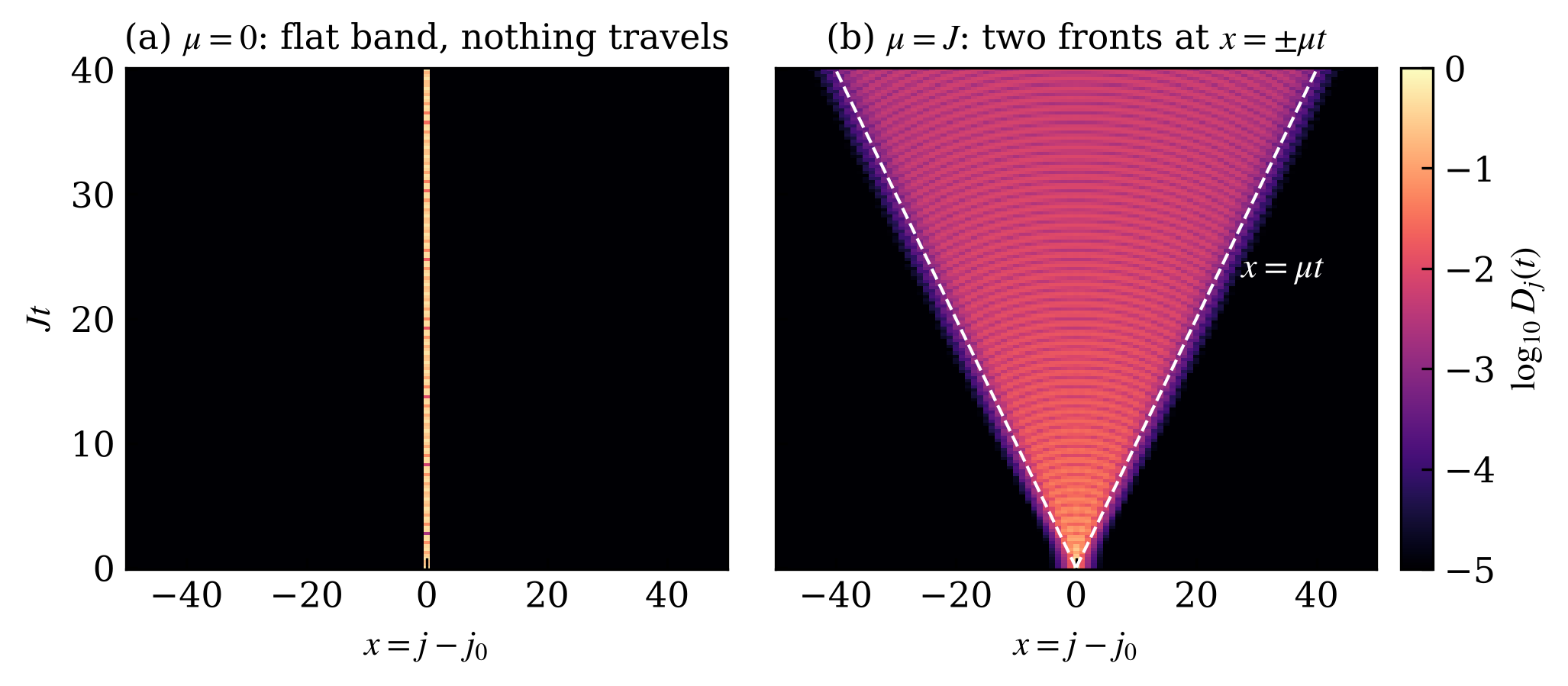}
		\caption{Exact response $D_j(t)$, Eq.~\eqref{eq:exact}, after one density measurement on a Kitaev ring of $160$ sites at $J=\Delta=1$ (logarithmic color scale). (a) $\mu=0$: flat band, $E(k)=2J$; the response stays at the measured site. (b) $\mu=J$: quasiparticles disperse and the response leaves along two fronts $x=\pm vt$, $v=\mu$, Eq.~\eqref{eq:kstar}.}
		\label{fig:flatband}
	\end{figure}

	\textit{What if nothing could move?} At $\mu=0$, $A^2=-4J^2$ on a ring, so $R(t)=\cos2Jt+A\sin(2Jt)/(2J)$ only rotates $a_j$ into $b_{j+1}$ within one bond. In the three-site example of Sec.~\ref{sec:parity}, $a_0$ rotates with $b_1$ and $a_1$ with $b_2$ by the angle $2Jt$, and Eq.~\eqref{eq:exact} gives $\Gamma_{\eta, a_1b_1}(t)=\eta\cos4Jt$, so $D_1(t)=\frac12|\cos4Jt|$: the response oscillates in place and never leaves. Figure~\ref{fig:flatband} contrasts this with the dispersing case $\mu=J$, whose fronts occupy the rest of these notes.
	
	\section{After the measurement, fronts}
	\label{sec:fronts}
	
	This section opens Eq.~\eqref{eq:exact} at large $t$: one propagator gives the position and width of the front (Sec.~\ref{sec:onewave}); two give its height, a measurement factor times an observable factor (Secs.~\ref{sec:pairs} and \ref{sec:factor}); a vanishing observable factor costs two powers of $\ell$ (Sec.~\ref{sec:blind}). The running example puts it together (Sec.~\ref{sec:example}), and Sec.~\ref{sec:whatif} changes one ingredient at a time.
	
	\subsection{One wave: where the front is, and why it widens as $t^{1/3}$}
	\label{sec:onewave}
	
\begin{figure}[!htb]
		\centering
		\includegraphics[width=\textwidth]{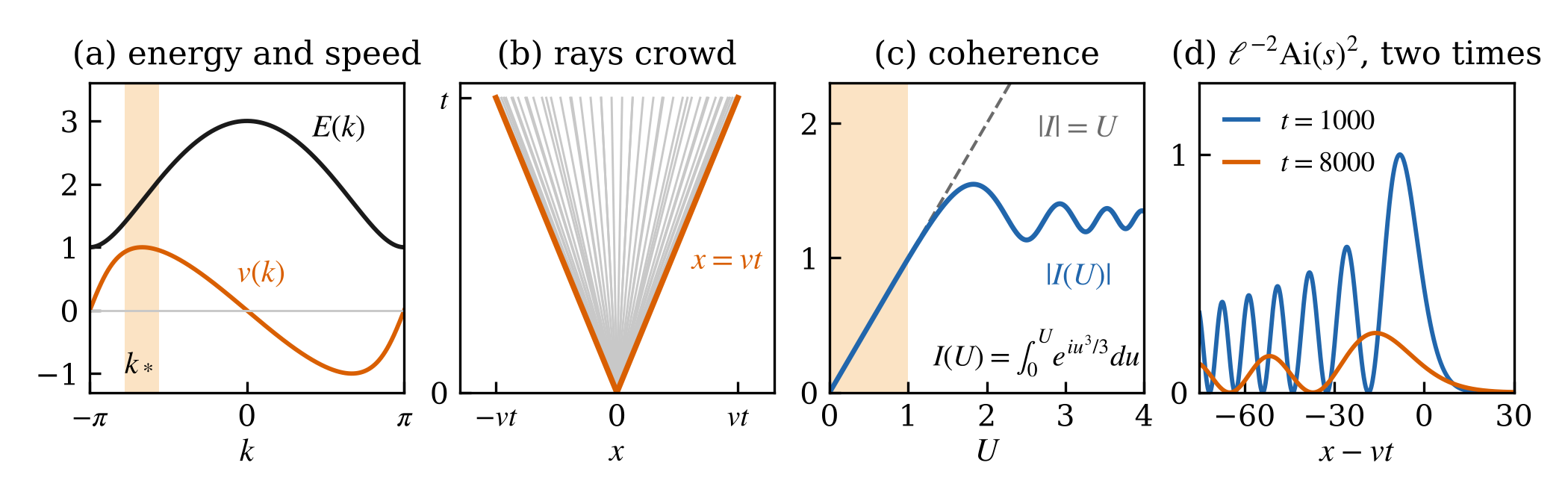}
		\caption{One wave near the front, $J=\Delta=\mu=1$, where $k_*=-2\pi/3$, $v=\mu$, and $E'''(k_*)=-\mu$, Eq.~\eqref{eq:kstar}. (a) $E(k)$ and $v(k)$; shading: the window $|k-k_*|<\ell^{-1}$ at $t=50$. (b) Rays $x=v(k)t$ for equally spaced momenta crowd at $x=\pm vt$: a caustic. (c) Why the window has the width $\ell^{-1}$: at the front the phase is $u^3/3$, and $|\int_0^Ue^{iu^3/3}du|$ grows like $U$ while all phases agree, up to $U\simeq1$, and then stops. (d) $\ell^{-2}\Ai(s)^2$ at two times eight apart: the width doubles, the height drops fourfold.}
		\label{fig:stationary}
	\end{figure}

	On a ring the propagator is diagonal in momentum. The block of $R(t)$ that takes the Majoranas $(a_l, b_l)$ of site $l$ to those of site $j$ is a $2\times2$ matrix,
	\begin{equation}
		R_{jl}(t)=\int_{-\pi}^\pi\frac{dk}{2\pi}e^{ik(j-l)}e^{A(k)t}, \qquad e^{A(k)t}=\cos E(k)t-\hat\Gamma_0(k)\sin E(k)t=\sum_{\tau=\pm}\mathcal P_\tau(k)e^{-i\tau E(k)t},
		\label{eq:Rk}
	\end{equation}
	where $\hat\Gamma_0(k)=\bigl(\begin{smallmatrix}0&g\\-\bar g&0\end{smallmatrix}\bigr)$ is the ground state at momentum $k$, so that $A(k)=-E(k)\hat\Gamma_0(k)$ and $\hat\Gamma_0^2=-1$, and $\mathcal P_\pm(k)=\frac12(1\mp i\hat\Gamma_0)$ project onto the two frequencies: each Majorana wave of momentum $k$ is a superposition of waves of frequency $+E(k)$ and $-E(k)$, as the main text says. To see where such waves arrive, take one of them with a smooth amplitude $y(k)$,
	\begin{equation}
		I(x, t)=\int_{-\pi}^\pi\frac{dk}{2\pi}y(k)e^{i[kx-E(k)t]}.
		\label{eq:I}
	\end{equation}
	At large $t$ the phase $kx-E(k)t$ oscillates rapidly in $k$ and neighboring momenta cancel, except where it is stationary, $x=E'(k)t$: a wave packet of momentum $k$ travels along the ray $x=v(k)t$ with the group velocity $v(k)=E'(k)$ (Fig.~\ref{fig:stationary}). For $x<vt$, with $v$ the largest speed, two momenta reach each point, each from a window $|k-k_{\mathrm s}|\lesssim(|E''|t)^{-1/2}$ in which the phase stays within about one radian, so with an amplitude of order $t^{-1/2}$. At $x=vt$ they merge at the fastest momentum $k_*$, where $E''(k_*)=0$, and the rays of a whole range of momenta pile up into a caustic. Beyond $x=vt$ no momentum arrives. Near the caustic we expand around $k_*$, $k=k_*+\kappa$:
	\begin{equation}
		E(k_*+\kappa)=E(k_*)+v\kappa+\tfrac16E'''(k_*)\kappa^3+\tfrac1{24}E''''(k_*)\kappa^4+\ldots, \qquad E'''(k_*)<0.
		\label{eq:Etaylor}
	\end{equation}
	The phase of Eq.~\eqref{eq:I} becomes $\chi+\kappa(x-vt)+\frac16|E'''(k_*)|t\kappa^3$, with the fast phase $\chi=k_*x-E(k_*)t$. The cubic term matters once $\kappa\sim t^{-1/3}$, so we measure momenta and positions in its units,
	\begin{equation}
		\kappa=\frac u\ell, \qquad \ell=\Bigl(\frac{|E'''(k_*)|t}2\Bigr)^{1/3}, \qquad s=\frac{x-vt}\ell,
		\label{eq:airyscale}
	\end{equation}
	and the phase becomes $\chi+su+u^3/3$, up to terms that vanish as $\ell\to\infty$. With $d\kappa=du/\ell$ and the Airy function,
	\begin{equation}
		\Ai(s)=\int_{-\infty}^\infty\frac{du}{2\pi}e^{i(su+u^3/3)}, \qquad\text{one finds}\qquad I(x, t)\simeq y(k_*)\,e^{i\chi}\,\ell^{-1}\Ai(s),
		\label{eq:onewave}
	\end{equation}
	the statement of the main text that $R_{jl}(t)\simeq\ell^{-1}\Ai(s)$ times a phase near the front. Width $\ell\propto t^{1/3}$ and height $\ell^{-1}$ follow from one count [Fig.~\ref{fig:stationary}(c)]: in the variable $u$ all phases agree for $|u|\lesssim1$, so a window of momenta of width $\ell^{-1}$ around $k_*$ adds up coherently, and an integrand of order one over this window gives $\ell^{-1}$. If instead the energy departs from its tangent as $C|\kappa|^z$, the same steps give $\ell\propto(|C|t)^{1/z}$, the exponent $z$ of Eq.~\eqref{eq:rule}; Sec.~\ref{sec:lr} meets $z\neq3$.
	
	\textit{Airy facts.} Differentiating under the integral gives $\Ai''(s)=s\Ai(s)$. Ahead of the front, $s\to+\infty$, $\Ai$ decays as $e^{-2s^{3/2}/3}$, since no ray arrives; behind it, it oscillates as $\sin(\frac23|s|^{3/2}+\frac\pi4)/(\sqrt\pi|s|^{1/4})$, the interference of the two rays that reach each point~\cite{ValleeSoares2004}. Integrating Eq.~\eqref{eq:onewave} over $s$ gives $\int\Ai\,ds=1$, and differentiating it $n$ times gives the rule we use most,
	\begin{equation}
		\int_{-\infty}^\infty\frac{du}{2\pi}u^ne^{i(su+u^3/3)}=(-i)^n\Ai^{(n)}(s).
		\label{eq:moments}
	\end{equation}
	The maximum of $\Ai^2$ sits at the first zero of $\Ai'$, $s=-1.0188$, the $s=-1.02$ at which Fig.~1(d) of the main text follows the heights.
	
	\textit{The widths of the main text.} For the nearest-neighbor Kitaev chain $|E'''(k_*)|=v$, Eq.~\eqref{eq:kstar}, so $\ell=(|\mu|t/2)^{1/3}$ in the topological and $\ell=(Jt)^{1/3}$ in the trivial phase. At $|\mu|=2J$ the fastest momentum sits at the kink, and Eq.~\eqref{eq:airyscale} applies on each side with $E=2J|\kappa|-\frac J{12}|\kappa|^3$, $|E'''|=J/2$, and $\ell=(Jt/4)^{1/3}$. All three phases have $z=3$. For the cluster chain the dictionary~\eqref{eq:dictionary} gives the widths $(\lambda t)^{1/3}$, $(t/4)^{1/3}$, and $t^{1/3}$ in cells, and doubling them gives the widths in sites of the main text, $(8\lambda t)^{1/3}$, $(2t)^{1/3}$, and $(8t)^{1/3}$ for $\lambda<1$, $=1$, and $>1$.
	
	\subsection{Two waves: the response is a sum over pairs}
	\label{sec:pairs}
	
\begin{figure}[!htb]
		\centering
		\includegraphics[width=\textwidth]{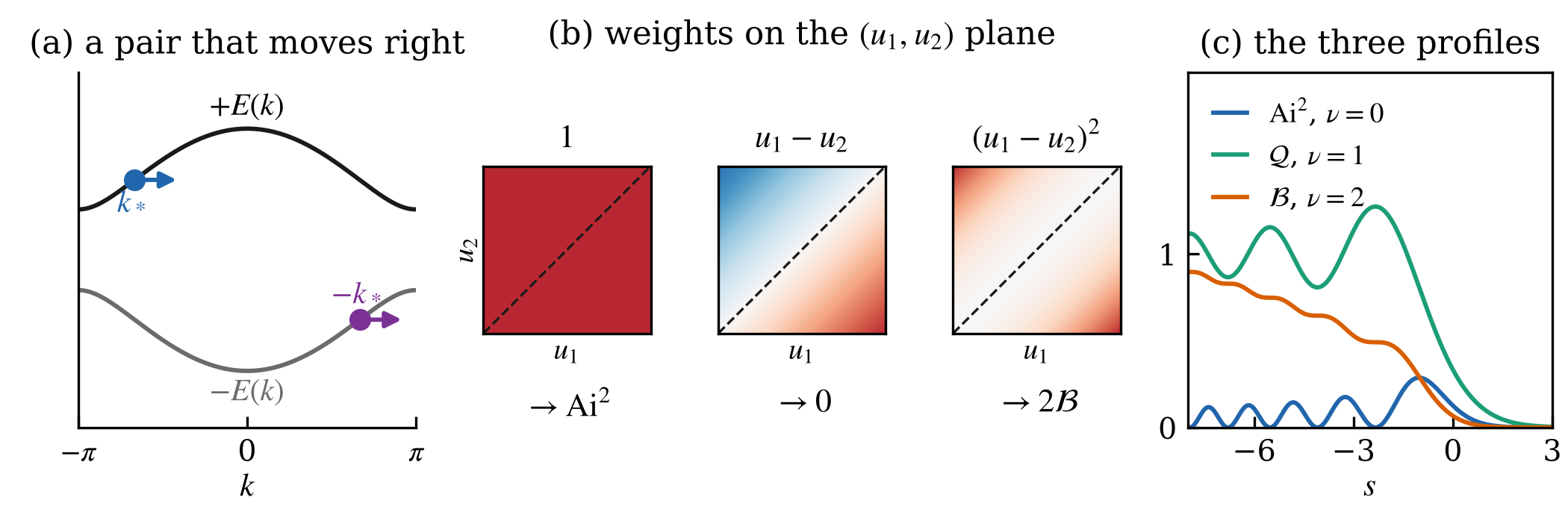}
		\caption{Two waves and their integrals. (a) A pair that builds the right-moving front at $J=\Delta=\mu=1$: a wave on the branch $+E$ near $k_*$ and one on $-E$ near $-k_*$; both move right, and their fast phases cancel. (b) Pair weights on the plane of the offsets $(u_1, u_2)$, red positive and blue negative. The phase $\Theta$ of Eq.~\eqref{eq:Theta} is unchanged by the mirror $u_1\leftrightarrow u_2$ (dashed), so an odd weight integrates to zero: a constant gives $\Ai^2$, $u_1-u_2$ gives zero, $(u_1-u_2)^2$ gives $2\mathcal B$, Eqs.~\eqref{eq:airytable} and \eqref{eq:B}. (c) The three profiles: $\Ai^2$ ($\nu=0$), $\mathcal Q$ ($\nu=1$, Sec.~\ref{sec:critical}), and $\mathcal B$ ($\nu=2$).}
		\label{fig:pairdict}
	\end{figure}

	Equation~\eqref{eq:exact} contains two propagators, one for each index of $S_\eta$. Inserting Eq.~\eqref{eq:Rk} twice, with momenta $k$ and $k'$,
	\begin{equation}
		\bigl[RS_\eta R^{\mathrm T}\bigr]_{jj'}=\int\!\!\int\frac{dk\,dk'}{(2\pi)^2}e^{ik(j-j_0)+ik'(j'-j_0)}\sum_{\tau, \tau'=\pm}e^{-i[\tau E(k)+\tau'E(k')]t}\mathcal P_\tau(k)\tilde S(k, k')\mathcal P_{\tau'}(k')^{\mathrm T},
		\label{eq:double}
	\end{equation}
	where $\tilde S(k, k')=\sum_{l, l'}e^{-ik(l-j_0)-ik'(l'-j_0)}S_{\eta, ll'}$ is the Fourier transform of the change, a $2\times2$ matrix. Each term is a pair of waves, one with momentum $k$ and frequency $\tau E(k)$, the other with $k'$ and $\tau'E(k')$. A wave of frequency $\tau E(k)$ moves with the velocity $\tau E'(k)$, and since $E'(-k)=-E'(k)$ it moves right at the front speed for $\tau=+$ near $k_*$ or $\tau=-$ near $-k_*$, so four pairs reach the right-moving front. In the pairs with opposite frequencies, $(\tau, \tau')=(+, -)$ near $(k_*, -k_*)$ and its mirror $(-, +)$, the fast phases cancel, $e^{i\chi}e^{-i\chi}=1$, and these pairs build the smooth front [Fig.~\ref{fig:pairdict}(a)]. In the pairs with equal frequencies, near $(k_*, k_*)$ and $(-k_*, -k_*)$, a factor $e^{\pm2i\chi}$ survives and oscillates from site to site with the wave number $2k_*$. Mirror pairs are complex conjugates, so two pairs need computing. Where the oscillation is present, the main text replaces $D_j$ by its root mean square $\overline D_j$ over whole periods of consecutive sites, three at $\mu=J$; otherwise $\overline D_j=D_j$.
	
	For the pair $(+, -)$ write $k=k_*+u_1/\ell$ and $k'=-k_*+u_2/\ell$. Since $E$ is even, $E(k')=E(k_*-u_2/\ell)$, and Eq.~\eqref{eq:Etaylor} turns the phase $k'x+E(k')t$ of the second wave into $-\chi+su_2+u_2^3/3$, the mirror image of the first. The pair carries the phase
	\begin{equation}
		\Theta(u_1, u_2)=(u_1+u_2)s+\frac{u_1^3+u_2^3}3,
		\label{eq:Theta}
	\end{equation}
	symmetric under $u_1\leftrightarrow u_2$. Let $W(k, k')$ be the weight of the pair, the number that the observable picks out of $\mathcal P_+\tilde S\mathcal P_-^{\mathrm T}$. Expanded near the fastest momenta, $W=W_0+W_1/\ell+W_2/\ell^2+\ldots$ with $W_n$ a polynomial of degree $n$ in $u_1$ and $u_2$, and since $dk\,dk'=du_1du_2/\ell^2$,
	\begin{equation}
		\delta O\simeq\ell^{-2}\int\!\!\int\frac{du_1du_2}{(2\pi)^2}\Bigl[W_0+\frac{W_1}\ell+\frac{W_2}{\ell^2}+\ldots\Bigr]e^{i\Theta(u_1, u_2)}+\text{conjugate pair}.
		\label{eq:leading}
	\end{equation}
	Each wave contributes $\ell^{-1}$, so the generic height is $\ell^{-2}$, and each power of a momentum offset costs another $\ell^{-1}$. By Eq.~\eqref{eq:moments} for each wave,
	\begin{equation}
		\int\!\!\int\frac{du_1du_2}{(2\pi)^2}u_1^{n_1}u_2^{n_2}e^{i\Theta}=(-i)^{n_1+n_2}\Ai^{(n_1)}(s)\Ai^{(n_2)}(s),
		\label{eq:airytable}
	\end{equation}
	and three cases do all the work [Fig.~\ref{fig:pairdict}(b)]. A constant weight gives $\Ai(s)^2$. A weight odd under $u_1\leftrightarrow u_2$, such as $u_1-u_2$, gives zero, because $\Theta$ is symmetric. The square of the odd weight gives the second profile of the main text,
	\begin{equation}
		\int\!\!\int\frac{du_1du_2}{(2\pi)^2}(u_1-u_2)^2e^{i\Theta}=-2\Ai''\Ai+2\Ai'^2=2\mathcal B(s), \qquad \mathcal B(s)=\Ai'(s)^2-s\Ai(s)^2=\int_s^\infty\Ai(u)^2du,
		\label{eq:B}
	\end{equation}
	by $\Ai''=s\Ai$; the integral form follows from $\mathcal B'=-\Ai^2$ and $\mathcal B(+\infty)=0$. Two more entries are needed below, $(u_1^2+u_2^2)\to-2s\Ai^2$ and $(u_1-u_2)(u_1^4-u_2^4)\to4i(\Ai'^2-2s\Ai^2)$, both from Eq.~\eqref{eq:airytable} with $\Ai'''=\Ai+s\Ai'$ and the next derivatives. The profile $\mathcal B$ is positive and vanishes ahead of the front; behind it, it grows as $|s|^{1/2}/\pi$ like a staircase, flat at the zeros of $\Ai$, where $\mathcal B'=-\Ai^2$ vanishes. The same function gives the density at the front of free fermions released from a filled half-chain~\cite{EislerRacz2013}.
	
	\subsection{The weight factorizes: measurement factor times observable factor}
	\label{sec:factor}
	
	Every observable of the main text is a bilinear $\Gamma_{a_jb_{j+d}}$, with $d=0$ for the density and $d=1$ for the bond energy; the stabilizer is the bond energy of its sublattice, Eq.~\eqref{eq:dictionary}. The observable enters through the phase
	\begin{equation}
		G(k)=e^{-ikd}g(k), \qquad G(k)=g(k)\ \text{for the density}, \qquad G(k)=e^{-ik}g(k)\ \text{for the bond energy},
		\label{eq:G}
	\end{equation}
	whose real part is the observable factor of Eqs.~\eqref{eq:uv} and \eqref{eq:complement}. The projector $\mathcal P_+(k)$ has rank one, $\mathcal P_+(k)=\mathbf u(k)\mathbf u(k)^\dagger$ with $\mathbf u(k)=(-ig(k), 1)^{\mathrm T}/\sqrt2$, and $\mathcal P_-(k')^{\mathrm T}=\mathbf u(-k')\mathbf u(-k')^\dagger$, so the middle of Eq.~\eqref{eq:double} is one number times a fixed matrix,
	\begin{equation}
		\mathcal P_+(k)\tilde S(k, k')\mathcal P_-(k')^{\mathrm T}=\mathcal S(k, k')\,\mathbf u(k)\mathbf u(-k')^\dagger, \qquad \mathcal S(k, k')=\mathbf u(k)^\dagger\tilde S(k, k')\mathbf u(-k'),
		\label{eq:Scal}
	\end{equation}
where $\mathcal S(k, k')$ says how strongly the measurement couples the two waves. Real couplings give $g(-k')=\overline{g(k')}$, so with rows and columns in the order $(a, b)$ the fixed matrix is
\begin{equation}
	\mathbf u(k)\mathbf u(-k')^\dagger=\frac12\begin{pmatrix}g(k)g(k')&-ig(k)\\ig(k')&1\end{pmatrix}.
	\label{eq:uu}
\end{equation}
Its entry $(a, b)$ lands on $(a_j, b_{j+d})$ in Eq.~\eqref{eq:double} with the phase $e^{ik'd}$, and its entry $(b, a)$ lands on $(b_{j+d}, a_j)$ with $e^{ikd}$; the common factor $e^{i(k+k')x}$ goes into $\Theta$. Since $\delta\Gamma$ is antisymmetric, half the difference of the two entries gives the weight of the pair exactly,
	\begin{equation}
		W_{+-}(k, k')=-\frac i4\mathcal S(k, k')\,e^{i(k+k')d}\bigl[G(k)+G(k')\bigr].
		\label{eq:Wpm}
	\end{equation}
	At $k=k_*$, $k'=-k_*$, the bracket is $2\re G(k_*)$, the observable factor. And since $S_\eta$ is real and antisymmetric, $\tilde S(k, -k)$ is anti-Hermitian, $\mathcal S(k_*, -k_*)$ is imaginary, and
	\begin{equation}
		\mathcal M=i\mathcal S(k_*, -k_*)
		\label{eq:Mfactor}
	\end{equation}
	is real: the measurement factor. With $W_0=-\frac i2\mathcal S(k_*, -k_*)\re G(k_*)$ in Eq.~\eqref{eq:leading}, the constant weight gives $\Ai^2$ and the conjugate pair doubles the real part:
	\begin{equation}
		\delta\Gamma_{a_jb_{j+d}}\simeq-\mathcal M\,\re G(k_*)\,\ell^{-2}\Ai(s)^2, \qquad s=\frac{x+d-vt}\ell.
		\label{eq:factor}
	\end{equation}
	The measurement enters only through $\mathcal M$, the observable only through $\re G(k_*)$: the factorization of the main text, how strongly $M$ creates the fastest quasiparticles times how strongly $O_j$ detects them. The equal-frequency pairs have a weight proportional to $G(k)-G(k')$, which vanishes at $k=k'$; its linear term is odd (Sec.~\ref{sec:blind}), so they start at order $\ell^{-4}$. Any other local quadratic observable works the same way: its pair weight is $\mathcal S(k, k')$ times a matrix element of the observable between $\mathbf u(-k')$ and $\mathbf u(k)$, which at $(k_*, -k_*)$ is its observable factor.
	
\textit{The parity measurement always creates the fastest quasiparticles.} For $M=ia_{j_0}b_{j_0}$, the three rules of Sec.~\ref{sec:parity} and Eq.~\eqref{eq:Gamma0ring} combine into one expression for the change, nonzero only between $a$ and $b$ Majoranas,
\begin{equation}
	S_{\eta, a_lb_{l'}}=-\frac\eta{2p_\eta}\bigl(g_{l-j_0}+\eta\delta_{lj_0}\bigr)\bigl(g_{j_0-l'}+\eta\delta_{l'j_0}\bigr)+2\eta\,\delta_{lj_0}\delta_{l'j_0},
	\label{eq:Sreal}
\end{equation}
which, with $\eta+m=2\eta p_\eta$, is rule (iii) for $l, l'\neq j_0$, the erasure of rule (ii) when one index is $j_0$, and $\eta-m$ at $l=l'=j_0$, as rule (i) requires. With $g=g(k)$ and $g'=g(k')$, its Fourier transform follows from $\sum_rg_re^{-ikr}=g(k)$, and Eq.~\eqref{eq:Scal} reads $\mathcal S=\frac12\bigl(i\bar g\tilde S_{ab}-i\bar g'\tilde S_{ba}\bigr)$, so
	\begin{equation}
		\tilde S(k, k')=\frac1{2p_\eta}\begin{pmatrix}0&\eta+2m-g-\bar g'-\eta g\bar g'\\-\eta-2m+\bar g+g'+\eta\bar gg'&0\end{pmatrix}, \qquad \mathcal S(k, k')=\frac i{2p_\eta}\bigl[m(\bar g+\bar g')-1-\bar g\bar g'\bigr].
		\label{eq:Sparity}
	\end{equation}
	At $k'=-k_*$, $\bar g'=g(k_*)$, and since $|g|=1$,
	\begin{equation}
		\mathcal M=\frac{1-m\re g(k_*)}{p_\eta}>0,
		\label{eq:Mparity}
	\end{equation}
	positive because $|m|<1$ whenever both outcomes occur and $|\re g|\leq1$: in every chain of these notes a parity measurement creates the fastest quasiparticles, and the observable factor alone decides whether the leading term survives. For the density, $d=0$, Eqs.~\eqref{eq:Wpm} and \eqref{eq:Sparity} give real weights for the two kinds of pair,
	\begin{equation}
		W_{+-}=\frac{m\bigl[1+\re(g\bar g')\bigr]-\re g-\re g'}{4p_\eta}, \qquad W_{++}=\frac\eta4|g-g'|^2.
		\label{eq:Wdensity}
	\end{equation}
	
	\subsection{When the observable is blind: two powers are lost}
	\label{sec:blind}
	
\begin{figure}[!htb]
		\centering
		\includegraphics[width=\textwidth]{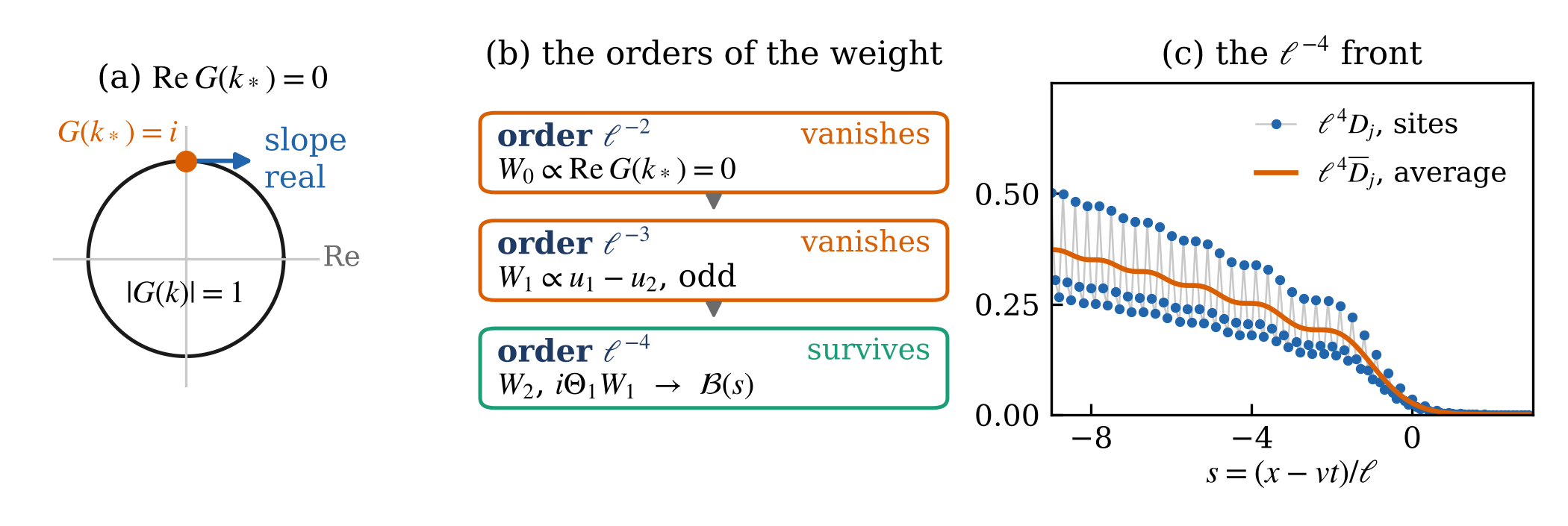}
		\caption{Why a blind observable loses two powers of $\ell$. (a) $G(k)$ lies on the unit circle; blind means $G(k_*)=\pm i$, where the circle is horizontal and $G'(k_*)$ is real, Eq.~\eqref{eq:Gfirst}. (b) The orders of the pair weight: the constant vanishes, the odd linear term integrates to zero, and the second order, with the quartic phase correction $\Theta_1$ of Eq.~\eqref{eq:Theta1}, survives with the profile $\mathcal B$. (c) The running example at $t=2000$, Eq.~\eqref{eq:Dtopo}: $\ell^4D_j$ (dots) oscillates with a period of three sites; its local average $\ell^4\overline D_j$ (line) is smooth.}
		\label{fig:cancel}
	\end{figure}

	Suppose $\re G(k_*)=0$, so $G(k_*)=\pm i$. Then $W_0$ vanishes: the observable is blind to the fastest quasiparticles, and, as in the main text, we call this loss of the leading term a cancellation. The front must then come from pairs whose momenta are offset from $k_*$ by about $\ell^{-1}$. Write $G=e^{i\varphi(k)}$. At $G(k_*)=\pm i$ the derivative $G'=i\varphi'G$ is real: at the top or bottom of the unit circle the tangent is horizontal [Fig.~\ref{fig:cancel}(a)]. With $G(k')=\overline{G(k_*-u_2/\ell)}$,
	\begin{equation}
		G(k)+G(k')=\frac{G'(k_*)}\ell\,(u_1-u_2)+O(\ell^{-2}),
		\label{eq:Gfirst}
	\end{equation}
	so $W_1\propto u_1-u_2$ is odd under the exchange of the two waves and integrates to zero, Eq.~\eqref{eq:airytable}. The orders $\ell^{-2}$ and $\ell^{-3}$ both vanish: losing the leading term costs two powers of $\ell$, not one, because the first correction is odd. At order $\ell^{-4}$ three things survive [Fig.~\ref{fig:cancel}(b)]: the quadratic terms of the weight, the odd linear term times the odd quartic correction of the phase, and the equal-frequency pairs. The quartic term of Eq.~\eqref{eq:Etaylor}, $tE''''(k_*)(k-k_*)^4/24$ in the main text, adds to the phase of the pair
	\begin{equation}
		\frac{\Theta_1}\ell=-\frac{E''''(k_*)}{12|E'''(k_*)|\,\ell}\bigl(u_1^4-u_2^4\bigr),
		\label{eq:Theta1}
	\end{equation}
	since $t\kappa^4/24=(2\ell^3/|E'''|)(u/\ell)^4/24$; it is odd and of order $\ell^{-1}$, and its product with $W_1$ is even. For a general blind observable, with $\varphi'$ and $\varphi''$ the derivatives of the phase of $G$ at $k_*$ and $\mathcal S$ expanded to first order, the smooth part is
	\begin{equation}
		\delta\Gamma_{a_jb_{j+d}}\simeq2\re\Bigl\{-\frac{G(k_*)}4\Bigl[\mathcal S_*\bigl((\varphi''+2\Xi)s\Ai^2-\Xi\Ai'^2\bigr)-\varphi'(\partial_1-\partial_2)\mathcal S\,\mathcal B\Bigr]\Bigr\}\ell^{-4}, \qquad \Xi=\frac{\varphi'E''''(k_*)}{3|E'''(k_*)|},
		\label{eq:secondgeneral}
	\end{equation}
	with $\mathcal S_*=\mathcal S(k_*, -k_*)$ and $\partial_1$, $\partial_2$ the derivatives with respect to the two momenta. It follows as in Sec.~\ref{sec:example}: expand Eq.~\eqref{eq:Wpm} to second order in the offsets, add $i\Theta_1W_1$, and integrate with Eqs.~\eqref{eq:airytable} and \eqref{eq:B} and the two entries below Eq.~\eqref{eq:B}; the terms odd under $u_1\leftrightarrow u_2$ drop. For the density it gives twice Eq.~\eqref{eq:topoint}, the pair plus its conjugate. The bracket of $\mathcal S_*$ is proportional to $\mathcal B=\Ai'^2-s\Ai^2$ exactly when $\varphi''=-\Xi$. This holds for the nearest-neighbor chains, where energy and observable phase come from the same circle, and the profile is pure $\mathcal B$; otherwise $\mathcal B$ mixes with $s\Ai^2$, as on the line $\theta_{\mathrm b}$ of Sec.~\ref{sec:lr}. The equal-frequency pairs give an oscillating term $\propto\ell^{-4}\mathcal B(s)\cos(2\chi+\chi_0)$, with $\chi_0$ a fixed phase.
	
	\textit{The rule of the main text.} Let $\nu$ be the lowest order in $\ell^{-1}$ at which the weight survives the integration over both momenta, counting the offsets and the small phases from the higher terms of $E(k)$. At an open end each wave also carries its first-site amplitude, Eq.~\eqref{eq:hardwall}; if it vanishes as $|k-k_*|^\sigma$, each wave loses a further $\ell^{-\sigma}$, and $\sigma=0$ in the bulk. Collecting the powers gives Eq.~\eqref{eq:rule}, Eq.~(3) of the main text, $\overline D_j\simeq\ell^{-2-\nu-2\sigma}\Phi(s)$, decaying as $t^{-(2+\nu+2\sigma)/z}$. The derivation assumes a single fastest momentum (one pair of waves on the ray $x=vt$), a transform $\tilde S$ bounded near the fastest momenta, and dominance of the contribution from two front waves. If instead $\mathcal M$ vanished, the leading term would vanish just as well and the front would again come from slightly slower quasiparticles, weaker and decaying faster. Section~\ref{sec:Zcrit} treats a string measurement with an unbounded transform. Section~\ref{sec:lr} includes multiple equally fast momenta and the edge case of Eq.~\eqref{eq:Dkpiedge}, where an initial correlation tail dominates the contribution from two front waves. A cancellation at a smooth maximum of $v(k)$ gives $\nu=2$ by the odd-linear-term argument; at a linear gap closing the argument changes, Sec.~\ref{sec:critical}.
	
	\subsection{The running example: one density measurement at $J=\Delta=\mu$}
	\label{sec:example}
	
	At $\mu=J$, Sec.~\ref{sec:speed} gives $k_*=-2\pi/3$, $v=J$, $E(k_*)=\sqrt3J$, $E'''(k_*)=-J$, so $\ell=(Jt/2)^{1/3}$; Eq.~\eqref{eq:Gamma0ring} gives $m=g_0=-0.2587$; and $g(k_*)=i$.
	
\begin{figure}[!htb]
		\centering
		\includegraphics[width=\textwidth]{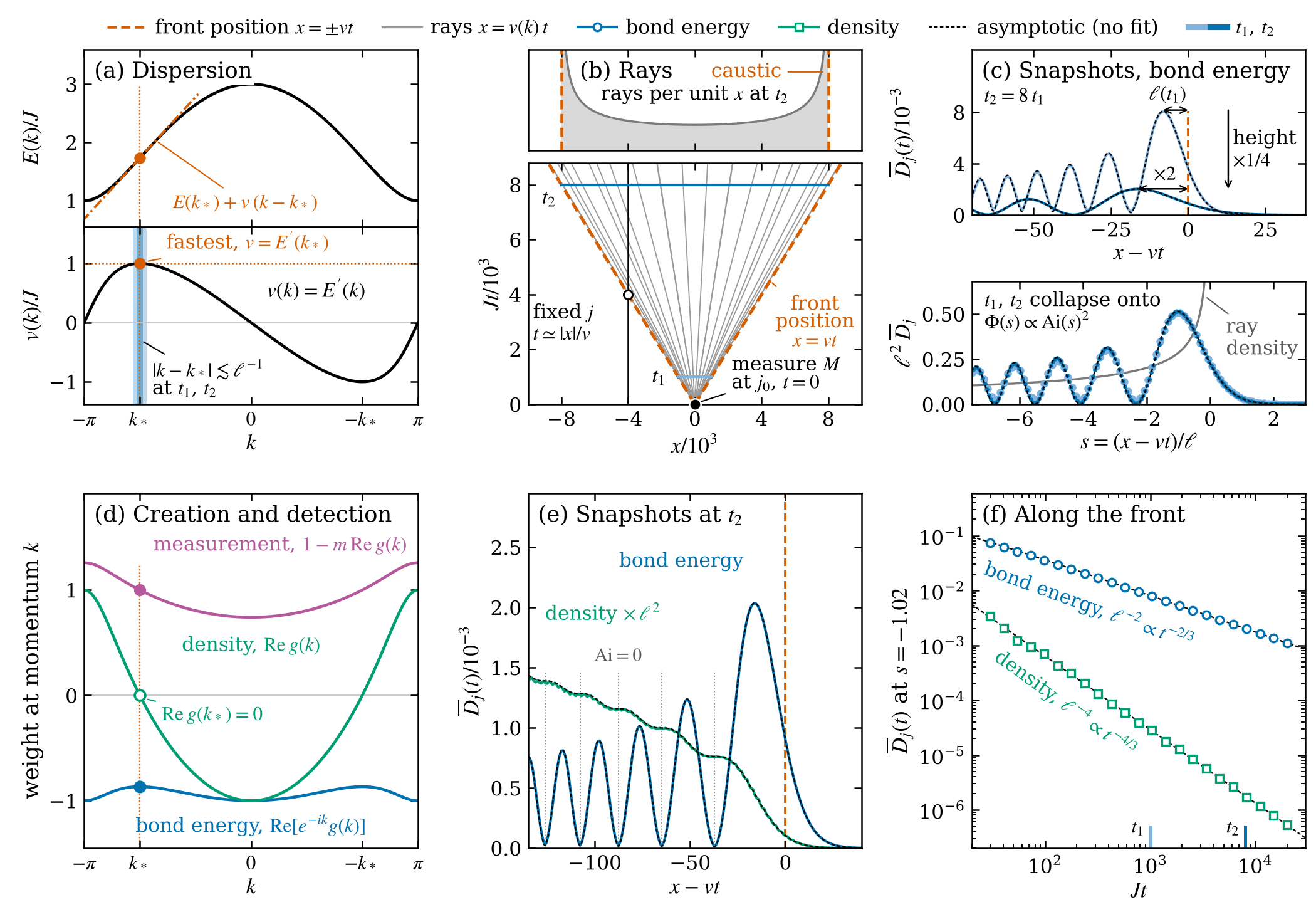}
		\caption{Anatomy of the running example, density measured at $j_0$, $J=\Delta=\mu=1$; exact data from Eq.~\eqref{eq:exact}, nothing fitted. (a) $E(k)$ with its tangent at $k_*$, and $v(k)$ with the windows $|k-k_*|\lesssim\ell^{-1}$ at $t_1=1000$ and $t_2=8000$. (b) Rays $x=v(k)t$ and their density, which diverges at the caustic $x=\pm vt$. (c) The bond-energy front at $t_1$ and $t_2$: $\ell^2\overline D_j$ collapses onto $1.7931\Ai(s)^2$, Eq.~\eqref{eq:DBtopo}. (d) The factors $1-m\re g(k)$, $\re g(k)$, and $\re[e^{-ik}g(k)]$; the density factor vanishes at $k_*$ (open marker). (e) The density at $t_2$, times $\ell^2$, beside the bond energy: its profile $\mathcal B$ rises behind the front like a staircase, flat at the zeros of $\Ai$ (dotted). (f) The heights at $s=-1.02$, $\ell^{-2}\propto t^{-2/3}$ for the bond energy and $\ell^{-4}\propto t^{-4/3}$ for the density, with the derived prefactors (dashed).}
		\label{fig:anatomy}
	\end{figure}

	\textit{The bond energy sees the fastest quasiparticles.} Its observable factor is $\re[e^{-ik_*}g(k_*)]=-E(k_*)/(2J)=-\sqrt3/2$, and $\mathcal M=1/p_\eta$ since $\re g(k_*)=0$. With $\langle B_j\rangle=-\Gamma_{a_jb_{j+1}}$, Eq.~\eqref{eq:factor} gives $\delta B_{j, \eta}=-(E(k_*)/2J)\ell^{-2}\Ai^2/p_\eta$, and with $\sum_\eta p_\eta/p_\eta^2=4/(1-m^2)$,
	\begin{equation}
		D^B_j(t)\simeq\frac{E(k_*)}{J\sqrt{1-m^2}}\,\ell^{-2}\Ai(s)^2, \qquad \ell=\Bigl(\frac{\mu t}2\Bigr)^{1/3},
		\label{eq:DBtopo}
	\end{equation}
	which is $1.7931\,\ell^{-2}\Ai(s)^2$ at $\mu=J$, with $s$ measured from the bond midpoint $x=j+\frac12-j_0$ (a shift of order $\ell^{-1}$ in $s$). From $t_1$ to $8t_1$ the width doubles and the height drops fourfold [Fig.~\ref{fig:anatomy}(c)].
	
	\textit{The density is blind and loses two powers.} We need $g$ near $k_*$ to second order. With $g=e^{i\varphi}$ and $\varphi'=\im(h'/h)$, one finds at $k_*$ that $\varphi'=1$ and $\varphi''=-c$ with $c=\mu/E(k_*)$, so
	\begin{equation}
		g(k_*+\kappa)=i\exp\bigl(i\kappa-\tfrac i2c\kappa^2\bigr)+O(\kappa^3)=i-\kappa+\tfrac12(c-i)\kappa^2+O(\kappa^3), \qquad c=\frac\mu{E(k_*)}.
		\label{eq:gexp}
	\end{equation}
	For the pair $(+, -)$, $\re g=-u_1/\ell+cu_1^2/2\ell^2$, $\re g'=u_2/\ell+cu_2^2/2\ell^2$, and $1+\re(g\bar g')=(u_1-u_2)^2/2\ell^2$ to this order, and the weight~\eqref{eq:Wdensity} becomes
	\begin{equation}
		W_{+-}=\frac{u_1-u_2}{4p_\eta\ell}+\frac{m(u_1-u_2)^2-c(u_1^2+u_2^2)}{8p_\eta\ell^2}+O(\ell^{-3}).
		\label{eq:Wtopo}
	\end{equation}
	The linear term is odd and drops. With $E''''(k_*)=3\mu^2/E(k_*)$ and $|E'''(k_*)|=\mu$, Eq.~\eqref{eq:Theta1} gives $\Theta_1=-\frac c4(u_1^4-u_2^4)$, and the second-order integrand $W_2+i\Theta_1W_1$ integrates, by Eq.~\eqref{eq:B} and the two entries below it, to
	\begin{equation}
		\frac{2m\mathcal B+2cs\Ai^2}{8p_\eta}+\frac c{4p_\eta}\bigl(\Ai'^2-2s\Ai^2\bigr)=\frac{m+c}{4p_\eta}\mathcal B(s).
		\label{eq:topoint}
	\end{equation}
	The quartic phase correction turns the $s\Ai^2$ of the quadratic weight into $\mathcal B$: small phase corrections are of the same order as weight corrections and cannot be dropped. The conjugate pair doubles Eq.~\eqref{eq:topoint}. The equal-frequency weight $W_{++}=\frac\eta4(u_1-u_2)^2/\ell^2$ gives $\eta\mathcal B\cos2\chi$. Altogether, with $\delta n_j=-\frac12\delta\Gamma_{a_jb_j}$, the outcome sums~\eqref{eq:sums}, and the local average $\cos^22\chi\to\frac12$,
	\begin{equation}
		\delta n_{j, \eta}\simeq-\frac{\ell^{-4}\mathcal B(s)}2\Bigl[\frac{m+c}{2p_\eta}+\eta\cos2\chi\Bigr], \qquad \overline D_j(t)\simeq\frac{\ell^{-4}\mathcal B(s)}2\Bigl[\frac{(m+c)^2}{1-m^2}+\frac12\Bigr]^{1/2}, \qquad s=\frac{x-\mu t}\ell.
		\label{eq:Dtopo}
	\end{equation}
	At $\mu=J$, $c=1/\sqrt3$ and $(m+c)^2/(1-m^2)=0.1088$, so $\overline D_j\simeq0.3901\,\ell^{-4}\mathcal B(s)=0.983\,(Jt)^{-4/3}\mathcal B(s)$. Since $4k_*\equiv-2\pi/3$, $D_j^2$ oscillates with a period of three sites, and a three-site average implements $\overline D_j$ [Fig.~\ref{fig:cancel}(c)]. At $s=-1.02$ the leading bond-to-density front-height ratio is about $4.51\ell^2$, or about $450$ at $Jt=2000$, where $\ell=10$: Fig.~1(d) of the main text and Fig.~\ref{fig:anatomy}(f) here.
	
	\subsection{What if: one change at a time}
	\label{sec:whatif}
	
	The running example contains the whole method: find $k_*$, $v$, and $\ell$ from the energy, the observable factor $\re G(k_*)$, and the measurement factor, then integrate. Every other front of the main text changes one ingredient.
	
	\subsubsection{The trivial phase: the two laws exchange}
	\label{sec:trivial}
	
	For $\mu>2J$ the fastest quasiparticles move at $v=2J$, $\ell=(Jt)^{1/3}$, and by Sec.~\ref{sec:speed} the density sees them, $\re g(k_*)=-E(k_*)/\mu$, while the bond energy is blind. For the density, Eqs.~\eqref{eq:factor} and \eqref{eq:Mparity} give $\delta n_{j, \eta}=\re g(k_*)[1-m\re g(k_*)]\ell^{-2}\Ai^2/(2p_\eta)$, and the outcome sums give
	\begin{equation}
		D_j(t)\simeq\frac{|\re g(k_*)|\bigl[1-m\re g(k_*)\bigr]}{\sqrt{1-m^2}}\ell^{-2}\Ai(s)^2, \qquad s=\frac{x-2Jt}{(Jt)^{1/3}},
		\label{eq:Dtriv}
	\end{equation}
	a $t^{-2/3}$ decay. For the blind bond energy, Eq.~\eqref{eq:secondgeneral} applies to the phase $e^{-ik}g$, with $\varphi'=-1$, $\varphi''=2J/E(k_*)$, and $E''''(k_*)=12J^2/E(k_*)$ at $k_*$, so $\Xi=-\varphi''$ and the profile is pure $\mathcal B$; the phase of $g$ itself is stationary at $\pm k_*$, so $(\partial_1-\partial_2)\mathcal S=0$. With $\mathcal S_*=-i[1-m\re g(k_*)]/p_\eta$,
	\begin{equation}
		D^B_j(t)\simeq\frac{2J\bigl[1-m\re g(k_*)\bigr]}{E(k_*)\sqrt{1-m^2}}\ell^{-4}\mathcal B(s), \qquad \ell=(Jt)^{1/3},
		\label{eq:DBtriv}
	\end{equation}
	a $t^{-4/3}$ decay. The two laws of the topological phase have exchanged places: which observable decays faster is decided by the phase, and within a phase by the observable.
	
	\subsubsection{The critical point: right movers on one side only}
	\label{sec:critical}
	
\begin{figure}[!htb]
		\centering
		\includegraphics[width=\textwidth]{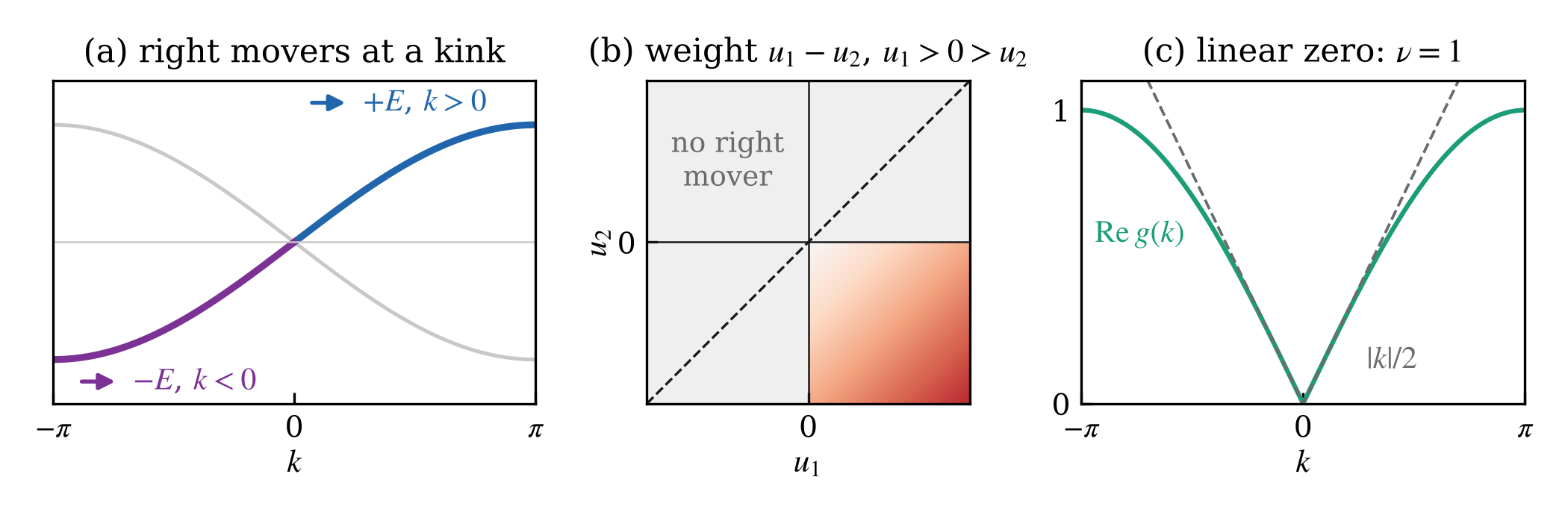}
		\caption{The front at a gap closing, $J=\Delta=1$, $\mu=-2J$, $E(k)=4J|\sin(k/2)|$. (a) Only $k>0$ on the branch $+E$ and $k<0$ on $-E$ move right. (b) The pair lives on the quadrant $u_1>0>u_2$ (red, weight $u_1-u_2$); the exchange $u_1\leftrightarrow u_2$ (dashed) maps it onto a quadrant with no right mover (gray), so the odd weight survives and gives $\ell^{-3}\mathcal Q(s)$, Eq.~\eqref{eq:Q}. (c) The density factor $\re g(k)=\sin(|k|/2)$ vanishes linearly: $\nu=1$.}
		\label{fig:onesided}
	\end{figure}

	At $\mu=-2J$ the fastest quasiparticles sit at the gap closing $k=0$, where $E=2J|k|-\frac J{12}|k|^3$ has a kink [Fig.~\ref{fig:onesided}(a)]. This is the $\mu=2J$ of Fig.~\ref{fig:circles}(b) after the map of Sec.~\ref{sec:speed}, which moves the closing from $k=\pi$ to $k=0$ and reverses the signs of $m$ and $m_{\mathrm e}$ (the $-8/(3\pi)$ below Eq.~\eqref{eq:me} becomes the $8/(3\pi)$ of Eq.~\eqref{eq:Dedgecrit}) but leaves $D_j$ unchanged. A wave on the branch $+E$ moves right only for $k>0$, one on $-E$ only for $k<0$. With $k=u_1/\ell$, $u_1>0$, and $k'=u_2/\ell$, $u_2<0$, the phase of each wave is again $su+u^3/3$, now with $\ell=(Jt/4)^{1/3}$. The pair lives on the quadrant $u_1>0>u_2$, and the exchange $u_1\leftrightarrow u_2$ maps it onto a quadrant with no right mover [Fig.~\ref{fig:onesided}(b)]: the symmetry that killed the odd linear term is gone.
	
	Here $g(k)=-ie^{ik/2}\sgn(k)$, Sec.~\ref{sec:ground}, so $\re g(k)=\sin(|k|/2)$ vanishes linearly: $\nu=1$ [Fig.~\ref{fig:onesided}(c)]. Since $\re g(0^\pm)=0$, $\mathcal M=1/p_\eta$, and Eq.~\eqref{eq:Wdensity} gives $W_{+-}=-(u_1-u_2)/(8p_\eta\ell)$. With the one-sided integrals $\mathcal A_n(s)=\int_0^\infty\frac{du}{2\pi}u^ne^{i(su+u^3/3)}$ for the first wave and their conjugates for the second, the weight $u_1-u_2$ integrates to a new profile,
	\begin{equation}
		\mathcal A_1\bar{\mathcal A}_0+\mathcal A_0\bar{\mathcal A}_1=\frac{\mathcal Q(s)}{2\pi}, \qquad \mathcal Q(s)=\int_s^\infty\Ai(u)\,du.
		\label{eq:Q}
	\end{equation}
	To see it, call the left side $F(s)$. Since $\mathcal A_1=-i\mathcal A_0'$, $F=2\im(\mathcal A_0'\bar{\mathcal A}_0)$ and $F'=2\im(\mathcal A_0''\bar{\mathcal A}_0)$. Integrating $\frac d{du}e^{i(su+u^3/3)}$ from $0$ to $\infty$ gives $\mathcal A_0''=s\mathcal A_0-i/(2\pi)$, the boundary term at $u=0$ being new. Hence $F'=-\re\mathcal A_0/\pi=-\Ai/(2\pi)$, since $\re\mathcal A_0=\Ai/2$, and $F(+\infty)=0$ gives the identity. With the conjugate pair and $\delta n=-\frac12\delta\Gamma_{a_jb_j}$,
	\begin{equation}
		\delta n_{j, \eta}\simeq\frac{\ell^{-3}\mathcal Q(s)}{16\pi p_\eta}, \qquad D_j(t)\simeq\frac{\ell^{-3}\mathcal Q(s)}{8\pi\sqrt{1-m^2}}, \qquad D^B_j(t)\simeq\frac{\ell^{-3}\mathcal Q(s)}{4\pi\sqrt{1-m^2}}, \qquad \ell=\Bigl(\frac{Jt}4\Bigr)^{1/3},
		\label{eq:Dcrit}
	\end{equation}
	with $m=2/\pi$ and $s=(x-2Jt)/\ell$; the bond energy, whose weight also vanishes linearly, follows in the same way. Both decay as $t^{-1}$. The profile $\mathcal Q$ is positive, vanishes ahead of the front, and oscillates about $\int\Ai=1$ behind it, with its largest value $1.274$ at the first zero of $\Ai$, $s=-2.338$ [Fig.~\ref{fig:pairdict}(c)]. At a linear closing the observable factor that vanishes as $|k-k_*|^\nu$ sets $\nu$ directly, as the main text says.
	
	\textit{The tail ahead of the critical front.} At $t=0$ the critical chain already responds as $1/x^2$, Eq.~\eqref{eq:tailt0}. Far ahead of the front, $x-vt\gg\ell$, only the shift term of Eq.~\eqref{eq:update} survives, the product of the propagated columns $R\Gamma_0e_{a_{j_0}}$ and $R\Gamma_0e_{b_{j_0}}$, whose entries $\int\frac{dk}{2\pi}y(k)e^{ikx\pm iE(k)t}$ have no stationary point there. Only the jump of $g$ at $k=0$ contributes, and one integration by parts on each side gives $\int\frac{dk}{2\pi}e^{ikx}\sin Et\simeq-vt/[\pi(x^2-v^2t^2)]$ and $\int\frac{dk}{2\pi}e^{ikx}g\cos Et\simeq ig(0^+)x/[\pi(x^2-v^2t^2)]$, and
	\begin{equation}
		D_j(t)\simeq\frac1{2\pi^2\sqrt{1-m^2}\,(x^2-v^2t^2)}\qquad(x-vt\gg\ell),
		\label{eq:tail}
	\end{equation}
	which at $t=0$ is Eq.~\eqref{eq:tailt0} [Fig.~\ref{fig:edgebond}(b)]. The tail grows by $x^2/(x^2-v^2t^2)$ as the front approaches, and at $x-vt\sim\ell$ it is of order $1/(vt\ell)$, a factor $\ell^{-1}$ below the front height $\ell^{-3}$, so it leaves the height law alone. It is proportional to $\eta/(2p_\eta)$, whose outcome average vanishes: no signal, only a correlation with the outcome inherited from the ground state.
	
	\subsubsection{An open end}
	\label{sec:edge}
	
\begin{figure}[!htb]
		\centering
		\includegraphics[width=\textwidth]{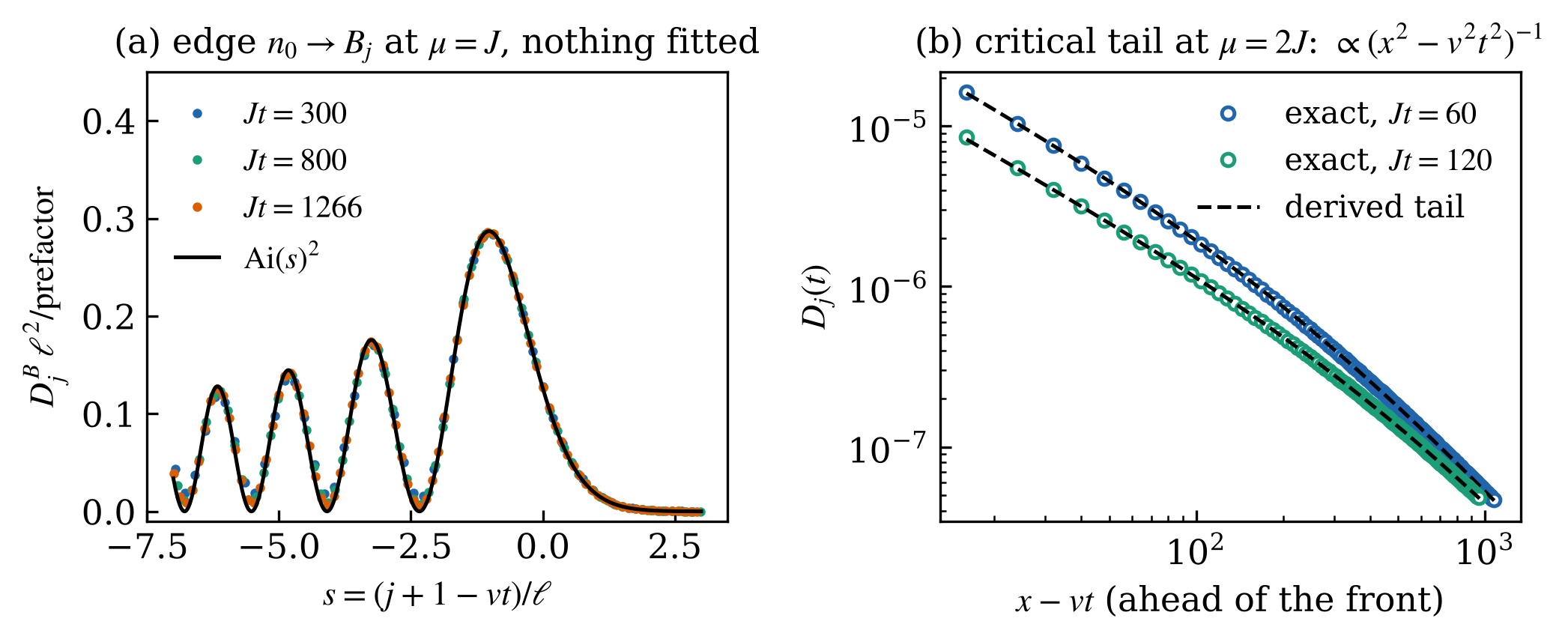}
		\caption{Two checks against the exact response, Eq.~\eqref{eq:exact}, $J=\Delta=1$. (a) The bond energy after measuring $n_0$ on an open chain of $3000$ sites at $\mu=J$, divided by the derived height and prefactor: three times fall on $\Ai(s)^2$, with a residual that shrinks as $\ell$ grows. (b) Ahead of the critical front, $\mu=2J$, on a ring of $N=2^{16}$ sites with the antiperiodic boundary condition $c_{j+N}=-c_j$ (circles), against Eq.~\eqref{eq:tail} (dashed): a correlation with the outcome that exists from $t=0$ and grows as the front approaches.}
		\label{fig:edgebond}
	\end{figure}

	Measure the density of the first site of a half-line, $j_0=0$. Expanding the change of Sec.~\ref{sec:parity} in the standing waves of Eq.~\eqref{eq:hardwall}, and the observable in their plane-wave parts far from the end, gives a double integral of the form of Eq.~\eqref{eq:double}, with the same waves and phases, positions measured from $j+1$, and weights that carry the first-site amplitudes:
	\begin{equation}
		W^{\mathrm e}_{\tau, -\tau}=\frac{\sin k\sin k'\,\overline{g(k')}\,\mathcal N(k, k')}{p_\eta}, \qquad \mathcal N=1+m_{\mathrm e}\bigl(\omega_k+\omega_{k'}\bigr)+\omega_k\omega_{k'}, \qquad \omega_k=\frac\mu{E(k)}.
		\label{eq:edgeweights}
	\end{equation}
	The observable enters only through $\overline{g(k')}$, which at $k'=-k_*$ is the bulk factor $g(k_*)$; only the measurement factor changes, to $-\sin^2k_*\,\mathcal N_*/p_\eta$ with $\mathcal N_*=(\omega_*+m_{\mathrm e})^2+1-m_{\mathrm e}^2>0$ and $\omega_*=\mu/E(k_*)$. At the fastest momentum $\sin k_*\neq0$, so $\sigma=0$ and every power is unchanged: the cancellation survives at an open end. The steps of the bulk give, in the trivial and topological phases,
	\begin{equation}
		D_j\simeq\frac{2|\re g(k_*)|\bigl[(|\re g(k_*)|+m_{\mathrm e})^2+1-m_{\mathrm e}^2\bigr]}{\sqrt{1-m_{\mathrm e}^2}}\ell^{-2}\Ai(s)^2, \qquad s=\frac{j+1-2Jt}{(Jt)^{1/3}},
		\label{eq:Dedgetriv}
	\end{equation}
	\begin{equation}
		\overline D_j\simeq\frac{\mu^2}{2J^2}\ell^{-4}\mathcal B(s)\Bigl[\frac{4(m_{\mathrm e}+c_{\mathrm e})^2}{1-m_{\mathrm e}^2}+\frac12\Bigr]^{1/2}, \qquad c_{\mathrm e}=\frac{3+c^2}{4c}, \qquad c=\frac\mu{E(k_*)}, \qquad \ell=\Bigl(\frac{\mu t}2\Bigr)^{1/3},
		\label{eq:Dedgetopo}
	\end{equation}
	the second from the factor $\sin k\sin k'\mathcal N/p_\eta$ expanded to first order in the offsets. At the closing $f^a_k\propto\sin k$ vanishes, but $f^b_k=\omega_kf^a_k$ stays finite: the Majorana $b_0$ carries the coupling, again $\sigma=0$, and the one-sided calculation gives twice the bulk front with $m\to m_{\mathrm e}$,
	\begin{equation}
		D_j(t)\simeq\frac{\ell^{-3}\mathcal Q(s)}{4\pi\sqrt{1-m_{\mathrm e}^2}}, \qquad m_{\mathrm e}=\frac8{3\pi}.
		\label{eq:Dedgecrit}
	\end{equation}
	The factor two holds at every linear closing of these notes. For the bond energy the second wave carries $e^{ik'}\overline{g(k')}$, the bulk bond phase at $k'=-k_*$, so it too keeps its bulk powers at the edge, with the prefactors $4(E(k_*)/2J)^3\mathcal N_*/\sqrt{1-m_{\mathrm e}^2}$ of $\ell^{-2}\Ai^2$ (topological), $1/(2\pi\sqrt{1-m_{\mathrm e}^2})$ of $\ell^{-3}\mathcal Q$ (closing), and $(8J/\mu)[4(m_{\mathrm e}+c_{\mathrm t})^2/(1-m_{\mathrm e}^2)+\frac12]^{1/2}$ of $\ell^{-4}\mathcal B$ (trivial), with $c_{\mathrm t}=(\mu^2+3E(k_*)^2)/(4\mu E(k_*))$. Figure~\ref{fig:edgebond}(a) tests the first, $2.2997$ at $\mu=J$. At the nearest-neighbor closing, the edge fronts have a relative correction $O(\ell^{-1})$, with a coefficient close to $2$, explaining why the edge data of Fig.~2(c) of the main text lie below $\mathcal Q$.
	
	\subsubsection{Deforming the band: $J\neq\Delta$}
	\label{sec:aniso}
	
	For $\Delta\neq J$ the point $h(k)=\mu+2J\cos k+2i\Delta\sin k$ runs on an ellipse [dashed in Fig.~\ref{fig:circle}], Eq.~\eqref{eq:complement} fails, and the phase is still topological for $|\mu|<2J$. Maximizing $E'^2$ to first order in $J-\Delta$ moves the fastest momentum by $\delta k_*=2(J-\Delta)E(k_*)/(J\mu)$. The numerator of $\re g=-(\mu+2J\cos k)/E$ is independent of $\Delta$ and has slope $2J\sin k_*=-E(k_*)$ at the unperturbed $k_*$, so
	\begin{equation}
		\re g(k_*)=-\frac{2(J-\Delta)E(k_*)}{J\mu}+O\bigl((J-\Delta)^2\bigr).
		\label{eq:rstar}
	\end{equation}
	The leading term of Eq.~\eqref{eq:factor} returns with this small weight beside the terms of Eq.~\eqref{eq:Dtopo}:
	\begin{equation}
		\overline D_j(t)^2\simeq\frac{\bigl[\re g(k_*)\,\ell^{-2}\Ai(s)^2-\frac12(m+c)\ell^{-4}\mathcal B(s)\bigr]^2}{1-m^2}+\frac{\ell^{-8}\mathcal B(s)^2}8.
		\label{eq:Daniso}
	\end{equation}
	The two smooth terms are equal when $|\re g(k_*)|\ell^2=\frac12|m+c|$, which with $\ell^3=\mu t/2$ defines the crossover time
	\begin{equation}
		t_\times=\frac2\mu\Bigl(\frac{|m+c|\,J\mu}{4E(k_*)|J-\Delta|}\Bigr)^{3/2}\propto J^{1/2}|J-\Delta|^{-3/2},
		\label{eq:tcross}
	\end{equation}
	$t_\times\approx0.020J^{1/2}|J-\Delta|^{-3/2}$ at $\mu=J$. Inside one phase the density front decays as $t^{-4/3}$ before $t_\times$ and as $t^{-2/3}$ after it. The exponent $\frac32$ is general: if a perturbation $\delta$ gives a blind observable a weight $\propto\delta$, the leading term wins once $\delta\,\ell^2\sim1$, at $t_\times\propto|\delta|^{-3/2}$. The cancellation belongs to the band structure, the circle, not to the phase.
	
	\subsubsection{The cluster chain measured in its field}
	\label{sec:clusterX}
	
	By the dictionary~\eqref{eq:dictionary}, measuring $X_{j_0}$ and observing $K_j$ is measuring the parity and observing the bond energy of a Kitaev chain with $J=\Delta=1$ and $\mu=-2\lambda$, in cells. In the cluster phase the fastest quasiparticles carry no field weight but a stabilizer weight; in the paramagnet, no stabilizer weight. With $\varrho(k)=|\lambda-e^{ik}|=E(k)/2$ and $\varrho=\varrho(k_*)$, Eqs.~\eqref{eq:DBtopo}, \eqref{eq:Dcrit}, and \eqref{eq:DBtriv} become, in cells,
	\begin{equation}
		D_j\simeq\frac{2\sqrt{1-\lambda^2}}{\sqrt{1-m^2}}\ell_{\mathrm{cell}}^{-2}\Ai^2, \qquad D_j\simeq\frac{\ell_{\mathrm{cell}}^{-3}\mathcal Q}{4\pi\sqrt{1-m^2}}, \qquad D_j\simeq\frac{1-m\re g(k_*)}{\varrho\sqrt{1-m^2}}\ell_{\mathrm{cell}}^{-4}\mathcal B,
		\label{eq:DX}
	\end{equation}
	for $\lambda<1$, $\lambda=1$, and $\lambda>1$, with $m=\langle X_j\rangle_0$, $\re g(k_*)=\varrho/\lambda$ in the paramagnet, widths $\ell_{\mathrm{cell}}=(\lambda t)^{1/3}$, $(t/4)^{1/3}$, and $t^{1/3}$ in cells, and $s=(q-q_0-2\min(\lambda, 1)t)/\ell_{\mathrm{cell}}$, with $q$ the cell and $q_0$ the cell of $j_0$. In sites $\ell=2\ell_{\mathrm{cell}}$, so $\ell_{\mathrm{cell}}^{-2-\nu}=2^{2+\nu}\ell^{-2-\nu}$. These are the laws of the Kitaev bond energy, $t^{-2/3}$, $t^{-1}$, and $t^{-4/3}$: with $K_j$ in the role of the bond energy, the cancellation has moved to the paramagnet.
	
	\subsubsection{The string measurement in the gapped phases}
	\label{sec:Zgapped}
	
	Now measure $Z_{j_0}$: by Sec.~\ref{sec:string} the correlations across the cut are erased, both outcomes respond alike, and $D_j=|\delta K_j|$. The cut sits at different places in the two sublattices [Fig.~\ref{fig:cluster}(b)]. In the sublattice of $j_0$, which carries the stabilizers with odd $j-j_0$, it separates $a_{j_0}$ from $b_{j_0}$ inside a cell, and the only Hamiltonian term across it is $-\lambda X_{j_0}$. In the other, which carries those with even $j-j_0$, it falls between two cells, and the only term across it is $-K_{j_0}$. The measurement sets the crossing term to zero and leaves the others unchanged, depositing the energy
	\begin{equation}
		\epsilon=\lambda\langle X_{j_0}\rangle_0\quad(j-j_0\ \text{odd}), \qquad \epsilon=\langle K_{j_0}\rangle_0\quad(j-j_0\ \text{even}),
		\label{eq:eps}
	\end{equation}
	positive by Sec.~\ref{sec:clusterground}. In a gapped phase the erased correlations decay exponentially and $\tilde S$ is bounded. On the line $k'=-k$ the transform counts the pairs at separation $r$ that straddle the cut: $N_r=|r|$ between cells and $N_r=|r+1|$ within a cell, in the mirrored Kitaev coordinate of Eq.~\eqref{eq:dictionary}. Thus $\tilde S_{ab}(k, -k)=-\sum_r N_r g_r e^{-ikr}$ and $\mathcal M(k)=\re[\bar g(k)\sum_r N_r g_r e^{-ikr}]$. To show that $\varrho(k)\mathcal M(k)$ does not depend on $k$, let $y_n=\int\frac{dk}{2\pi}e^{ikn}/\varrho(k)$ be the Fourier coefficients of $1/\varrho$. Since $g(k)=(\lambda-e^{ik})/\varrho(k)$ and $\varrho^2\,d(1/\varrho)/dk=-\lambda\sin k/\varrho$,
	\begin{equation}
		g_r=\lambda y_r-y_{r+1}, \qquad 2n(1+\lambda^2)y_n=\lambda\bigl[(2n+1)y_{n+1}+(2n-1)y_{n-1}\bigr].
		\label{eq:yrec}
	\end{equation}
	Since $\varrho\bar g=\lambda-e^{-ik}$, the product is $\varrho\mathcal M=\re\bigl[(\lambda-e^{-ik})\sum_rN_rg_re^{-ikr}\bigr]$, and for both $N_r=|r|$ and $N_r=|r+1|$ its Fourier coefficient at $n\neq0$ is half the difference of the two sides of Eq.~\eqref{eq:yrec}, so it vanishes. The coefficient at $n=0$ is $-g_{-1}=\langle K_{j_0}\rangle_0$ between cells and $\lambda g_0=\lambda\langle X_{j_0}\rangle_0$ within a cell, the deposited energy of Eq.~\eqref{eq:eps}. Hence $\mathcal M(k)=\epsilon/\varrho(k)$. In the cluster phase $\re G(k_*)=-\varrho(k_*)$, so the product of the two factors is the deposited energy itself, and
	\begin{equation}
		D_j(t)\simeq\epsilon\,\ell_{\mathrm{cell}}^{-2}\Ai(s)^2=4\epsilon\,\ell^{-2}\Ai(s)^2, \qquad s=\frac{x-4\lambda t}\ell,
		\label{eq:DZgapped}
	\end{equation}
	up to a shift of $s$ of order $\ell^{-1}$: the laws of $X_{j_0}$, as the main text says. In the paramagnet the stabilizer is blind, and a second cancellation happens: the derivative of $\mathcal M(k)=\epsilon/\varrho(k)$ along $k'=-k$ compensates the curvature term of Eq.~\eqref{eq:secondgeneral}, so only the equal-frequency pairs remain at order $\ell^{-4}$, and the stabilizers oscillate under an $\ell^{-4}\mathcal B$ envelope,
	\begin{equation}
		\overline D_j(t)\simeq\frac{\lambda\,\epsilon'}{2\sqrt2(\lambda^2-1)}\ell_{\mathrm{cell}}^{-4}\mathcal B(s), \qquad \epsilon'=\langle K_{j_0}\rangle_0\ (j-j_0\ \text{odd}), \quad \epsilon'=\langle X_{j_0}\rangle_0\ (j-j_0\ \text{even}),
		\label{eq:DZpara}
	\end{equation}
	with a period of $\pi/k_*$ cells, three cells at $\lambda=2$, over which the overline averages.
	
	\textit{A term that lifts the cancellation.} The five-site term $-\lambda'\sum_jZ_jX_{j+1}X_{j+2}X_{j+3}Z_{j+4}$ commutes with both symmetries and by Eq.~\eqref{eq:JW} equals $-i\lambda'\sum_jb_ja_{j+4}$, a coupling between cells $q$ and $q+2$ of one sublattice. For $|\lambda'|<\lambda-1$ the paramagnet stays gapped, but energy and observable phase no longer come from one circle. To first order in $\lambda'$ the fastest quasiparticle acquires the stabilizer weight $\re G(k_*)=4\varrho\lambda'/\lambda$, and by the rule of Sec.~\ref{sec:aniso} the $t^{-4/3}$ decay turns into $t^{-2/3}$ after $t_\times\propto|\lambda'|^{-3/2}$. At $\lambda=2$, where $\langle X\rangle_0=0.93422$ and $\langle K\rangle_0=0.25866$, comparison with the rms envelope gives $t_\times\simeq0.0021|\lambda'|^{-3/2}$ for odd and $0.28|\lambda'|^{-3/2}$ for even $j-j_0$.
	
	\subsubsection{The string measurement at the critical point}
	\label{sec:Zcrit}
	
\begin{figure}[!htb]
		\centering
		\includegraphics[width=\textwidth]{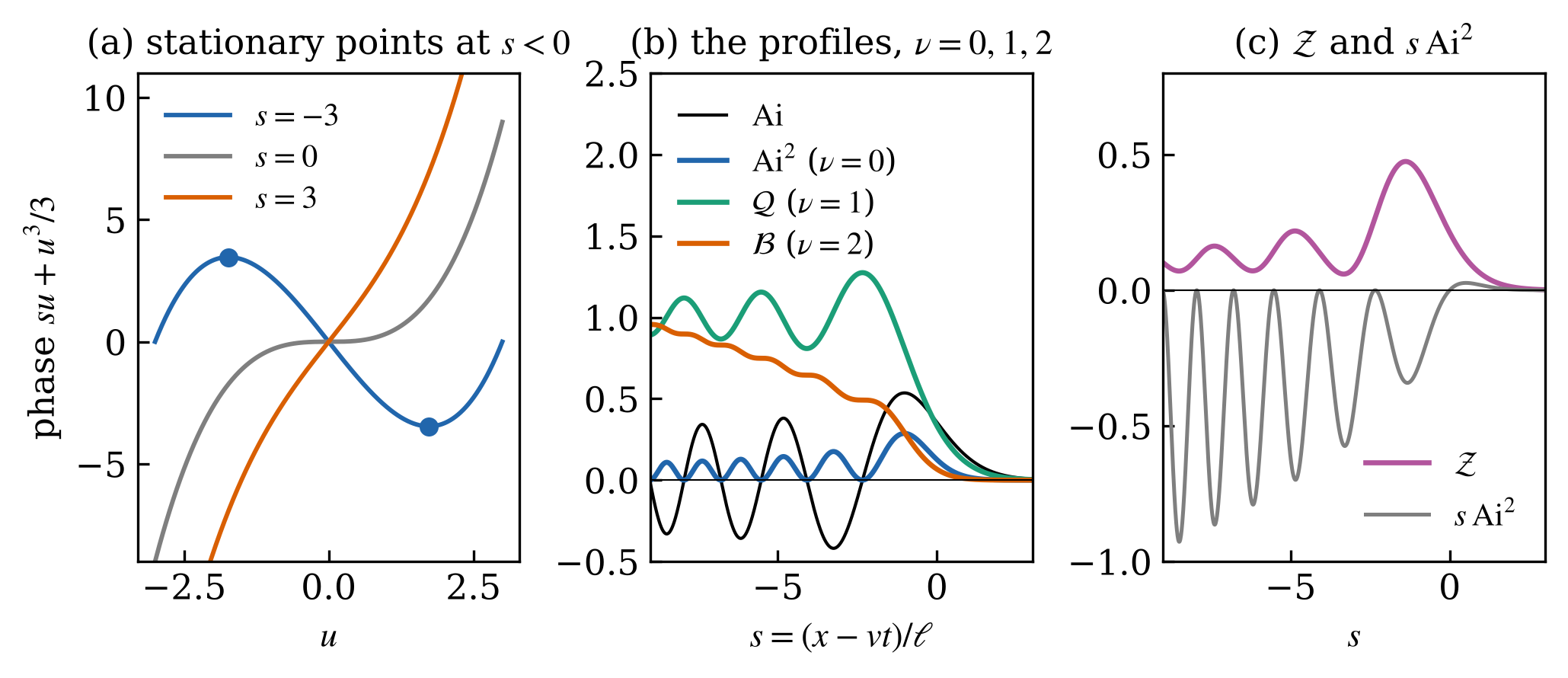}
		\caption{The Airy phase and the profiles. (a) The phase $su+u^3/3$ has two stationary points for $s<0$ (dots), which merge at $s=0$ and disappear for $s>0$: behind the front two rays arrive at each point, ahead of it none. (b) $\Ai$ and the profiles $\Ai^2$ ($\nu=0$), $\mathcal Q$ ($\nu=1$), and $\mathcal B$ ($\nu=2$). (c) The critical string profile $\mathcal Z$ of Eq.~\eqref{eq:DZcrit}, and $s\Ai^2$, which mixes into $\mathcal B$ when the condition below Eq.~\eqref{eq:secondgeneral} fails.}
		\label{fig:profiles}
	\end{figure}

	At $\lambda=1$ the erased correlations decay only as the inverse distance, their number at separation $r$ grows as $|r|$, and $\sum_r|r|g_re^{-ikr}$ diverges: the transform is not bounded, and Eq.~\eqref{eq:rule} does not apply. We evaluate Eq.~\eqref{eq:exact} directly. At $\lambda=1$, with $\gamma_{2q}=a_q$ and $\gamma_{2q+1}=b_q$ in one sublattice, Eq.~\eqref{eq:Hsector} becomes $H_p=-i\sum_\iota\gamma_\iota\gamma_{\iota+1}$, a uniform chain of Majoranas, in which $X$ and $K$ are the products $i\gamma_\iota\gamma_{\iota+1}$ on even and odd links. Then $A_{\iota, \iota\pm1}=\mp2$, a plane wave $e^{ik\iota}$ has $A=-4i\sin k$, and
	\begin{equation}
		R_{\iota\iota'}(t)=\int_{-\pi}^\pi\frac{dk}{2\pi}e^{ik(\iota-\iota')-4it\sin k}=\mathrm J_{\iota-\iota'}(4t), \qquad \Gamma_{0, \iota\iota'}=-\frac2{\pi(\iota-\iota')}\quad(\iota-\iota'\ \text{odd}),
		\label{eq:uniform}
	\end{equation}
	and $\Gamma_{0, \iota\iota'}=0$ for even $\iota-\iota'$. In both sublattices the cut separates two neighbors $\gamma_{c-1}$ and $\gamma_c$, so one calculation covers every $j$. The change is $S_{\iota_1\iota_2}=2/[\pi(\iota_1-\iota_2)]$ for $\iota_1<c\leq\iota_2$, and Eq.~\eqref{eq:exact} is an exact double sum of Bessel functions,
	\begin{equation}
		\delta K=\sum_{\iota_1<c\leq\iota_2}\frac2{\pi(\iota_1-\iota_2)}\bigl[\mathrm J_{\iota-\iota_1}\mathrm J_{\iota+1-\iota_2}-\mathrm J_{\iota-\iota_2}\mathrm J_{\iota+1-\iota_1}\bigr], \qquad \iota_1-\iota_2\ \text{odd},
		\label{eq:Zsum}
	\end{equation}
	which is how Fig.~2(d) of the main text was computed. Here $E(k)=4\sin k$ has its largest slope $v=4$ at $k=0$, with $E'''=-4$, so Eq.~\eqref{eq:onewave} gives the known behavior of the Bessel function $\mathrm J_n$ near its turning point, $\mathrm J_n(4t)\simeq\ell^{-1}\Ai(s_n)$ with $\ell=(2t)^{1/3}$ and $s_n=(n-4t)/\ell$, and $\mathrm J_{n+1}\simeq\ell^{-1}[\Ai(s_n)+\ell^{-1}\Ai'(s_n)]$. In the bracket the leading products cancel, which is the antisymmetry of $S$, and what remains is $\ell^{-3}[\Ai(s_1)\Ai'(s_2)-\Ai'(s_1)\Ai(s_2)]=\ell^{-3}(s_1-s_2)K_{\mathrm{Ai}}(s_1, s_2)$, with the Airy kernel of Fig.~\ref{fig:cluster}(c) and $s_{1, 2}=(\iota-\iota_{1, 2}-4t)/\ell$. The prefactor $2/[\pi(\iota_1-\iota_2)]=-2/[\pi\ell(s_1-s_2)]$ cancels $s_1-s_2$, each pair occupies the area $\ell^{-2}$ in the plane of $(s_1, s_2)$, and half of the pairs have odd $\iota_1-\iota_2$, so the sum becomes an integral over the quadrant $s_1>s>s_2$. With $K_{\mathrm{Ai}}(s_1, s_2)=\int_0^\infty\Ai(s_1+u)\Ai(s_2+u)du$ the integrals over $s_1$ and $s_2$ give $\mathcal Q(s+u)$ and $1-\mathcal Q(s+u)$, and
	\begin{equation}
		D_j(t)=|\delta K_j|\simeq\frac{\mathcal Z(s)}{\pi\ell^2}, \qquad \mathcal Z(s)=\int_s^\infty\mathcal Q(u)\bigl[1-\mathcal Q(u)\bigr]du, \qquad \ell=(2t)^{1/3}, \qquad s=\frac{x-4t}\ell,
		\label{eq:DZcrit}
	\end{equation}
	Eq.~(8) of the main text. The powers count as follows: the antisymmetric bracket gives $\ell^{-3}$ instead of the $\ell^{-2}$ of two propagators, each erased correlation its inverse distance $\ell^{-1}$, and the two sums over a range $\ell$ give $\ell^2$; in all $\ell^{-2}$. The profile $\mathcal Z$ is positive, as a numerical evaluation confirms, has its maximum $0.474$ at $s=-1.384$, where $\mathcal Q$ first reaches one, decays ahead of the front like $\mathcal Q$, and falls off as $1/(\pi|s|^{1/2})$ far behind it [Fig.~\ref{fig:profiles}(c)]. Measuring $Z_{j_0}$ instead of $X_{j_0}$ thus changes the critical decay from $t^{-1}$ to $t^{-2/3}$ at the same Hamiltonian and width.
	
	\subsubsection{The ends of the cluster chain and the record of the outcome}
	\label{sec:clusteredge}
	
\begin{figure}[!htb]
		\centering
		\includegraphics[width=\textwidth]{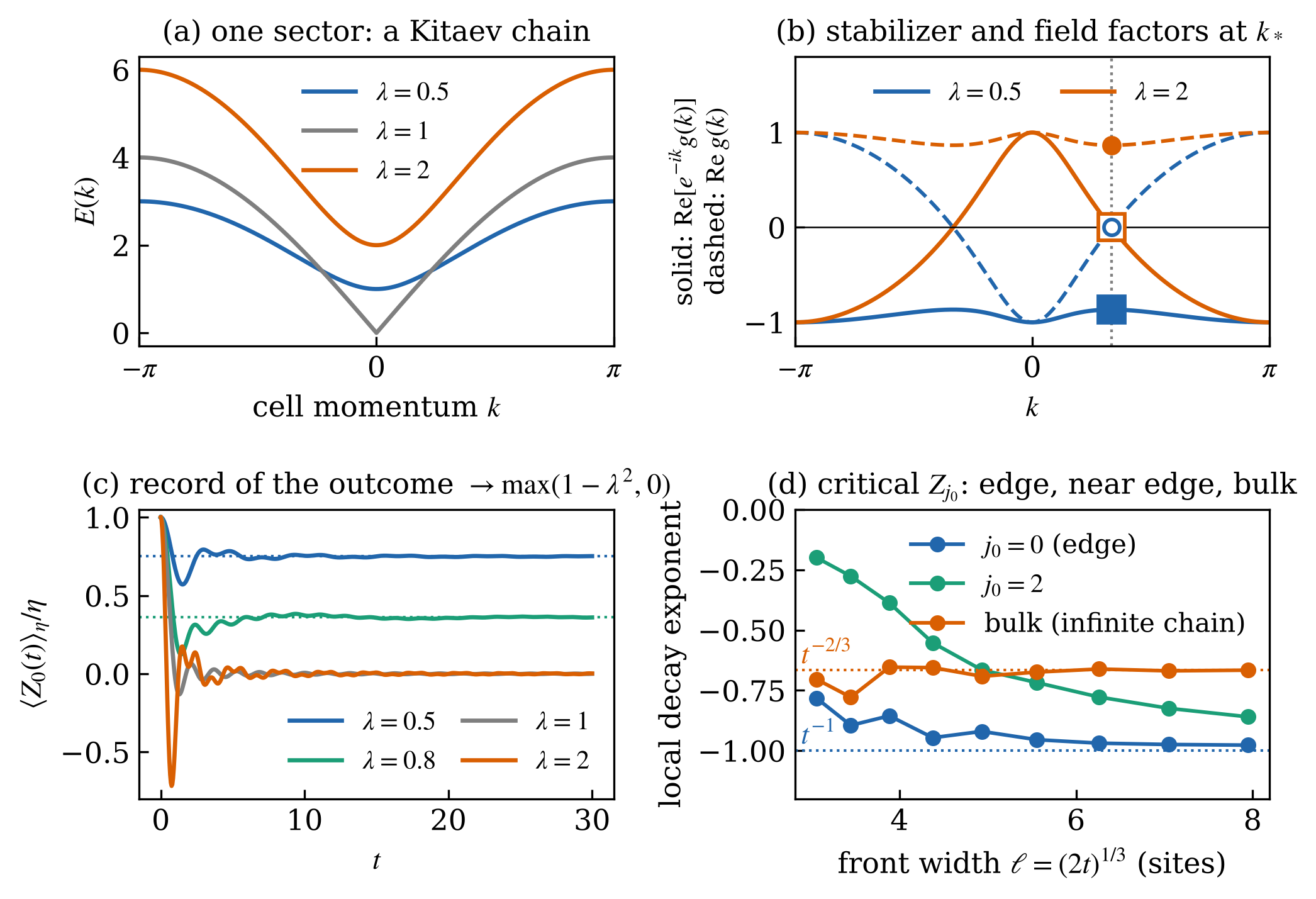}
		\caption{The cluster chain. (a) The energy $E(k)=2|\lambda-e^{ik}|$ of one sublattice, $k$ per cell, at $\lambda=0.5$, $1$, and $2$. (b) The stabilizer factor $\re[e^{-ik}g(k)]$ (solid) and the field factor $\re g(k)$ (dashed) at $\lambda=0.5$ and $2$; at the fastest momentum the field factor vanishes in the cluster phase and the stabilizer factor in the paramagnet (open markers). (c) The record of the outcome after measuring $Z_0$ on a half-line of $300$ cells: $\langle Z_0(t)\rangle_\eta/\eta$ approaches $1-\lambda^2$ for $\lambda<1$ (dotted) and zero otherwise, Eq.~\eqref{eq:memory}. (d) The critical string measurement on an open chain: the local decay exponent $\Delta\log D_{\mathrm{peak}}/\Delta\log t$ is computed between consecutive sampled times and plotted against $\ell$, where $D_{\mathrm{peak}}(t)$ is the maximum response $D_j(t)$ across the front. It tends to $-1$ for $j_0=0$ (Sec.~\ref{sec:clusteredge}) and to $-\frac23$ in the bulk, Eq.~\eqref{eq:DZcrit}; for $j_0=2$ it crosses from one to the other as $\ell$ grows past $j_0$.}
		\label{fig:clusterfronts}
	\end{figure}

	At the left end the string is empty, so both $X_0$ and $Z_0=a_0$ can be measured; both act on the even sublattice. Its standing waves are those of Eq.~\eqref{eq:hardwall} in the mirror, with $a$ and $b$ exchanged, and the amplitude of a mode on $a_0=Z_0$ is proportional to its group velocity, $|f^a_k|^2=v(k)^2/(2\pi)$ in cells. The measurement of $Z_0$ erases one row and one column, Eq.~\eqref{eq:SZ0}; its transform is bounded, unlike that of the bulk $Z_{j_0}$ at $\lambda=1$, and its measurement factor $v(k)v(k')/2$ is nonzero at the fastest momenta. Both edge measurements therefore follow the bulk $X_{j_0}$ laws in all three phases, with $\sigma=0$, as the main text says; at $\lambda=1$ the first site turns the $t^{-2/3}$ of the bulk string measurement back into $t^{-1}$ [Fig.~\ref{fig:clusterfronts}(d)]. The prefactors follow from Sec.~\ref{sec:edge} with $m_{\mathrm e}=\langle X_0\rangle_0=(2\lambda/\pi)\int_0^\pi\sin^2k/\varrho(k)\,dk$. In cells, in the cluster phase,
	\begin{equation}
		D^{X_0}_j\simeq\frac{4\sqrt{1-\lambda^2}\bigl(1-2m_{\mathrm e}\lambda\sqrt{1-\lambda^2}\bigr)}{\sqrt{1-m_{\mathrm e}^2}}\ell_{\mathrm{cell}}^{-2}\Ai^2, \qquad D^{Z_0}_j\simeq4\lambda^2\sqrt{1-\lambda^2}\,\ell_{\mathrm{cell}}^{-2}\Ai^2;
		\label{eq:Dedgecl}
	\end{equation}
	at $\lambda=1$, with $m_{\mathrm e}=8/(3\pi)$, $D^{X_0}_j\simeq\ell_{\mathrm{cell}}^{-3}\mathcal Q/(2\pi\sqrt{1-m_{\mathrm e}^2})$, twice the bulk front with $m\to m_{\mathrm e}$, and $D^{Z_0}_j\simeq\ell_{\mathrm{cell}}^{-3}\mathcal Q/(2\pi)$; in the paramagnet, $\overline D^{X_0}_j\simeq(4/\lambda)\ell_{\mathrm{cell}}^{-4}\mathcal B[4(m_{\mathrm e}-c_{\mathrm C})^2/(1-m_{\mathrm e}^2)+\frac12]^{1/2}$ with $c_{\mathrm C}=(4\lambda^2-3)/(4\lambda\varrho)$, and $D^{Z_0}_j\simeq2\ell_{\mathrm{cell}}^{-4}\mathcal B/\sqrt{\lambda^2-1}$.
	
	\textit{The record of the outcome.} After $Z_0=a_0$ is measured, $\langle a_0\rangle_\eta=\eta$ (Sec.~\ref{sec:string}), and $\langle Z_0(t)\rangle_\eta=\eta R_{a_0a_0}(t)$. In the modes of the half-line, the zero mode of Sec.~\ref{sec:cluster}, with weight $1-\lambda^2$ on $a_0$, does not evolve, while every band mode oscillates at its own frequency:
	\begin{equation}
		\langle Z_0(t)\rangle_\eta=\eta\Bigl[\max(1-\lambda^2, 0)+\frac1{2\pi}\int_0^\pi v(k)^2\cos\bigl(E(k)t\bigr)dk\Bigr].
		\label{eq:memory}
	\end{equation}
	At $t=0$ the bracket is one; at $\lambda=1$ it is exactly $\mathrm J_1(4t)/(2t)$. The integral dephases as $t^{-3/2}$, so $\langle Z_0(t)\rangle_\eta\to\eta(1-\lambda^2)$ in the cluster phase and $\to0$ otherwise [Fig.~\ref{fig:clusterfronts}(c)]: the edge keeps a one-site record of the outcome while the stabilizer front leaves.
	
	\section{Long-range couplings}
	\label{sec:lr}
	
	Now let every site couple to every other one: $f_r=r^{-\alpha}$ in Eq.~\eqref{eq:HK}, with $J=\Delta$ and $\alpha>1$, so that the total coupling of one site, $\sum_rr^{-\alpha}$, is finite. The density is measured and observed. The four steps of Secs.~\ref{sec:eq} to \ref{sec:fronts} go through with two changes: the energy has a new shape near $k=0$, and so have the standing waves near an open end. The measurement rules of Sec.~\ref{sec:measure} use only $\Gamma_0$, so Eqs.~\eqref{eq:update}, \eqref{eq:exact}, and \eqref{eq:Mparity} hold as they stand, and so do the weights of Eqs.~\eqref{eq:Wdensity} and \eqref{eq:edgeweights}, with the $g(k)$ of this section and, at an edge, $\sin k$ replaced by $\sqrt{\pi/2}f^a_k$. Inverse-square couplings, $\alpha=2$, are the known threshold below which free particles have no maximal speed~\cite{Buyskikh2016, Cevolani2016, Tran2020}; the long-range Kitaev chain itself is studied in Refs.~\cite{Vodola2014, Vodola2016, Lepori2016, Lepori2017jstat, Viyuela2016, VanRegemortel2016, Jager2020}.
	
	\subsection{Equilibrium: the polylogarithm}
	\label{sec:polylog}
	
	At $J=\Delta$, Eq.~\eqref{eq:hk} becomes
	\begin{equation}
		h(k)=\mu+2J\Li_\alpha(e^{ik}), \qquad \Li_\alpha(z)=\sum_{r\geq1}\frac{z^r}{r^\alpha}, \qquad E_\alpha(k)=|h(k)|, \qquad g(k)=-\frac{h(k)}{E_\alpha(k)},
		\label{eq:Ealpha}
	\end{equation}
	Eq.~(9) of the main text, with the polylogarithm $\Li_\alpha$. At $z=\pm1$, $\Li_\alpha(1)=\zeta(\alpha)$, the Riemann zeta function, and $\Li_\alpha(-1)=-\eta_{\mathrm D}(\alpha)$, with the Dirichlet eta function $\eta_{\mathrm D}(\alpha)=\sum_{r\geq1}(-1)^{r-1}r^{-\alpha}=(1-2^{1-\alpha})\zeta(\alpha)$ (the subscript keeps it apart from the outcome $\eta$).
	
\begin{figure}[!htb]
		\centering
		\includegraphics[width=\textwidth]{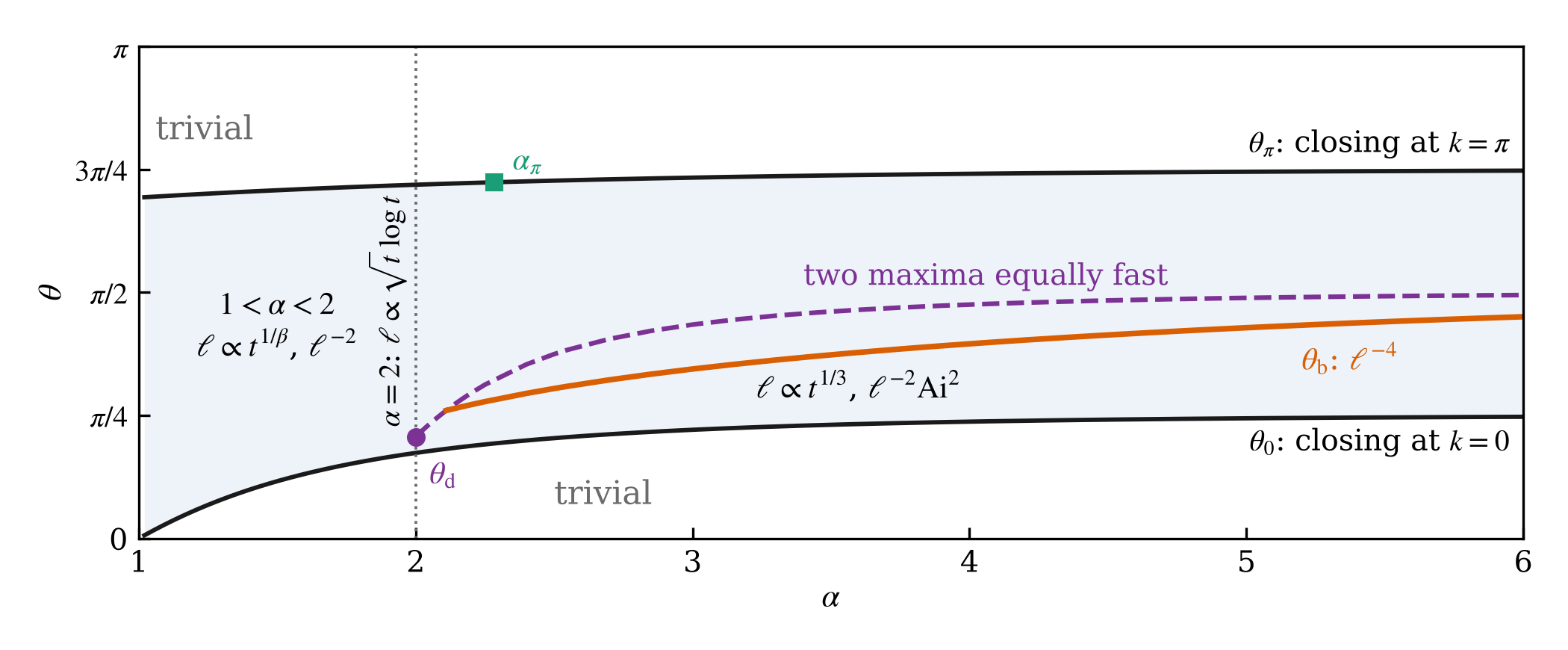}
		\caption{The long-range chain at $J=\sin\theta$, $\mu=-2\cos\theta$. Shaded: topological, between $\theta_0(\alpha)$ (gap closes at $k=0$) and $\theta_\pi(\alpha)$ (at $k=\pi$), Eq.~\eqref{eq:LRphase}. For $1<\alpha<2$, $\ell\propto t^{1/\beta}$ with $\beta=\alpha-1$; on the dotted line $\alpha=2$ the band-edge waves lead for $\theta>\theta_{\mathrm d}=0.6467$ (dot), with $\ell\propto\sqrt t\log t$; for $\alpha>2$ an interior maximum leads, $\ell\propto t^{1/3}$. Dashed: two interior maxima equally fast. Orange: the line $\theta_{\mathrm b}(\alpha)$, from $\alpha\simeq2.110$, on which the density is blind and decays as $\ell^{-4}$. Square: $\alpha_\pi=2.283$; on $\theta_\pi$ above it the waves at $k=\pi$ are the fastest.}
		\label{fig:lrmap}
	\end{figure}

	\textit{The phases.} The curve $h(k)$ is no longer a circle, but it still meets the real axis only twice: with the Gamma function, $r^{-\alpha}=\Gamma(\alpha)^{-1}\int_0^\infty w^{\alpha-1}e^{-rw}dw$ the sum over $r$ is geometric, and
	\begin{equation}
		\im\Li_\alpha(e^{ik})=\sum_{r\geq1}\frac{\sin kr}{r^\alpha}=\frac{\sin k}{\Gamma(\alpha)}\int_0^\infty\frac{w^{\alpha-1}e^{-w}dw}{1-2e^{-w}\cos k+e^{-2w}}>0\qquad(0<k<\pi).
		\label{eq:imLi}
	\end{equation}
	So $h$ is real only at $k=0$ and $\pi$, where it equals $\mu+2J\zeta(\alpha)$ and $\mu-2J\eta_{\mathrm D}(\alpha)$, and the curve winds once around the origin, as the circle of a topological chain does, exactly when these have opposite signs:
	\begin{equation}
		-2J\zeta(\alpha)<\mu<2J\eta_{\mathrm D}(\alpha)=2J(1-2^{1-\alpha})\zeta(\alpha),
		\label{eq:LRphase}
	\end{equation}
	the topological interval of the main text. Since $E_\alpha=|h|$ can vanish only where $h$ is real, the gap closes only at the ends of the interval, at $k=0$ for $\mu=-2J\zeta(\alpha)$ and at $k=\pi$ for $\mu=2J\eta_{\mathrm D}(\alpha)$; elsewhere the chain is gapped. The examples use $J=\sin\theta$, $\mu=-2\cos\theta$, $0<\theta<\pi$, topological between $\theta_0=\arctan[1/\zeta(\alpha)]$ and $\theta_\pi=\pi-\arctan[1/\eta_{\mathrm D}(\alpha)]$, which tend to $\pi/4$ and $3\pi/4$ ($|\mu|<2J$) as $\alpha\to\infty$ (Fig.~\ref{fig:lrmap}).
	
	\textit{The term $|k|^{\alpha-1}$.} Expanding $e^{ikr}$ in powers of $kr$ produces the moments $\sum_rr^{n-\alpha}$, finite only for $n<\alpha-1$; the far part of the sum behaves like $\int^\infty r^{-\alpha}e^{ikr}dr$, which contains a power $|k|^{\alpha-1}$ that no Taylor series has. The result is~\cite{DLMFPolylog}
	\begin{equation}
		\Li_\alpha(e^{ik})=\Gamma(1-\alpha)(-ik)^{\alpha-1}+\sum_{n\geq0}\zeta(\alpha-n)\frac{(ik)^n}{n!}, \qquad (-ik)^{\alpha-1}=|k|^{\alpha-1}e^{-i\pi(\alpha-1)\sgn(k)/2},
		\label{eq:polyexp}
	\end{equation}
	for noninteger $\alpha$ and $|k|<2\pi$, with $\zeta$ continued to all arguments; at integer $\alpha$ the first term merges with one term of the sum into a logarithm. Near $k=\pi$, analyticity gives the Taylor series $\Li_\alpha(-e^{i\kappa})=-\sum_{n\geq0}\eta_{\mathrm D}(\alpha-n)(i\kappa)^n/n!$, where $k=\pi+\kappa$ and $\eta_{\mathrm D}$ is understood by analytic continuation. Differentiating the series gives $\frac d{dk}\Li_\alpha(e^{ik})=i\Li_{\alpha-1}(e^{ik})$.
	
	\textit{The speed at the band edge.} From $E_\alpha^2=|h|^2$, $E_\alpha'=\re(\bar hh')/E_\alpha$: as in Sec.~\ref{sec:speed}, only the part of the tangent $h'$ along $h$ changes the energy. At $k\to0$ with the gap open $h$ is real, so only $\re h'=-2J\sum_r\sin(kr)/r^{\alpha-1}$ counts, and by Eq.~\eqref{eq:polyexp} with $\alpha\to\alpha-1$
	\begin{equation}
		|E_\alpha'(0^+)|=2J\lim_{k\to0^+}\sum_{r\geq1}\frac{\sin kr}{r^{\alpha-1}}=\infty\ (1<\alpha<2), \qquad \pi J\ (\alpha=2), \qquad 0\ (\alpha>2),
		\label{eq:v0}
	\end{equation}
	where at $\alpha=2$ the sum is the Fourier series $\sum_r\sin(kr)/r=(\pi-k)/2$. This line organizes Sec.~\ref{sec:lr} [Fig.~\ref{fig:longrange}(b)]: for $1<\alpha<2$ the band-edge waves are infinitely fast; at $\alpha=2$ they move at $\pi J$ and compete with the interior of the band; for $\alpha>2$ they stand still, and an interior maximum of the speed leads, as for nearest neighbors.
	
	\subsection{Before the measurement: the first site of a half-line}
	\label{sec:firstsite}
	
	At $J=\Delta$ each $a_j$ couples only to $b$'s on its right, so on the half-line $j\geq0$ the matrix $h$ is upper triangular, $h_{jj}=\mu$, $h_{j, j+r}=2Jf_r$. The squared energies of the $a$ modes are the eigenvalues of $hh^{\mathrm T}$ (Sec.~\ref{sec:halfline}), and $(hh^{\mathrm T})_{jl}=\sum_{n\geq\max(j, l)}h_{jn}h_{ln}$ depends only on $j-l$, because every row of $h$ is complete from its diagonal on: the end produces no correction. So $hh^{\mathrm T}=T(E_\alpha^2)$, the half-infinite Toeplitz matrix of $E_\alpha(k)^2$, where $T(f)_{jl}=f_{j-l}$ for a function $f(k)=\sum_nf_ne^{ink}$. We label its eigenvectors $\psi_k$ by the momentum $k\in(0, \pi)$ of their bulk part, normalized as in Eq.~\eqref{eq:hardwall}, and again $f^b_k=\mu f^a_k/E_\alpha(k)$, because the first column of $h$ holds only $\mu$.
	
The first-site amplitudes follow from one matrix element, $[(\Lambda-T(E_\alpha^2))^{-1}]_{00}=\int_0^\pi dk\,(f^a_k)^2/[\Lambda-E_\alpha(k)^2]$ for $\Lambda$ off the spectrum. Its closed form follows here from scalar Wiener--Hopf factorization~\cite{BottcherSilbermann1999}:
	\begin{equation}
		\bigl[(\Lambda-T(E_\alpha^2))^{-1}\bigr]_{00}=\exp\Bigl[-\int_{-\pi}^\pi\frac{dk}{2\pi}\log\bigl(\Lambda-E_\alpha(k)^2\bigr)\Bigr].
		\label{eq:szego}
	\end{equation}
	To prove it, split $\log(\Lambda-E_\alpha^2)=L_-+L_0+L_+$ into its parts with negative, zero, and positive Fourier components, and write $\Lambda-E_\alpha^2=\xi_-\xi_+$ with $\xi_\pm=e^{L_0/2+L_\pm}$. Since $[T(fg)-T(f)T(g)]_{jl}=\sum_{n<0}f_{j-n}g_{n-l}$ vanishes when $f$ has no positive or $g$ no negative Fourier components, $T(\Lambda-E_\alpha^2)=T(\xi_-)T(\xi_+)$, its inverse is $T(1/\xi_+)T(1/\xi_-)$, and the $00$ entry is $(1/\xi_+)_0(1/\xi_-)_0=e^{-L_0}$. Now let $\Lambda$ approach a level of the band from above: $\log(\Lambda+i0-E_\alpha^2)=\log|E_\alpha^2-\Lambda|+i\pi$ where $E_\alpha^2>\Lambda$, and $-\frac1\pi\im$ of the resolvent is the weight of the level on the first site. If the level $\Lambda=E_\alpha(k)^2$ is reached at one momentum $k>0$, the fraction of the zone where $E_\alpha^2>\Lambda$ is $k/\pi$ or $1-k/\pi$, and per unit $k$
	\begin{equation}
		(f^a_k)^2=\frac1\pi e^{-\mathcal I(E_\alpha(k)^2)}\sin k\,\Bigl|\frac{dE_\alpha^2}{dk}\Bigr|, \qquad \mathcal I(\Lambda)=\int_{-\pi}^\pi\frac{dk'}{2\pi}\log\bigl|E_\alpha(k')^2-\Lambda\bigr|.
		\label{eq:bweight}
	\end{equation}
	For nearest neighbors $\int\frac{dk'}{2\pi}\log|\cos k'-\cos k|=-\log2$ gives $e^{-\mathcal I}=1/(2J|\mu|)$, and Eq.~\eqref{eq:bweight} returns the hard wall $(2/\pi)\sin^2k$ of Eq.~\eqref{eq:hardwall}.
	
	If a regular energy level is reached at several momenta, several independent channels share that energy. Write $a(k)=E_\alpha(k)^2$. Let the arcs $a<\Lambda$ run counterclockwise from $z_i^-=e^{ik_i^-}$, where $a'<0$, to $z_i^+=e^{ik_i^+}$, where $a'>0$. The channel with one outgoing wave at $k_i^+$ has first-site weight per unit momentum
	\begin{equation}
		c_i(\Lambda)=\frac1{2\pi}\frac{\prod_j|z_i^+-z_j^-|}{\prod_{j\neq i}|z_i^+-z_j^+|}, \qquad \bigl|f^a_{k_i^+}\bigr|^2=e^{-\mathcal I(\Lambda)}c_i(\Lambda)|a'(k_i^+)|.
		\label{eq:channelweights}
	\end{equation}
	The opposite direction follows by complex conjugation. The positive factors obey $\sum_i c_i=\sin(\pi\Upsilon)/\pi$, where $\Upsilon$ is the fraction of the zone with $a>\Lambda$, so the channel densities $e^{-\mathcal I(\Lambda)}c_i$ sum to the density per unit squared energy obtained from Eq.~\eqref{eq:szego}. For one arc, $c_1=\sin|k|/\pi$, recovering Eq.~\eqref{eq:bweight}. In Eq.~\eqref{eq:lredge}, $|f^a_{k_*}|^2$ is the weight of the fastest outgoing channel. At a regular interior fastest momentum this weight is finite and nonzero, so $\sigma=0$.
	
	\textit{The exponent $\sigma$.} The factor $e^{-\mathcal I}$ is smooth and positive, so the amplitude near the fastest momentum follows $\sin k\,|dE_\alpha^2/dk|$. (Open chains of $4000$ sites at $\theta=1.3$ confirm the band-edge powers below: $(f^a_k)^2$ has the fitted slopes $0.54$, $0.97$, and $2.000$ in $\log k$ for $\alpha=1.5$, $2$, and nearest neighbors.) (i) At an interior fastest momentum both factors are finite: $\sigma=0$. (ii) At a gapped band edge with $1<\alpha<2$, $E_\alpha^2-E_\alpha(0)^2\propto|k|^\beta$ with $\beta=\alpha-1$, Eq.~\eqref{eq:cbeta} below, so the amplitude vanishes as $|k|^{\beta/2}$: $\sigma=\beta/2$. (iii) At $\alpha=2$, $dE_2^2/dk\to\mp2\pi JE_2(0)$, so $(f^a_k)^2\propto|k|$ and $\sigma=\frac12$. (iv) At a gap closing $f^a_k$ vanishes with $E_\alpha$, but $f^b_k=\mu f^a_k/E_\alpha$ stays finite and $b_0$ carries the coupling: $\sigma=0$, as at the nearest-neighbor closing of Sec.~\ref{sec:edge}. Its value is universal: at a closing $\mu+2J\sum_rf_rz^r$ has no zeros inside the unit disk, so by Jensen's formula the circle average of $\log|h|$ equals its value $\log|\mu|$ at $z=0$, $e^{-\mathcal I(0)}=\mu^{-2}$, and $(f^b_k)^2\to2/\pi$ at a linear closing and $2\beta/\pi$ at the closing $E_\alpha\propto|k|^\beta$ of Sec.~\ref{sec:lrfronts}. The nearest-neighbor band edge, $(f^a_k)^2\propto k^2$, would give $\sigma=1$, but it never carries the fastest waves.
	
	\textit{A continuum picture.} Near a gapped fractional band edge, $T(E_\alpha^2)-E_\alpha(0)^2$ acts on long waves as a constant times the fractional Laplacian with symbol $|k|^\beta$. On the full line, the fractional Laplacian multiplies each Fourier mode $e^{ikj}$ by $|k|^\beta$. The half-line operator is defined by extending a function as zero to negative positions before applying the full-line operator. Let $F_\varpi(y)$ be its eigenfunction with eigenvalue one for the symbol $|k|^{2\varpi}$, normalized to a sine of unit amplitude far from the wall. A complete definition is given by its Laplace transform~\cite{Kwasnicki2011}:
	\begin{equation}
		\begin{aligned}
			W_\varpi(u)&=\frac{u^{2\varpi}-1}{u^2-1}, \qquad W_\varpi(1)=\varpi, \\
			\Omega_\varpi(p)&=\exp\Bigl[\frac p\pi\int_0^\infty\frac{\log W_\varpi(u)}{u^2+p^2}\,du\Bigr], \\
			\widetilde F_\varpi(p)&=\int_0^\infty e^{-py}F_\varpi(y)\,dy=\frac{\sqrt\varpi}{(1+p^2)\Omega_\varpi(p)}, \qquad p>0, \quad 0<\varpi\leq1.
		\end{aligned}
		\label{eq:fractionalstates}
	\end{equation}
	The function approaches $\sin[y+(1-\varpi)\pi/4]$ far from the wall and is proportional to $y^\varpi$ near it; $F_1(y)=\sin y$. Thus $\psi_k(j)\simeq\sqrt{2/\pi}F_{\beta/2}(kj)$ at a gapped fractional band edge, and the $F$ in Eq.~\eqref{eq:Dfracedge} is $F_{\beta/2}$. At the fractional closing at $k=0$, the functions used below are
	\begin{equation}
		F_{\mathrm a}=F_\beta, \qquad \widetilde F_{\mathrm b}(p)=-p^\beta\widetilde F_\beta(p), \qquad F_{\mathrm b}(0)=-\sqrt\beta.
		\label{eq:fractionalpartner}
	\end{equation}
	The second relation is the continuum limit of $\psi^b=h^{\mathrm T}\psi^a/E$; at $\beta=1$ it gives $F_{\mathrm b}(y)=-\cos y$. Oscillatory integrals in this section are understood with a factor $e^{-\varepsilon u}$ on a half-line, or $e^{-\varepsilon|u|}$ on the full line, followed by $\varepsilon\downarrow0$.
	
	\subsection{After the measurement: the fronts}
	\label{sec:lrfronts}
	
\begin{figure}[!htb]
		\centering
		\includegraphics[width=\textwidth]{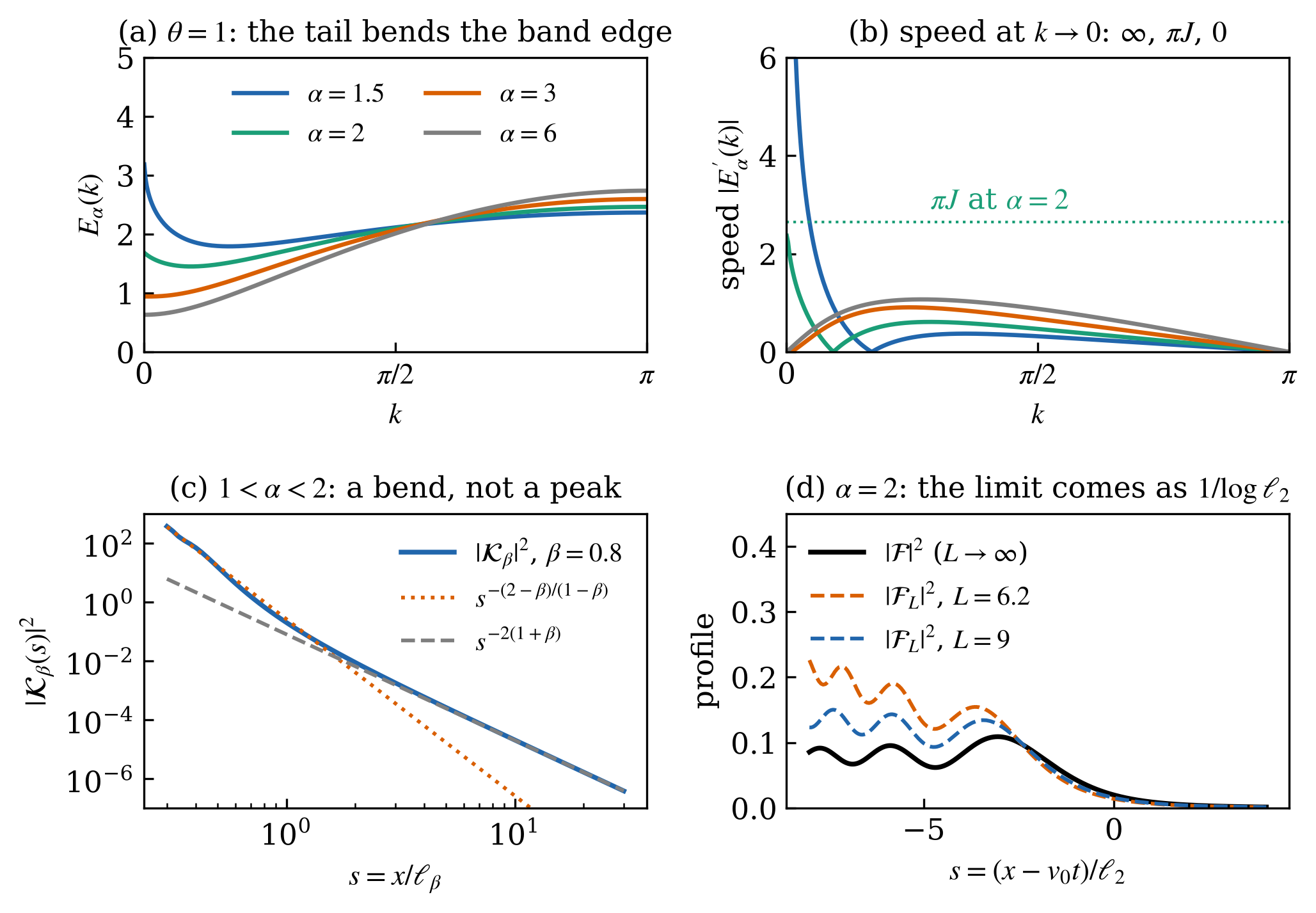}
		\caption{The long-range chain at $\theta=1$. (a) $E_\alpha(k)$ for $\alpha=1.5$, $2$, $3$, and $6$. (b) The speed $|E_\alpha'(k)|$: at $k\to0$ it diverges for $\alpha=1.5$, tends to $\pi J$ at $\alpha=2$ (the curve stays visibly below because of corrections $k\log^2k$), and vanishes for $\alpha>2$, Eq.~\eqref{eq:v0}. (c) $|\mathcal K_\beta|^2$ of Eq.~\eqref{eq:Kbeta} at $\beta=0.8$ with its two power laws, which cross near $s=1.64$: a bend, not a peak. (d) At $\alpha=2$, $|\mathcal F_L|^2$ at $L=6.2$ and $9$ and its limit $|\mathcal F|^2$, Eq.~\eqref{eq:Fresnel}, approached only as $1/L$.}
		\label{fig:longrange}
	\end{figure}

	\textit{Gapped, $1<\alpha<2$: no finite speed.} With the gap at $k=0$ open, $h(0)=\mu+2J\zeta(\alpha)$ is real and $g(0)=-\sgn h(0)=\pm1$. With $\beta=\alpha-1$, Eq.~\eqref{eq:polyexp} gives $h(k)=h(0)+2J\Gamma(-\beta)|k|^\beta e^{-i\pi\beta\sgn(k)/2}+O(|k|)$, and to first order the modulus changes by the part of the correction along $h(0)$:
	\begin{equation}
		E_\alpha(k)\simeq E_\alpha(0)+c_\beta|k|^\beta, \qquad c_\beta=-2Jg(0)\Gamma(-\beta)\cos\frac{\pi\beta}2\neq0.
		\label{eq:cbeta}
	\end{equation}
	The group velocity $\beta|c_\beta||k|^{\beta-1}$ diverges as $k\to0$ [Fig.~\ref{fig:longrange}(a), (b)]: $z=\beta<1$, no finite speed. Waves with $t|c_\beta||k|^\beta\lesssim1$ still add in phase, and with $k=u/\ell_\beta$, $s=x/\ell_\beta$,
	\begin{equation}
		\int_{-\pi}^\pi\frac{dk}{2\pi}e^{ikx+iE_\alpha(k)t}\simeq e^{iE_\alpha(0)t}\ell_\beta^{-1}\mathcal K_\beta(s), \qquad \mathcal K_\beta(s)=\int_{-\infty}^\infty\frac{du}{2\pi}e^{ius+i|u|^\beta}, \qquad \ell_\beta=(|c_\beta|t)^{1/\beta},
		\label{eq:Kbeta}
	\end{equation}
	for $c_\beta>0$; for $c_\beta<0$ the kernel is conjugated, and only $|\mathcal K_\beta|^2$ enters. The width grows faster than $t$, so at $x\sim\ell_\beta$ only the band edge contributes, and Eq.~\eqref{eq:rule} holds with $v=0$. The kernel has no peak [Fig.~\ref{fig:longrange}(c)]: for $|s|\ll1$ the stationary point $|u|=(\beta/|s|)^{1/(1-\beta)}$ gives $|\mathcal K_\beta|^2\simeq\beta^{(2-\beta)/(1-\beta)}/[2\pi\beta(1-\beta)|s|^{(2-\beta)/(1-\beta)}]$; for $|s|\gg1$ the nonanalytic point $u=0$ gives $|\mathcal K_\beta|\simeq\Gamma(1+\beta)\sin(\pi\beta/2)/(\pi|s|^{1+\beta})$, in physical units $Jt/x^\alpha$, the direct coupling $Jx^{-\alpha}$ acting for a time $t$. The front is a bend between two powers, $6$ and $3.6$ at $\alpha=1.8$, crossing at $|s|=1.64$. At the band edge $g\to g(0)=\pm1$: pure particles or holes, which the density sees fully. Equation~\eqref{eq:factor} with $\re G=g(0)$, $\mathcal M=[1-mg(0)]/p_\eta$, and $\Ai^2\to|\mathcal K_\beta|^2$ gives the peakless front of the main text,
	\begin{equation}
		\delta n_{j, \eta}\simeq-\frac{m-g(0)}{2p_\eta}\ell_\beta^{-2}|\mathcal K_\beta(s)|^2, \qquad D_j(t)\simeq\frac{1-mg(0)}{\sqrt{1-m^2}}\ell_\beta^{-2}|\mathcal K_\beta(s)|^2,
		\label{eq:Dfracbulk}
	\end{equation}
	decaying as $t^{-2/\beta}$, tested in Fig.~2(e) of the main text at $\alpha=1.8$, $\theta=1$. At an edge both waves sit near the band edge, where $\omega_k\simeq\omega_0=\mu/E_\alpha(0)$ and $g\simeq g(0)$, so the weight of Eq.~\eqref{eq:edgeweights} is a constant times $f^a_kf^a_{k'}$ and the response is the square of one propagator from the first site,
	\begin{equation}
		\delta\Gamma_{a_jb_j}\simeq-g(0)\frac{\mathcal N_0}{2p_\eta}\Bigl|\int_0^\pi dk\,f^a_k\psi_k(j)e^{iE_\alpha(k)t}\Bigr|^2, \qquad \mathcal N_0=(\omega_0+m_{\mathrm e})^2+1-m_{\mathrm e}^2,
		\label{eq:edgesquare}
	\end{equation}
	with $m_{\mathrm e}=-\int_0^\pi dk\,f^a_kf^b_k$ as in Eq.~\eqref{eq:me}. Each wave carries $f^a_k\propto k^{\beta/2}$ and loses $\ell_\beta^{-\beta/2}$, so the height is $\ell_\beta^{-2-\beta}$, $\sigma=\beta/2$; with Eq.~\eqref{eq:bweight} and the continuum waves,
	\begin{equation}
		D_j(t)\simeq\frac{\mathcal N_0\,\ell_\beta^{-2-\beta}}{2\sqrt{1-m_{\mathrm e}^2}}|\mathcal K^{\mathrm e}_\beta(s)|^2, \qquad \mathcal K^{\mathrm e}_\beta(s)=\frac2\pi\bigl[\beta E_\alpha(0)|c_\beta|e^{-\mathcal I(E_\alpha(0)^2)}\bigr]^{1/2}\int_0^\infty du\,u^{\beta/2}F(us)e^{iu^\beta}, \qquad s=\frac j{\ell_\beta}.
		\label{eq:Dfracedge}
	\end{equation}
	
	\textit{Closing the gap at $k=0$ for $1<\alpha<2$.} At $\mu=-2J\zeta(\alpha)$, $h(k)\simeq2J\Gamma(-\beta)(-ik)^\beta$: $E_\alpha\simeq C_{\mathrm f}|k|^\beta$ with $C_{\mathrm f}=2J|\Gamma(-\beta)|$, the same width law $\ell_\beta=(C_{\mathrm f}t)^{1/\beta}$, and $g(k)=e^{-i\pi\beta\sgn(k)/2}$. The density weight $\re g=\cos(\pi\beta/2)$ does not vanish, $\nu=0$, and $f^b_k$ is finite, $\sigma=0$: bulk and edge both decay as $\ell_\beta^{-2}$, as the main text says. Since $g$ differs on the two sides of the band edge, Eq.~\eqref{eq:Wdensity} weighs pairs on the same side and on opposite sides differently; with $c_{\mathrm f}=\cos(\pi\beta/2)$ and $\mathcal K_\pm(s)=\int_0^\infty\frac{du}{2\pi}e^{\pm ius-iu^\beta}$,
	\begin{equation}
		D_j\simeq\ell_\beta^{-2}\Bigl[\frac{\Phi_1^2}{1-m^2}+\frac{\Phi_2^2}4\Bigr]^{1/2}, \quad \Phi_1=2(m-c_{\mathrm f})\re(\mathcal K_+\bar{\mathcal K}_-)+c_{\mathrm f}(mc_{\mathrm f}-1)\bigl(|\mathcal K_+|^2+|\mathcal K_-|^2\bigr), \quad \Phi_2=4\sin^2\frac{\pi\beta}2\re(\mathcal K_+\mathcal K_-).
		\label{eq:critfracbulk}
	\end{equation}
	At the edge, with the $a$ and $b$ parts $F_{\mathrm a}$, $F_{\mathrm b}$ of the standing waves of $(-d^2/dj^2)^\beta$, the same steps give $D_j\simeq2\beta\ell_\beta^{-2}|\re(\mathcal K_{\mathrm a}\bar{\mathcal K}_{\mathrm b})|/(\pi^2\sqrt{1-m_{\mathrm e}^2})$ with $\mathcal K_{\mathrm{a, b}}(s)=\int_0^\infty du\,F_{\mathrm{a, b}}(us)e^{iu^\beta}$ and $F_{\mathrm b}(0)=-\sqrt\beta$, the continuum form of $(f^b_k)^2\to2\beta/\pi$.
	
	\textit{Inverse-square couplings, $\alpha=2$.} For $0<k<2\pi$, $\re\Li_2(e^{ik})=\pi^2/6-\pi k/2+k^2/4$ exactly and $\im\Li_2(e^{ik})=-\int_0^k\log[2\sin(u/2)]du=k[1+\log(1/k)]+O(k^3)$, so with the gap at $k=0$ open
	\begin{equation}
		E_2(k)=E_2(0)+g(0)\pi J|k|+Bk^2\bigl[1+\log(1/|k|)\bigr]^2-g(0)\frac J2k^2+\ldots, \qquad B=\frac{2J^2}{E_2(0)}.
		\label{eq:E2}
	\end{equation}
	The band-edge speed is $v_0=\pi J$, and $E_2$ departs from its tangent by $Bk^2\log^2(1/|k|)$. Take $g(0)=-1$, that is $\theta>\theta_0$ (the other case conjugates the kernels); the band edge is a kink, and as in Sec.~\ref{sec:critical} the right-moving waves come from one side only. With $k=u/\ell_2$ and $L=\log\ell_2$ the departure is $Bt\ell_2^{-2}u^2(L+1-\log u)^2$, of order one for $u\sim1$ when $BtL^2=\ell_2^2$:
	\begin{equation}
		\frac{\ell_2}{\log\ell_2}=\sqrt{Bt}, \qquad \ell_2\simeq\tfrac12\sqrt{Bt}\log(Bt),
		\label{eq:l2}
	\end{equation}
	Eq.~(10) of the main text. Since $\ell/\log\ell$ has its minimum $e$ at $\ell=e$, there are two solutions for $\sqrt{Bt}>e$, and the width is the larger; inserting $\ell_2\simeq\sqrt{Bt}$ on the right of $\ell_2=\sqrt{Bt}\log\ell_2$ gives the second form. With $s=(x-\pi Jt)/\ell_2$ the rescaled phase is $us+u^2[(1+(1-\log u)/L)^2+E_2(0)/(4JL^2)]$, quadratic as $L\to\infty$, so the profile is a Fresnel integral. With the weight of the gapped band edge, Eq.~\eqref{eq:Dfracbulk},
	\begin{equation}
		D_j(t)\simeq\frac{1-mg(0)}{\sqrt{1-m^2}}\ell_2^{-2}|\mathcal F_L(s)|^2, \qquad \mathcal F(s)=\int_0^\infty\frac{du}{2\pi}e^{i(us+u^2)}=\frac{e^{-is^2/4}}{2\pi}\int_{s/2}^\infty e^{iw^2}dw,
		\label{eq:Fresnel}
	\end{equation}
	where $\mathcal F_L$ is $\mathcal F$ with the full rescaled phase at finite $L$, integrated over the band-edge window with a smooth cutoff well below $u=\ell_2$. For the numerical curves in Fig.~\ref{fig:longrange}(d), we multiply the integrand by $e^{-(u/60)^2}$ and use a rectangular sum over $1\,200\,001$ uniformly spaced points from $u=10^{-6}$ to $u=120$, including both endpoints, with spacing $\Delta u=(120-10^{-6})/1\,200\,000$. The two finite-$L$ curves use $L=6.2$ and $L=9$. Ahead of the front $\mathcal F\simeq i/(2\pi s)$; behind it $|\mathcal F|^2\to1/(4\pi)$. The corrections are of relative order $1/L$, so $|\mathcal F_L|^2\to|\mathcal F|^2$ only as $\log t\to\infty$: at $L=6.2$ the first maximum of $|\mathcal F_L|^2$ still exceeds that of $|\mathcal F|^2$ by a factor $1.4$ [Fig.~\ref{fig:longrange}(d)]. Hence Fig.~2(f) of the main text ($L=6.24$, $6.65$, $7.05$) compares the data with $|\mathcal F_L|^2$ at each time, using the cutoff $e^{-(u/50)^8}$ and trapezoidal quadrature on $0\leq u\leq150$ with $\Delta u=0.004$. At an edge, case (iii) gives $(f^a_k)^2\simeq2JE_2(0)e^{-\mathcal I(E_2(0)^2)}k$ and $D_j\simeq2\pi JE_2(0)e^{-\mathcal I(E_2(0)^2)}\mathcal N_0\ell_2^{-3}|\mathcal F^{\mathrm e}(s)|^2/\sqrt{1-m_{\mathrm e}^2}$, with $\mathcal F^{\mathrm e}(s)=\int_0^\infty\frac{du}{2\pi}u^{1/2}e^{i(us+u^2)}$ and $s=(j+1-\pi Jt)/\ell_2$: the height $\ell_2^{-3}$ of Table~I.
	
	The band-edge waves lead only if no interior momentum is faster; comparing $\pi J$ with the largest $|E_2'|$ inside the band, a root search gives $\theta_{\mathrm d}=0.6467$, above which the band edge leads. Since $\theta_0=0.5462<\theta_{\mathrm d}<\theta_\pi=2.2591$, the width law switches from $t^{1/3}$ to $\sqrt t\log t$ inside the topological phase: it follows the fastest waves, not the phase. At $\theta_{\mathrm d}$ an Airy wave from the interior maximum $k_{\mathrm d}$ and a Fresnel wave from the band edge share the speed $v_0=\pi J$. Their widths are $\ell_{\mathrm A}=(|E_2'''(k_{\mathrm d})|t/2)^{1/3}$ and $\ell_2$, with $\ell_2/\ell_{\mathrm A}\propto t^{1/6}\log t$. Put $s_{\mathrm A}=(x-v_0t)/\ell_{\mathrm A}$, $s=(x-v_0t)/\ell_2$, $\chi_{\mathrm A}=k_{\mathrm d}x-E_2(k_{\mathrm d})t$, and $r_{\mathrm d}=\re g(k_{\mathrm d})$. The two scalar waves and the resulting density response are
	\begin{equation}
		\begin{aligned}
			I_{\mathrm A}&=e^{i\chi_{\mathrm A}}\ell_{\mathrm A}^{-1}\Ai(s_{\mathrm A}), \qquad I_{\mathrm B}=e^{iE_2(0)t}\ell_2^{-1}\mathcal F_L(s), \\
			\mathcal X_{\mathrm d}&=-r_{\mathrm d}(1-mr_{\mathrm d})|I_{\mathrm A}|^2+(1+m)|I_{\mathrm B}|^2+(1+m)(1-r_{\mathrm d})\re(I_{\mathrm A}I_{\mathrm B}), \\
			\mathcal Y_{\mathrm d}&=2(1+r_{\mathrm d})\re(I_{\mathrm A}\overline{I_{\mathrm B}}), \qquad D_j(t)\simeq\Bigl[\frac{\mathcal X_{\mathrm d}^2}{1-m^2}+\frac{\mathcal Y_{\mathrm d}^2}4\Bigr]^{1/2}.
		\end{aligned}
		\label{eq:twowidthdensity}
	\end{equation}
	These expressions follow by inserting both waves and their conjugates into Eq.~\eqref{eq:double}, with the weights of Eq.~\eqref{eq:Wdensity}, before taking the outcome root mean square. The terms containing both waves retain their relative phases and are the interference contribution. Since the two waves depend on $s_{\mathrm A}$ and $s$, the front needs two scaling variables.
	
	\textit{Gapped, $\alpha>2$: an interior maximum, and the circle deformed.} The band-edge speed vanishes, Eq.~\eqref{eq:v0}, so an interior maximum $k_*$ of $|E_\alpha'|$ leads, generically with $E_\alpha'''(k_*)\neq0$: an Airy front, $\ell=(|E_\alpha'''(k_*)|t/2)^{1/3}$, with $\sigma=0$ at an edge by case (i). What changes is the cancellation: the long-range tail deforms the circle behind Eq.~\eqref{eq:complement}, and generically $\re g(k_*)\neq0$. The density keeps its leading term, Eq.~\eqref{eq:Dtriv} with this $\re g(k_*)$, and decays as $t^{-2/3}$; at an edge, by Eqs.~\eqref{eq:edgeweights} and \eqref{eq:bweight},
	\begin{equation}
		D_j(t)\simeq\frac{\pi(f^a_{k_*})^2\mathcal N_*|\re g(k_*)|}{\sqrt{1-m_{\mathrm e}^2}}\ell^{-2}\Ai(s)^2, \qquad \mathcal N_*=(\omega_*+m_{\mathrm e})^2+1-m_{\mathrm e}^2,
		\label{eq:lredge}
	\end{equation}
	which for nearest neighbors is Eq.~\eqref{eq:Dedgetriv}. Where $|\re g(k_*)|$ is small the $\ell^{-4}$ terms compete until $\ell^2|\re g(k_*)|\sim1$, as in Sec.~\ref{sec:aniso}. The cancellation survives on a line $\theta_{\mathrm b}(\alpha)$ in the topological phase where $\re g(k_*)=0$ (Fig.~\ref{fig:lrmap}); a root search gives its start at $\alpha\simeq2.110$, $\theta\simeq0.817$, and $\theta_{\mathrm b}=0.95761$, $1.08213$, $1.24305$, and $1.41591$ at $\alpha=2.5$, $3$, $4$, and $6$, approaching $\pi/2$ as $\alpha\to\infty$. On the line $\nu=2$, but energy and observable no longer come from one circle, the condition $\varphi''=-\Xi$ below Eq.~\eqref{eq:secondgeneral} fails, and $\mathcal B$ mixes with $s\Ai^2$. At $\alpha=3$, where $k_*=0.91781$, $v=0.73079$, $E_\alpha'''(k_*)=-1.0801$, $E_\alpha''''(k_*)=2.9327$, $g(k_*)=-i$, and $m=0.3271$, Eq.~\eqref{eq:secondgeneral} and the equal-frequency pairs give
	\begin{equation}
		\overline D_j(t)\simeq\ell^{-4}\Bigl[\bigl(0.3178\,\mathcal B-0.0691\,s\Ai^2\bigr)^2+\bigl(0.4308\,\mathcal B\bigr)^2\Bigr]^{1/2}, \qquad \ell=(0.5400\,t)^{1/3}, \qquad s=\frac{x-0.73079\,t}\ell.
		\label{eq:Dblind3}
	\end{equation}
	For a single regular fastest momentum on the line $\theta_{\mathrm b}$, the edge density profile and prefactor also follow explicitly. Define
	\begin{equation}
		\mathcal A_{\mathrm e}(k)=\frac\pi2|f^a_k|^2\bigl[1+2m_{\mathrm e}\omega_k+\omega_k^2\bigr], \qquad \omega_k=\frac\mu{E_\alpha(k)}, \qquad \varphi(k)=\arg g(k).
		\label{eq:edgeblindweight}
	\end{equation}
	With $\Xi$ from Eq.~\eqref{eq:secondgeneral}, the result is
	\begin{equation}
		\overline D_j(t)\simeq\ell^{-4}\Biggl[\frac{\Bigl[(\mathcal A_{\mathrm e}'\varphi'+\mathcal A_{\mathrm e}\Xi)\mathcal B(s)-\mathcal A_{\mathrm e}(\varphi''+\Xi)s\Ai(s)^2\Bigr]^2}{1-m_{\mathrm e}^2}+\frac{\pi^2}2|f^a_{k_*}|^4(\varphi'\omega')^2\mathcal B(s)^2\Biggr]^{1/2}.
		\label{eq:Dblindedge}
	\end{equation}
	All coefficients and derivatives are evaluated at $k_*$, holding the Hamiltonian parameters and $m_{\mathrm e}$ fixed. Use the Airy scale of Eq.~\eqref{eq:airyscale}, with $x=j+1$. The first square comes from the opposite-frequency pairs. The second comes from the equal-frequency carrier after averaging its squared cosine to $1/2$. Boundary-phase derivatives multiply $(u_1-u_2)(u_1+u_2)$ and integrate to zero, while a constant boundary phase drops out of the average. For nearest neighbors this expression reproduces Eq.~\eqref{eq:Dedgetopo}.
	
	On the dashed line of Fig.~\ref{fig:lrmap} two interior maxima are equally fast, and two Airy fronts of different widths share one ray.
	
	\textit{Closing the gap at $k=0$ for $\alpha>2$.} At $\mu=-2J\zeta(\alpha)$, $h=2J[\Li_\alpha(e^{ik})-\zeta(\alpha)]$, and $|E_\alpha'|\leq|h'|=2J|\Li_{\alpha-1}(e^{ik})|\leq2J\zeta(\alpha-1)$, a bound attained by $h\simeq2iJ\zeta(\alpha-1)k$ as $k\to0^+$: the front moves at $v_{\mathrm c}=2J\zeta(\alpha-1)$. With $h/(2J)=\mathcal X+i\mathcal Y$ for $k>0$, Eq.~\eqref{eq:polyexp} gives
	\begin{equation}
		\mathcal X=\Gamma(1-\alpha)\cos\frac{\pi\beta}2k^\beta-\frac{\zeta(\alpha-2)}2k^2+\ldots, \qquad \mathcal Y=\zeta(\alpha-1)k-\Gamma(1-\alpha)\sin\frac{\pi\beta}2k^\beta-\frac{\zeta(\alpha-3)}6k^3+\ldots,
		\label{eq:XY}
	\end{equation}
	and $E_\alpha=2J[\mathcal Y+\mathcal X^2/(2\mathcal Y)+\ldots]$. For $2<\alpha<4$ the term $k^\beta$ of $\mathcal Y$ is the lowest power above one, so $E_\alpha\simeq v_{\mathrm c}k-C_{\mathrm c}k^{\alpha-1}$ with $C_{\mathrm c}=2J\Gamma(1-\alpha)\sin[\pi(\alpha-1)/2]>0$ and $\ell=(C_{\mathrm c}t)^{1/z_{\mathrm c}}$, $z_{\mathrm c}=\alpha-1$: the exponent $1/z_{\mathrm c}$ falls continuously from $1$ to $\frac13$. At $\alpha=4$ the term $k^3\log(1/k)$ gives $\ell\propto(t\log t)^{1/3}$; for $\alpha>4$ the analytic $k^3$ term leads, with $C_{\mathrm c}=2J[\zeta(\alpha-3)/6-\zeta(\alpha-2)^2/(8\zeta(\alpha-1))]>0$ and $\ell\propto t^{1/3}$; hence $z_{\mathrm c}=\min(\alpha-1, 3)$. Since $g\to\mp i$ for $k\to0^\pm$, the density weight vanishes at the closing,
	\begin{equation}
		\begin{aligned}
			&\re g(k)=-\frac{\mathcal X}{(\mathcal X^2+\mathcal Y^2)^{1/2}}\simeq r_{\mathrm c}|k|^{\nu_{\mathrm c}}, \qquad \nu_{\mathrm c}=\min(\alpha-2, 1), \\
			&r_{\mathrm c}=-\frac{\Gamma(1-\alpha)\cos(\pi\beta/2)}{\zeta(\alpha-1)}\ (2<\alpha<3), \qquad r_{\mathrm c}=\frac{\zeta(\alpha-2)}{2\zeta(\alpha-1)}\ (\alpha>3),
		\end{aligned}
		\label{eq:rcrit}
	\end{equation}
	with $r_{\mathrm c}=[\log(1/|k|)+\frac32]/[2\zeta(2)]$ at $\alpha=3$: the fastest quasiparticles are again equal mixtures of particle and hole. The pair weight of Eq.~\eqref{eq:Wdensity} becomes $-[\re g(k)+\re g(k')]/(4p_\eta)$, and with the one-sided integrals $\mathcal A^{\mathrm c}_\nu(s)=\int_0^\infty\frac{du}{2\pi}u^\nu e^{i(us+u^{z_{\mathrm c}})}$ of Sec.~\ref{sec:critical},
	\begin{equation}
		\delta\Gamma_{a_jb_j}\big|_{\mathrm{front}}\simeq-\frac{\ell^{-2-\nu_{\mathrm c}}}{p_\eta}\Phi_{\mathrm f}(s), \qquad \Phi_{\mathrm f}=r_{\mathrm c}\re\bigl(\mathcal A^{\mathrm c}_{\nu_{\mathrm c}}\bar{\mathcal A}^{\mathrm c}_0\bigr), \qquad D_j\simeq\frac{\ell^{-2-\nu_{\mathrm c}}|\Phi_{\mathrm f}(s)|}{\sqrt{1-m^2}},
		\label{eq:k0front}
	\end{equation}
	the height $\ell^{-2-\nu_{\mathrm c}}$ of the main text. The jump of $g$ at $k=0$ also gives the propagated correlations an algebraic tail (Sec.~\ref{sec:critical}), of size $1/(4\pi v_{\mathrm c}t)$ near the front, and its product with the front wave, the right-moving front times the tail of the left-moving wave, is of order $\ell^{-1}t^{-1}\propto\ell^{-\alpha}$. For $2<\alpha<3$, where $2+\nu_{\mathrm c}=\alpha$, it changes the profile but not the power:
	\begin{equation}
		D_j(t)\simeq\ell^{-\alpha}\Bigl[\frac{(\Phi_{\mathrm f}+m\Phi_{\mathrm d})^2}{1-m^2}+\Phi_{\mathrm d}^2\Bigr]^{1/2}, \qquad \Phi_{\mathrm d}=\kappa_{\mathrm f}\im\mathcal A^{\mathrm c}_0, \qquad \kappa_{\mathrm f}=\frac{C_{\mathrm c}}{2\pi v_{\mathrm c}}=-\frac1{4\Gamma(\alpha)\zeta(\alpha-1)\sin(\pi\alpha/2)}>0.
		\label{eq:k0profile}
	\end{equation}
	For $3<\alpha<4$ the tail is smaller than the front term by a relative factor $\ell^{3-\alpha}$, and for $\alpha>4$ by $\ell^{-1}$. At the $k=0$ closing with $\alpha>2$, the edge amplitude obeys $f^b_k\to-\sqrt{2/\pi}$, case (iv), and the one-sided calculation gives $D_j(t)\simeq2\ell^{-2-\nu_{\mathrm c}}|\Phi_{\mathrm f}(s)|/\sqrt{1-m_{\mathrm e}^2}$, with $s=(j+1-v_{\mathrm c}t)/\ell$. For $2<\alpha<3$, replace $\Phi_{\mathrm f}(s)$ by $\Phi_{\mathrm f}(s)+\kappa_{\mathrm f}\re\mathcal A^{\mathrm c}_0(s)$ in this edge expression. Bulk and edge therefore share the power $\ell^{-2-\nu_{\mathrm c}}$; the logarithmic cases $\alpha=3$ and $\alpha=4$ are specified below.
	At $\alpha=3$ the finite limit of $C_{\mathrm c}$ and the leading logarithmic observable factor give
	\begin{equation}
		C_{\mathrm c}=\frac{\pi J}2, \qquad \ell=\Bigl(\frac{\pi Jt}2\Bigr)^{1/2}, \qquad \Phi_{\mathrm f}(s)\simeq\frac{\log\ell}{2\zeta(2)}\re\bigl(\mathcal A^{\mathrm c}_1\bar{\mathcal A}^{\mathrm c}_0\bigr), \qquad z_{\mathrm c}=2.
		\label{eq:k0alpha3}
	\end{equation}
	Thus the height is $\ell^{-3}\log\ell$; the nonlogarithmic front and tail terms are smaller by a relative factor $O(1/\log\ell)$. At the edge the leading result is doubled with $m\to m_{\mathrm e}$. At $\alpha=4$, for $k>0$,
	\begin{equation}
		E_\alpha(k)=v_{\mathrm c}k-\frac J3k^3\log(1/k)+O(k^3), \qquad \ell^3=\frac{Jt}3\log\ell, \qquad r_{\mathrm c}=\frac{\zeta(2)}{2\zeta(3)}.
		\label{eq:k0alpha4}
	\end{equation}
	The width is the larger solution, replacing $(C_{\mathrm c}t)^{1/z_{\mathrm c}}$ at this value of $\alpha$. Its scaled phase approaches $su+u^3$ with corrections of order $1/\log\ell$, so the kernels $\mathcal A^{\mathrm c}_\nu$ with $z_{\mathrm c}=3$ give the leading profile.
	
	\textit{Closing the gap at $k=\pi$.} At $\mu=2J\eta_{\mathrm D}(\alpha)$, with $k=\pi+\kappa$, the Taylor series of Sec.~\ref{sec:polylog} gives
	\begin{equation}
		E_\alpha\simeq v_\pi|\kappa|-C_\pi|\kappa|^3, \qquad v_\pi=2J\eta_{\mathrm D}(\alpha-1)=2J(1-2^{2-\alpha})\zeta(\alpha-1), \qquad \re g\simeq-r_\pi|\kappa|, \qquad r_\pi=\frac{\eta_{\mathrm D}(\alpha-2)}{2\eta_{\mathrm D}(\alpha-1)},
		\label{eq:kpikin}
	\end{equation}
	with $C_\pi=2J[\eta_{\mathrm D}(\alpha-3)/6-\eta_{\mathrm D}(\alpha-2)^2/(8\eta_{\mathrm D}(\alpha-1))]$: a linear closing, analytic around it, whose density weight vanishes linearly. For nearest neighbors $\eta_{\mathrm D}\to1$, $v_\pi=2J$, $C_\pi=J/12$, $r_\pi=\frac12$: the critical chain of Sec.~\ref{sec:critical}. These waves launch the front only if no momentum is faster: near $k=\pi$ this needs $C_\pi>0$, true for $\alpha>2.2444$, and on the whole zone a root search gives $\alpha>\alpha_\pi=2.283$. Then the one-sided analysis of Sec.~\ref{sec:critical} applies with $\ell=(3C_\pi t)^{1/3}$, and the edge doubles it, case (iv):
	\begin{equation}
		D_j\simeq\frac{r_\pi\ell^{-3}\mathcal Q(s)}{4\pi\sqrt{1-m^2}}, \qquad D_j\simeq\frac{r_\pi\ell^{-3}\mathcal Q(s)}{2\pi\sqrt{1-m_{\mathrm e}^2}}\ \text{(edge)}, \qquad s=\frac{x-v_\pi t}\ell,
		\label{eq:Dkpi}
	\end{equation}
	with $x\to j+1$ at the edge; for $r_\pi=\frac12$ these are Eqs.~\eqref{eq:Dcrit} and \eqref{eq:Dedgecrit}. Figure~2(c) of the main text tests the bulk front at $\alpha=3$.
	
	Below $\alpha_\pi$ an interior maximum or, for $\alpha\leq2$, the band edge at $k=0$ is faster, and the gapped rows of Table~I apply: the jump of $g$ at $k=\pi$ adds only an algebraic tail of relative size $\ell^{1+\nu}/t\to0$, except for $1<\alpha<2$, where in the bulk it adds $\eta/(2\pi^2x^2)$ to $p_\eta\delta\Gamma_{a_jb_j}$, a new profile at the same height, and at an edge it changes the height. The correlation $\Gamma_{0, a_jb_0}=-\int_0^\pi dk\,f^b_k\psi_k(j)$ has an endpoint at $k=\pi$, where $f^b_k$ stays finite, $(f^b_\pi)^2=2/\pi$ by case (iv), and one integration by parts gives
	\begin{equation}
		\bigl|\langle ia_jb_0\rangle_0\bigr|=\bigl|\Gamma_{0, a_jb_0}\bigr|\simeq\frac2{\pi j}\qquad(j\gg1),
		\label{eq:Tedgepi}
	\end{equation}
	while the endpoint $k=0$, where $f^b_k$ vanishes, gives a faster decay. This frozen correlation multiplies the edge front wave $w_j(t)=\int_0^\pi dk\,f^a_k\cos(E_\alpha(k)t)\psi^b_k(j)$, of size $\ell_\beta^{-1-\beta/2}$, and with $\sum_\eta(\omega_0-\eta)^2/(4p_\eta)=\mathcal N_0/(1-m_{\mathrm e}^2)$,
	\begin{equation}
		D_j(t)\simeq\frac{\sqrt{\mathcal N_0}}{\pi\sqrt{1-m_{\mathrm e}^2}}\frac{|w_j(t)|}j\propto\ell_\beta^{-2-\beta/2},
		\label{eq:Dkpiedge}
	\end{equation}
	since $j\sim\ell_\beta$: one front wave times a frozen correlation, instead of two front waves, raises the edge height from $\ell_\beta^{-2-\beta}$ to $\ell_\beta^{-2-\beta/2}$. On open chains at $\alpha=1.5$ and $1.8$ the exact $|\Gamma_{0, a_jb_0}|\,j$ lies between $0.617$ and $0.669$ for $50\leq j\leq400$ ($2/\pi=0.637$), and the exact edge response lies within a factor $0.8$ to $1.5$ of Eq.~\eqref{eq:Dkpiedge}, depending on the phase of a beat at the frequency $E_\alpha(0)$.
	
	\textit{How fast the limits are reached.} The curves in Fig.~2 of the main text approach these leading profiles with corrections that vanish as regime-dependent powers of $\ell$; these powers can be noninteger in the fractional regime. At $\alpha=2$ the expansion parameter is $1/\log\ell_2$, hence the slow approach in Fig.~2(f).

	\section{How the figures are computed, and where each result is derived}
	\label{sec:numerics}
	
	\textit{The exact response.} Every exact curve evaluates Eq.~\eqref{eq:exact}; by Eq.~\eqref{eq:update} this needs $R(t)$ applied to four vectors, $e_{a_{j_0}}$, $e_{b_{j_0}}$, $\Gamma_0e_{a_{j_0}}$, and $\Gamma_0e_{b_{j_0}}$, each one fast Fourier transform on a ring, Eq.~\eqref{eq:Rk}. The two fronts must not meet around the ring, $N>2vt$, and where $\ell_\beta$ grows faster than $t$ we use $N=2^{22}$. At a gap closing a periodic ring contains the momentum where $E$ vanishes, an exact zero mode; an antiperiodic ring, $c_{j+N}=-c_j$, shifts every momentum by $\pi/N$ and avoids it. Nearest-neighbor half-lines use the standing waves of Eq.~\eqref{eq:hardwall} with $Q=2^{20}$ momenta, long-range ones the singular value decomposition of $h$. The bulk $Z_{j_0}$ at $\lambda=2$ uses an open chain of $N=3000$ sites with $j_0=750$, since on a ring the sign flip of Sec.~\ref{sec:string} would need a second cut, and the critical $Z_{j_0}$ uses the infinite-chain sum of Eq.~\eqref{eq:Zsum}. The numerical parameters of Figs.~1 and 2 of the main text are listed in Table~\ref{tab:mainparameters}. Except for the first-site measurements and the specified open chain, the measured site is in the bulk. In Fig.~1 and in the nearest-neighbor Kitaev and cluster curves of Fig.~2, $J=\Delta=1$. The long-range closing at $k=\pi$ also uses $J=\Delta=1$, while the other long-range curves use $J=\Delta=\sin\theta$ and $\mu=-2\cos\theta$. Times below are $t$, except where $Jt$ is written explicitly.
	
	\begin{table}[!htb]
		\centering
		\caption{Parameters of Figs.~1 and 2 of the main text. Ring sizes $N$ count physical sites, so a cluster ring has $N/2$ cells. A half-line is evaluated with $Q$ standing-wave momenta. Ring boundary conditions are periodic unless marked antiperiodic. All times in each listed tuple are used.}
		\label{tab:mainparameters}
		\begin{ruledtabular}
			\begin{tabular}{p{0.08\textwidth}p{0.30\textwidth}p{0.25\textwidth}p{0.27\textwidth}}
				Figure & Protocol and parameters & Times & Computation \\
				\hline
				1 & Kitaev $n_{j_0}\to n_j, B_j$, $\mu=J$ & $Jt=1000, 8000$ & Ring $N=2^{16}$ \\
				1(d) & Heights at $s=-1.02$ & $Jt=30(20000/30)^{r/22}$, $r=0, \ldots, 22$ & Same ring \\
				2(a) & Kitaev $n_{j_0}, n_0\to n_j$, $\mu=3$ & $(1600, 6400, 12800)$ & Ring $N=2^{18}$; half-line $Q=2^{20}$ \\
				2(a) & Cluster $X_{j_0}, X_0, Z_0\to K_j$, $\lambda=0.5$ & $(1600, 6400, 12800)$ & Ring $N=2^{17}$; half-line $Q=2^{20}$ \\
				2(b) & Kitaev $n_{j_0}, n_0\to n_j$, $\mu=1$ & $(1600, 6400, 12800)$ & Ring $N=2^{18}$; half-line $Q=2^{20}$ \\
				2(b) & Cluster $X_{j_0}, X_0, Z_0\to K_j$, $\lambda=2$ & $(800, 3200, 6400)$ & Ring $N=2^{17}$; half-line $Q=2^{20}$ \\
				2(b) & Cluster $Z_{j_0}\to K_j$, $\lambda=2$ & $(150, 300, 500)$ & Open chain $N=3000$, $j_0=750$ \\
				2(c) & Kitaev $n_{j_0}\to n_j$, $\mu=2$ & $(1600, 6400, 12800)$ & Antiperiodic ring $N=2^{18}$ \\
				2(c) & Cluster $X_{j_0}\to K_j$, $\lambda=1$ & $(800, 3200, 6400)$ & Antiperiodic ring $N=2^{17}$ \\
				2(c) & Kitaev $n_0\to n_j$, $\mu=2$; cluster $X_0, Z_0\to K_j$, $\lambda=1$ & $(6400, 12800, 25600)$ & Half-lines $Q=2^{20}$ \\
				2(c) & Long-range $n_{j_0}\to n_j$, $\alpha=3$, $\mu=2\eta_{\mathrm D}(3)$ & $(2000, 8000, 32000)$ & Antiperiodic ring $N=2^{18}$ \\
				2(d) & Cluster $Z_{j_0}\to K_j$, $\lambda=1$ & $(1600, 3200, 6400)$ & Infinite-chain Bessel sum \\
				2(e) & Long-range $n_{j_0}\to n_j$, $\alpha=1.8$, $\theta=1$ & $(128, 256, 512)$ & Ring $N=2^{22}$ \\
				2(f) & Long-range $n_{j_0}\to n_j$, $\alpha=2$, $\theta=1$ & $(8000, 16000, 32000)$ & Ring $N=2^{19}$ \\
			\end{tabular}
		\end{ruledtabular}
	\end{table}
	
	For Fig.~1, positions are $x=j-j_0$ for the density and $x=j+\frac12-j_0$ for the bond energy. The overline is the root mean square over three consecutive sites of both observables. In Fig.~2, the Kitaev curves at $\mu=J$ use the same three-site average; the bulk $Z_{j_0}$ and edge $X_0$ cluster curves at $\lambda=2$ use a three-cell average. Each exact curve in Fig.~2 is divided by its derived height and prefactor, with no fitted parameter; Table~\ref{tab:entries} gives the corresponding equations.
	
	\begin{table}[!htb]
		\caption{Every entry of Table~I of the main text: width, exponents of Eq.~\eqref{eq:rule}, profile, and the equations that derive the law with its prefactor for a bulk and a first-site (edge) measurement; Fig.~2 of the main text divides each exact front by the height and prefactor of the listed equation. A dash: Eq.~\eqref{eq:rule} does not apply (Sec.~\ref{sec:Zcrit}). Cluster widths in sites; $\sigma$ of the edge in the lower block, where $\beta=\alpha-1$, $z_{\mathrm c}=\min(\alpha-1, 3)$, $\nu_{\mathrm c}=\min(\alpha-2, 1)$, and $\nu=2$ on the line $\theta_{\mathrm b}$. At the closing at $k=0$, $\alpha=3$ has the additional height factor $\log\ell$ of Eq.~\eqref{eq:k0alpha3}, and the width at $\alpha=4$ is defined by Eq.~\eqref{eq:k0alpha4}.}
		\label{tab:entries}
		\begin{ruledtabular}
			\begin{tabular}{lllcclll}
				Measured $\to$ observed & Phase & $\ell$ & $\nu$ & $\sigma$ & Profile & Bulk & Edge \\
				\hline
				Kitaev $n\to n$ & topological & $(\mu t/2)^{1/3}$ & $2$ & $0$ & $\mathcal B$ & \eqref{eq:Dtopo} & \eqref{eq:Dedgetopo} \\
				& critical & $(Jt/4)^{1/3}$ & $1$ & $0$ & $\mathcal Q$ & \eqref{eq:Dcrit} & \eqref{eq:Dedgecrit} \\
				& trivial & $(Jt)^{1/3}$ & $0$ & $0$ & $\Ai^2$ & \eqref{eq:Dtriv} & \eqref{eq:Dedgetriv} \\
				Kitaev $n\to B$ & topological & $(\mu t/2)^{1/3}$ & $0$ & $0$ & $\Ai^2$ & \eqref{eq:DBtopo} & Sec.~\ref{sec:edge} \\
				& critical & $(Jt/4)^{1/3}$ & $1$ & $0$ & $\mathcal Q$ & \eqref{eq:Dcrit} & Sec.~\ref{sec:edge} \\
				& trivial & $(Jt)^{1/3}$ & $2$ & $0$ & $\mathcal B$ & \eqref{eq:DBtriv} & Sec.~\ref{sec:edge} \\
				Cluster $X$; $X_0$, $Z_0\to K$ & cluster & $(8\lambda t)^{1/3}$ & $0$ & $0$ & $\Ai^2$ & \eqref{eq:DX} & \eqref{eq:Dedgecl} \\
				& critical & $(2t)^{1/3}$ & $1$ & $0$ & $\mathcal Q$ & \eqref{eq:DX} & Sec.~\ref{sec:clusteredge} \\
				& paramagnet & $(8t)^{1/3}$ & $2$ & $0$ & $\mathcal B$ & \eqref{eq:DX} & Sec.~\ref{sec:clusteredge} \\
				Cluster $Z\to K$ & cluster & $(8\lambda t)^{1/3}$ & $0$ & $0$ & $\Ai^2$ & \eqref{eq:DZgapped} & \eqref{eq:Dedgecl} \\
				& critical (bulk) & $(2t)^{1/3}$ & -- & -- & $\mathcal Z$ & \eqref{eq:DZcrit} & n/a \\
				& critical (edge) & $(2t)^{1/3}$ & $1$ & $0$ & $\mathcal Q$ & n/a & Sec.~\ref{sec:clusteredge} \\
				& paramagnet & $(8t)^{1/3}$ & $2$ & $0$ & $\mathcal B$ & \eqref{eq:DZpara} & Sec.~\ref{sec:clusteredge} \\
				\hline
				Long-range $n\to n$ & gapped, $1<\alpha<2$ & $(|c_\beta|t)^{1/\beta}$ & $0$ & $\beta/2$ & $|\mathcal K_\beta|^2$ & \eqref{eq:Dfracbulk} & \eqref{eq:Dfracedge} \\
				& gapped, $\alpha=2$, $k\to0$ & $\ell_2$ of \eqref{eq:l2} & $0$ & $\frac12$ & $|\mathcal F_L|^2$ & \eqref{eq:Fresnel} & Sec.~\ref{sec:lrfronts} \\
				& gapped, $\alpha\geq2$, interior & $(|E_\alpha'''|t/2)^{1/3}$ & $0$ & $0$ & $\Ai^2$ & \eqref{eq:Dtriv} & \eqref{eq:lredge} \\
				& closing at $k=0$, $1<\alpha<2$ & $(C_{\mathrm f}t)^{1/\beta}$ & $0$ & $0$ & Eq.~\eqref{eq:critfracbulk} & \eqref{eq:critfracbulk} & Sec.~\ref{sec:lrfronts} \\
				& closing at $k=0$, $\alpha>2$ & $(C_{\mathrm c}t)^{1/z_{\mathrm c}}$ & $\nu_{\mathrm c}$ & $0$ & $\Phi_{\mathrm f}$ & \eqref{eq:k0front}, \eqref{eq:k0profile} & Sec.~\ref{sec:lrfronts} \\
				& closing at $k=\pi$, $\alpha>\alpha_\pi$ & $(3C_\pi t)^{1/3}$ & $1$ & $0$ & $\mathcal Q$ & \eqref{eq:Dkpi} & \eqref{eq:Dkpi} \\
			\end{tabular}
		\end{ruledtabular}
	\end{table}
	
	\textit{What is derived and what is checked.} Every power in Table~\ref{tab:entries} is derived, and the profiles and prefactors are given above in terms of explicit integrals and derivatives, including the edge result on $\theta_{\mathrm b}$ in Eq.~\eqref{eq:Dblindedge} and the two-width response in Eq.~\eqref{eq:twowidthdensity}. The boundary weights are given by Eqs.~\eqref{eq:bweight} and \eqref{eq:channelweights}; $\theta_{\mathrm d}$, $\theta_{\mathrm b}$, and $\alpha_\pi$ are numerical roots. The exact response tests the upper block except bulk $Z_{j_0}$ in the cluster phase (Figs.~1 and 2 of the main text, Figs.~\ref{fig:edgebond} and \ref{fig:clusterfronts}), the gapped bulk rows for $1<\alpha<2$ and $\alpha=2$, and the bulk closing at $k=\pi$; the other long-range entries are derived only.
	
	\textit{Experiments.} The experimental ingredients are discussed in the main text: mid-circuit measurement in neutral-atom arrays~\cite{Deist2022, Graham2023, Lis2023}, trapped ions~\cite{Pino2021}, and superconducting processors~\cite{Koh2023}; transverse-field Ising dynamics on ion and Rydberg platforms~\cite{Monroe2021, Browaeys2020} and, in kicked form, on a qubit chain~\cite{Mi2022}; site-resolved readout~\cite{Gross2021, Fukuhara2013}; and two- and three-site Kitaev chains in coupled quantum dots~\cite{Dvir2023, tenHaaf2025}.